\documentclass[useAMS,usenatbib, usegraphicx]{mn2e}

\usepackage{verbatim} 
\usepackage[usenames,dvipsnames]{color}
\usepackage{amssymb, amsmath}
\usepackage{multicol}
\usepackage{longtable}
\usepackage{rotating}
\usepackage{pdflscape}
\usepackage{url}
\usepackage{times}

\usepackage{graphicx}
\usepackage{epstopdf}

\usepackage{enumitem}
\usepackage[justification=centering]{caption}
\usepackage[]{subfigure}

\usepackage{natbib}
\bibpunct{(}{)}{;}{a}{}{,}

\usepackage{array}
\newcolumntype{x}[1]{>{\centering\arraybackslash\hspace{0pt}}m{#1}}

\makeatletter
\def\url@leostyle{%
  \@ifundefined{selectfont}{\def\UrlFont{\sf}}{\def\UrlFont{\small\ttfamily}}}
\makeatother
\DeclareRobustCommand{\ion}[2]{%
\relax\ifmmode
\ifx\testbx\f@series
{\mathbf{#1\,\mathsc{#2}}}\else
{\mathrm{#1\,\mathsc{#2}}}\fi
\else\textup{#1\,{\mdseries\textsc{#2}}}%
\fi}

\title[GAMA: Bivariate Functions of SF Galaxies]{\color{black}Galaxy And Mass Assembly (GAMA): Bivariate functions of H$\alpha$ star forming galaxies}

\color{black} \author[Gunawardhana \it{et.\,al}]{M.\,L.\,P.\,Gunawardhana$^{1,2,3}$\thanks{E-mail: madusha.gunawardhana@durham.ac.uk}, A.\,M.\,Hopkins$^{2}$\thanks{ahopkins@aao.gov.au}, E.\,N.\,Taylor$^{4}$, J.\,Bland--Hawthorn$^{1}$ 
\newauthor
P.\,Norberg$^{3}$, I.K.\,Baldry$^{5}$, J.\,Loveday$^6$, M.S.\,Owers$^{2}$, S.M.\,Wilkins$^{6}$, M.\,Colless$^{7}$
\newauthor
M.J.I.\,Brown$^{8}$, S.P.\,Driver$^{9,10}$, M.\,Alpaslan$^{10}$, S.\,Brough$^{2}$, M.\,Cluver$^{2}$, S.\,Croom$^{1}$
\newauthor
L.\,Kelvin$^{11}$, M.\,A.\,Lara-L\'opez$^2$, J.\,Liske$^{12}$, A.\,R.\,L\'opez--S\'anchez$^{13,2}$, A.\,S.\,G.\,Robotham$^{9}$ \\
$^{1}$Sydney Institute for Astronomy (SIfA), School of Physics, University of Sydney, NSW 2006, Australia\\
$^{2}$Australian Astronomical Observatory, PO Box 915, North Ryde, NSW 1670, Australia\\
$^{3}$Institute for Computational Cosmology, Department of Physics, Durham University, South Road, Durham, DH1 3LE, UK\\
$^{4}$School of Physics, The University of Melbourne, Parkville, VIC 3010, Australia\\ 
$^{5}$Astrophysics Research Institute, Liverpool John Moores University,Twelve Quays House, Egerton Wharf, Birkenhead, CH41 1LD, UK\\
$^{6}$Astronomy Centre, University of Sussex, Falmer, Brighton, BN1 9QH\\
$^{7}$Research School of Astronomy $\&$ Astrophysics, Australian National University, Cotter Road, Weston Creek, ACT 2611, Australia\\
$^{8}$School of Physics, Monash University, Clayton, VIC 3800, Australia\\ 
$^{9}$International Centre for Radio Astronomy Research (ICRAR), University of Western Australia, Crawley, WA 6009, Australia \\
$^{10}$Scottish Universities' Physics Alliance (SUPA), School of Physics and Astronomy, University of St Andrews, North Haugh, St Andrews, KY16 9SS, UK\\
$^{11}$Institut f\"{u}r Astro- und Teilchenphysik, Universit\"{a}t Innsbruck, Technikerstra{\ss}e 25, 6020 Innsbruck, Austria \\
$^{12}$European Southern Observatory, Karl-Schwarzschild-Str. 2, 85748 Garching, Germany\\
$^{13}$Department of Physics and Astronomy, Macquarie University, NSW 2109, Australia
}

\begin{document}
\color{black}
\date{Accepted date. Received date; in original form date}

\pagerange{\pageref{firstpage}--\pageref{lastpage}} \pubyear{2002}

\maketitle

\label{firstpage}

\begin{abstract}
{\color{black} {We present bivariate luminosity and stellar mass functions of
H$\alpha$ star forming galaxies drawn from the Galaxy And Mass
Assembly (GAMA) survey. While optically deep spectroscopic observations of GAMA over a wide sky area enable the detection of a large number of $0.001<SFR_{H\alpha}$ (M$_{\odot}$\,yr$^{-1}$)$<100$ galaxies}, the requirement for an H$\alpha$ detection in targets selected from an $r$--band magnitude limited survey leads to an incompleteness due to missing optically faint {star forming} galaxies. {Using $z<0.1$ bivariate distributions as a reference we model the higher--$z$ distributions, thereby approximating a correction for the missing optically faint star forming galaxies to the local  SFR and $\mathcal{M}$ densities. Furthermore, we obtain the $r$--band LFs and stellar mass functions of \textit{H$\alpha$ star forming} galaxies from the bivariate LFs}. As our sample is selected on the basis of detected H$\alpha$ emission, a direct tracer of on--going star formation, this sample represents a true star forming galaxy sample, and is drawn from both photometrically classified blue and red sub--populations, though mostly from the blue population. {On average 20--30\% of red galaxies at all stellar masses are star forming, implying that these galaxies may be dusty star forming systems.}}
\end{abstract}


\section{Introduction}

The observed univariate luminosity function (LF) is one of the fundamental measures of galaxy properties. It is usually one of the first results to be measured from galaxy surveys {\citep[e.g.][]{Davis1982, Loveday92, Lin96, Norberg02, Blanton03, Loveday12}}.
The importance of the LF, defined as the co--moving source density with luminosity (or magnitude) $L+\Delta L$, extends to all areas of astronomy. In an observational context, it is used to quantify the mean space density of galaxies per unit luminosity and the evolution of statistical properties of a galaxy sample across cosmic time {\citep[e.g.][]{Ly2007, Dale2010, Westra10, Gunawardhana13, Drake2013, Sobral13, Sobral2014}}. In theoretical modelling, the LF is a key ingredient needed to constrain the dark matter halo formation \citep[e.g.][]{Croton06, Bower08}. In an era of multi--wavelength legacy surveys with intrinsically complicated multi--band selections \citep[e.g.][Galaxy And Mass Assembly survey]{Driver11}, understanding the effects of selection and systematic biases on the shape of the LF is imperative in obtaining reliable LF measurements. 

Simply due to the existence of detection limits, no single survey can directly detect all sources to provide a complete and unbiased galaxy sample. The detection probability of an object is a function of a number of parameters, both external (e.g.\,survey selection, area and depth, star--galaxy separation, observing conditions and redshift, spectroscopic and target completeness) and intrinsic to the object (e.g.\,surface brightness, size and colour). As luminosity is strongly correlated with both sets of factors, the least luminous objects in any magnitude--limited survey have the poorest detection probabilities \citep{Geller12}, thus occupying a relatively small volume. The low luminosity galaxies, although they do not dominate the luminosity budget of the universe, greatly outnumber the luminous giants. As other studies have emphasised \citep{Petrosian98}, measuring the evolution of the slope of the faint end of the LF is a challenge. This arises because of the preferential bias against faint galaxies due to surface brightness limits \citep{Sprayberry96, Dalcanton97, Geller12}, galaxy morphologies \citep{Marzke98, Tempel11}, spectral types \citep{Folkes99, Madgwick02}, environment \citep{Xia06, Tempel09, Zandivarez11} and colour \citep{Blanton01} as well as external issues \citep{Driver05, Loveday12}.

Any galaxy sample selected based on a parameter other than the primary survey selection criteria is biased as a result of the dual sample and survey selection. \cite{Gunawardhana13} and \cite{Westra10} present the H$\alpha$ univariate LFs and determine the evolution of H$\alpha$ star formation rate density (SFRD) in the local universe using H$\alpha$ star forming (SF) galaxy samples drawn respectively from the $r$--band magnitude--limited Galaxy And Mass Assembly (GAMA) and Smithsonian Hectospec Lensing surveys. These studies show that their lowest redshift ($z<0.1$) samples are in fact the most complete and span the largest range in both intrinsic H$\alpha$ luminosity (L$_{H\alpha}$) and $r$--band absolute magnitude (M$_r$), {e.g.\,the GAMA $z<0.1$ sample probes $30.5\lesssim\log L_{H\alpha}\,(W) \lesssim36$, $-24\lesssim M_{r}\lesssim-10$ \citep{Gunawardhana13} and $7\lesssim \log \mathcal{M}/M_{\odot}<12$, where $\mathcal{M}$ is the galaxy stellar mass \citep{Baldry12}}. With increasing redshift, however, the sample completeness drops in the sense that a fraction of optically faint star forming galaxies are missing from the higher--$z$ sub--samples and this fraction increases with increasing redshift. As a consequence the final SFRDs based on bivariately selected samples are underestimated, manifesting as an apparent lack of evolution  with redshift in contrast to current observations {\citep{HB06, Perez08, Karim2011, Sobral2014}}. {In comparison, the LFs based on narrowband surveys do not suffer from the same bias as their targets are selected using the quantity they aim to measure \citep[see][for a discussion on the advantages and disadvantages of broadband and narrowband surveys]{Gunawardhana13}. In case of a bivariately selected sample, as is the case in this study, to recover} the missing contribution from optically faint star forming galaxies requires studying how the selection biases influence the LF.  

The bivariate LF \citep{Phillipps86} provides a powerful method of studying the luminosity density in different epochs inclusive of selection biases (e.g.\,bivariate brightness distributions, bivariate luminosity and size distribution). There is a rich collection of literature on using bivariate LFs to explore the space density of galaxies as a function of both survey selection wavelength and surface brightness limits \citep[e.g.][]{Driver99, Cross01, Blanton01, Driver05}, galaxy size \citep[e.g.][]{Sodre93, deJong00, deJong04, Cameron07}, radio luminosity \citep[e.g.][]{Sadler89, Ledlow96, Mauch07}, S{\'e}rsic index, stellar mass and spectral type \citep[e.g.][]{Ball06}, colour \citep[e.g.][]{Baldry04} and in pairs of various galaxy properties \citep[e.g.][]{Blanton03b, Driver06} as well as bivariate ultraviolet/infrared LFs \citep[e.g.][]{Saunders90, Takeuchi12}.    

In this followup paper to \cite{Gunawardhana13}, hereafter paper~I, we explore the GAMA bivariate L$_{H\alpha}$--M$_r$ and L$_{H\alpha}$--$\mathcal{M}$ functions. %%%%% 
{Paper I presents the local star formation history as traced by H$\alpha$ emitters contained within photometrically selected GAMA galaxies. The GAMA H$\alpha$ LFs probe a wider range in luminosity than other results to-date and demonstrate a Gaussian--like drop in number density ($\Phi$) at high luminosities, rather than the exponential drop characteristic of \cite{Schechter76} function. In paper~I we conclude that a \cite{Saunders90} functional form, widely used to characterise radio and infrared LFs, is now also required to give a better description of H$\alpha$ LFs. Despite the relatively large range in luminosity probed by the GAMA H$\alpha$ LFs up to $z<0.34$, the intrinsic SFR densities based on these LFs show a distinct lack of evolution in SFR density with increasing redshift. This apparent lack of evolution in SFRD is primarily caused by the sample selection. For GAMA, we find that the $r$--band apparent magnitude limit of the survey, along with the subsequent requirement for H$\alpha$ detection, leads to an incompleteness due to missing bright H$\alpha$ sources that are fainter than the $r$--band selection limit. While local estimates of SFR density measurements from a range of SFR--sensitive wavelengths (e.g.\,H$\alpha$, [\ion{O}{ii}], [\ion{O}{iii}], H$\beta$) based on narrowband surveys and slitless spectroscopy data show an evolution with redshift \citep[e.g.][]{Jones01, Shioya08, Sobral13}, those based on broadband surveys are almost always underestimated \citep[e.g.\,paper~I;][]{Westra10} as a consequence of the bivariate sample selection.}
{The primary aims of this investigation are to: (a) model the low redshift bivariate function to use as a reference to account for the missing optically faint star forming galaxies at higher--$z$ {($z<0.34$)}, (b) measure the (moderate) redshift evolution of stellar mass and SFR densities, (c) characterise the univariate M$_r$ LFs and {stellar mass functions (SMF)} of H$\alpha$ SF galaxies and compare the results with LFs and SMFs of photometrically classified blue galaxies, and (d) explore the characteristics of photometrically classified blue and red star forming sub--populations in GAMA. }

{The layout of this paper is as follows. We briefly describe the GAMA survey, our sample selection criteria and selection biases in \S\,\ref{sec:data}. In \S\,\ref{sec:construction}, we describe the derivation of the bivariate functions, taking into account different survey selection criteria. The resulting bivariate functions are presented in \S\,\ref{sec:biLF} and \S\,\ref{sec:biLFs_SMF}. These sections also include the univariate functions obtained from integrating the bivariate functions over L$_{H\alpha}$ axis.  Finally, in \S\,\ref{sec:funcfits} we detail of the functional forms used to fit the bivariate functions, and in \S\,\ref{sebsec:rlumdensity} and \S\,\ref{sec:SFH_mass}, we infer SFR and $\mathcal{M}$ densities by integrating the best--fitting  functional forms to the bivariate functions. Our results are discussed in \S\ref{sec:discuss} and we conclude in \S\ref{sec:conclude}.}

The assumed cosmological parameters are H$_0=70$ km\,s$^{-1}$\,Mpc$^{-1}$, $\Omega_M = 0.3$, and $\Omega_{\Lambda} = 0.7$. All magnitudes are presented in the AB system. A \cite{Chabrier2003} IMF, {commonly used to derive stellar masses in the literature,} is used to derive the stellar mass measurements used in this study \citep{Taylor11} and {the same IMF used in paper~I \citep[i.e.][IMF]{Baldry2003} is used to estimate SFRs. As these two IMF forms are sufficiently similar, no significant systematic effect introduced by adopting them}. To avoid confusion we state in the figure caption which IMF is used to obtain the results shown. 
\vspace{-0.5cm}

\section{The GAMA survey and bivariate sample selection} \label{sec:data}

Our study utilises data from the GAMA\footnote{\url{http://www.gama-survey.org/}} survey \citep{Baldry10, Robotham10, Driver11}. GAMA is a spectroscopic survey undertaken at the Anglo Australian Telescope (AAT) with 2--degree Field (2dF) fibre feed and the AAOmega multi--object spectrograph. AAOmega provides a resolution of $3.2$\,\AA\,full width at half--maximum with complete spectral coverage from $3700$ to $8900$\,\AA\, \citep{Sharp06, Driver11, Hopkins13}. GAMA--I covers three equatorial fields {at approximately 09hr, 12hr and 15hr, hereafter G09, G12 and G15, of $48$\,deg$^2$ each. G09 and G15 are limited to a depth of $r_{AB}<19.4$ while G12 extends to $r_{AB}<19.8$.} 

For the current analysis, we use GAMA--I spectroscopic data consisting of GAMA, SDSS, 2--degree Field Galaxy Redshift Survey \citep[][2dFGRS]{Colless01} and Millennium Galaxy Catalogue \citep[][MGC]{Driver05} sources. 
A detailed description of the data, the sample selection and the measurement of physical properties of galaxies such as SFRs is given in \S\,2 of paper~I. Briefly, the emission line measurements used for this investigation are measured from each flux calibrated spectrum, assuming a single Gaussian profile and a common redshift and line width within an adjacent set of lines (e.g.\,H$\alpha$, the [\ion{N}{ii}]~$\lambda\lambda$\,6548, 6583 and [\ion{S}{ii}] doublets), and simultaneously fitting the continuum local to the set of lines. This method of measuring fluxes does not take into account the effects of the underlying stellar absorption on Balmer line fluxes. To correct for this effect, we apply a constant correction to H$\alpha$ and H$\beta$ fluxes following the prescription of \cite{Hopkins03}. A comprehensive discussion of GAMA spectroscopic measurement process is presented in \cite{Hopkins13}, and further analyses on the effects of the assumption of a constant stellar absorption correction are presented in paper~I and in \cite{Gunawardhana11}. 

The derivation of stellar absorption, aperture and dust obscuration corrected H$\alpha$ luminosities (L$_{H\alpha}$) is described in \S\,3 of paper~I. The stellar masses ($\mathcal{M}$) and absolute magnitudes $k$--corrected to $z=0$ (M$_r$) have been derived using the stellar template spectrum that best fits $u,\,g,\,i,\,r,\,z$ {GAMA} photometry \citep[StellarMassesv08,][]{Taylor11}. {We have not attempted to correct the rest frame $r$--band continuum flux for H$\alpha$ emission contamination as the contamination is $<2\%$ for over $98\%$ of the lowest redshift data (H$\alpha$ spectral line is redshifted out of the $r$--band filter at $z\sim0.05$), and does not change the results we present in this analysis}. 
{Additionally, we compute absolute magnitudes based on {SDSS} Petrosian $r$--band magnitudes at $z=0.1$, hereafter M$^{0.1}_r$ \citep[\textsc{kcorrect V4\_2,}][]{Blanton07} to compare our results with previous GAMA studies \citep[e.g.][]{Loveday12, Baldry12}.}

{We use the same sample of galaxies as in paper~I}. Briefly, our sample includes all emission--line galaxies with H$\alpha$ fluxes (F$_{H\alpha}$) greater than a detection limit of $25\times10^{-20}$\,W/m$^2$, and H$\alpha$ emission signal--to--noise greater than $3$ {that are classified as star forming based on the prescription of \cite{Kewley02}}. {The emission line measurements for galaxies in the three GAMA regions that have been observed previously in earlier spectroscopic campaigns (e.g.\,SDSS, 2dFGRS, 6dFGRS) are either taken from their respective survey databases or accounted for through spectroscopic incompleteness corrections if the respective spectra are not flux calibrated (e.g.\,2dFGRS spectra). The sample properties and trends with redshift are explored in S\,2 and \S\,3 of paper~I. To summarise, the sample used for the calculation of $f_{V,H\alpha}$ includes the following galaxies: \begin{enumerate}[label=\arabic*., leftmargin=*]
	\itemsep4pt
	\item GAMA observations with observed H$\alpha$ emission above the detection limit of $25\times10^{-20}$W/m$^2$. AGNs are removed from the sample using BPT diagnostics and the \cite{Kewley02} prescription. The two line classifications \citep[paper~I;][]{Brough11} are used to recognise AGNs if one of the four spectral line measurements required for a BPT diagnostics is unavailable. A small subset of GAMA galaxies where none of these methods can be employed is included in the sample, as these are more likely to be star forming than AGNs \citep{Cid11}.

	\item SDSS observations with H$\alpha$ emission and continuum signal--to--noise greater than 3. The AGNs are removed as described above using SDSS line measurements available from MPA--JHU database\footnote{\url{http://www.mpa-garching.mpg.de/SDSS/DR7/raw_data.html}}.
	
\end{enumerate}
}

{
Figures\,\ref{fig:Halpha_in_Mr} and \ref{fig:Halpha_in_Mr_distribution} demonstrate the limitations of a bivariately selected sample. Figure\,\ref{fig:Halpha_in_Mr} shows the M$_r$ distributions of all GAMA galaxies in four different redshift ranges compared to the distributions of galaxies with reliable H$\alpha$ detections ({i.e.\,F$_{H\alpha}>25\times10^{-20}$\,W/m$^2$}) and galaxies with F$_{H\alpha}>1\times10^{-18}$\,W/m$^2$ (i.e.\,the flux limit used in the V$_{\rm max}$ calculations; \S\,4 of paper~I). The distributions of blue and red {sub-populations, as defined in Eq.\,\ref{eq:color} of this paper and in \cite{Loveday12},} of galaxies within each distribution are shown in the insets. The distributions of H$\alpha$ detected galaxies at each redshift range comprise both blue and red galaxies, though blue galaxies dominate the distributions. Also, the $z<0.1$ and $0.1<z<0.15$ distributions of the H$\alpha$ detected galaxies show a clear blue--red bimodal distribution, while the lack of such a trend in the higher redshift distributions is likely partly due to the difficulty in reliably measuring the H$\alpha$ feature in low signal--to--noise weak--line systems at higher redshifts, and partly due to the smaller luminosity range probed at higher redshifts. Even though the distribution of H$\alpha$ detected sample is bimodal, the sample, after imposing an H$\alpha$ flux limit for the estimation of V$_{\rm max}$ (paper~I), is biased against red weak--line galaxies at all redshifts, {\color{black}which are likely to be dusty star forming galaxies with low signal--to--noise spectra (\S\ref{subsec:Baldryvsus})}. 

Figure\,\ref{fig:Halpha_in_Mr_distribution} further demonstrates how the bivariate selection acts to limit the range of L$_{H\alpha}$ and M$_r$ probed by the LFs with increasing redshift. 
\begin{figure*}
	\begin{center}
		\includegraphics[scale=0.4]{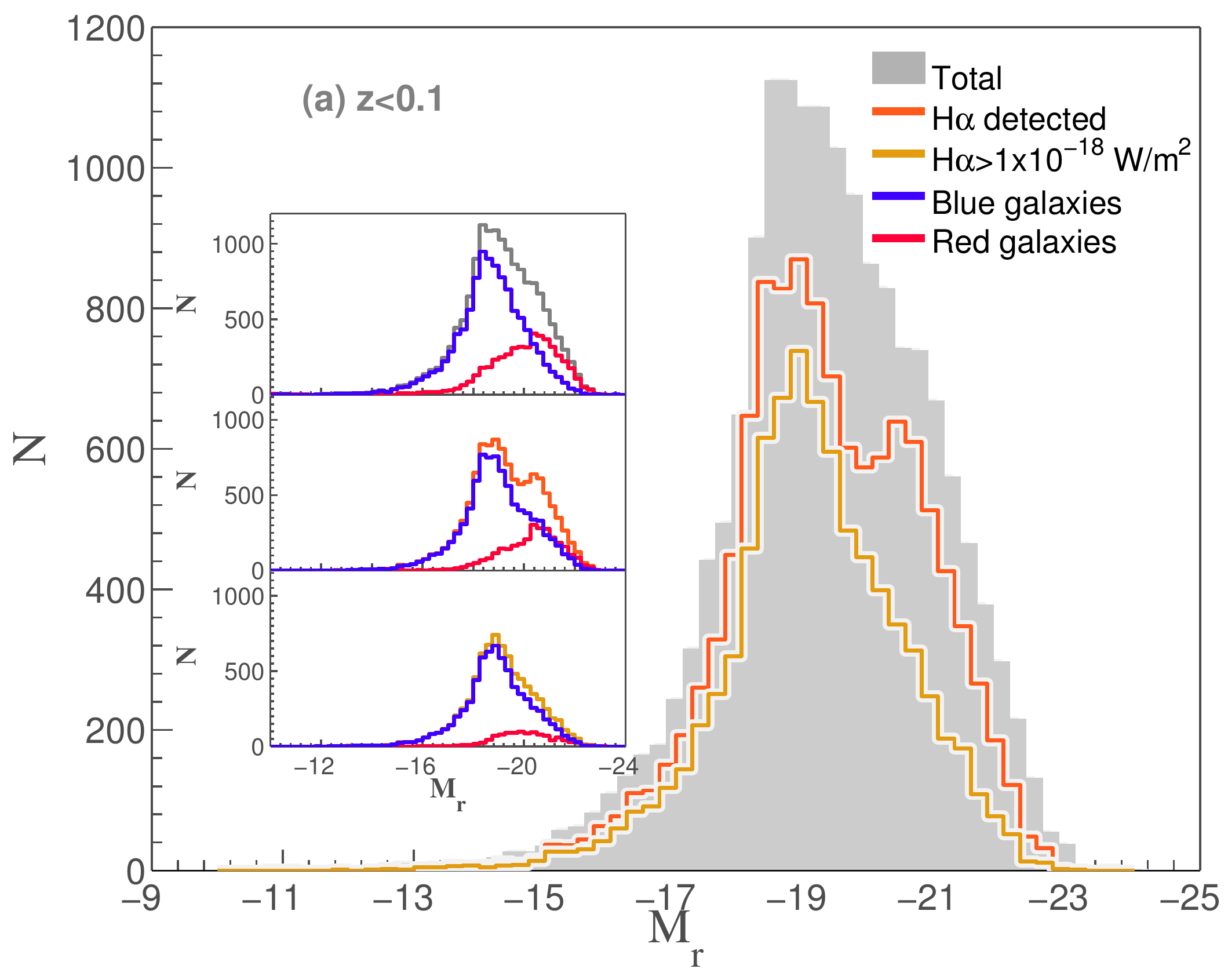}
		\includegraphics[scale=0.4]{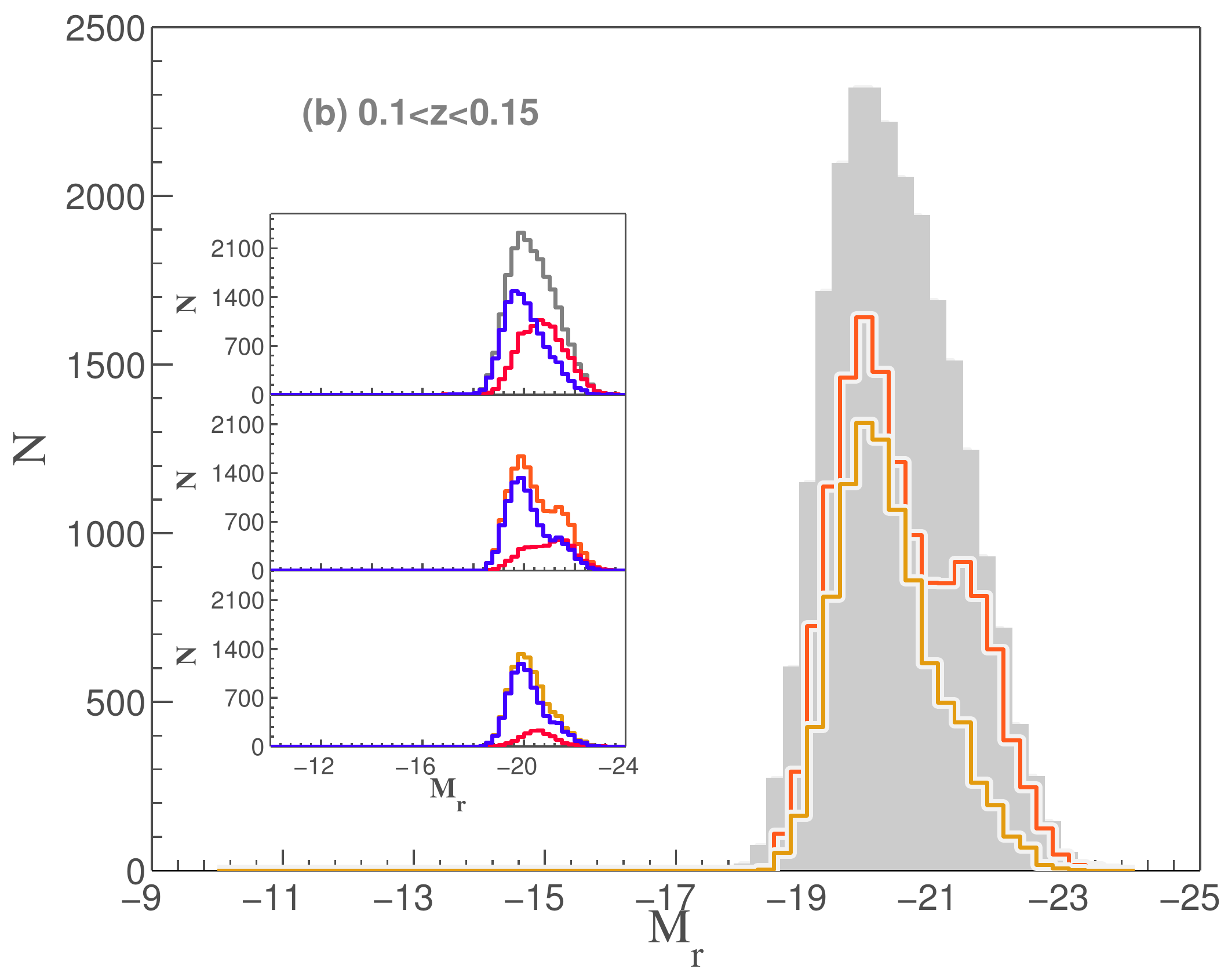}
		\includegraphics[scale=0.4]{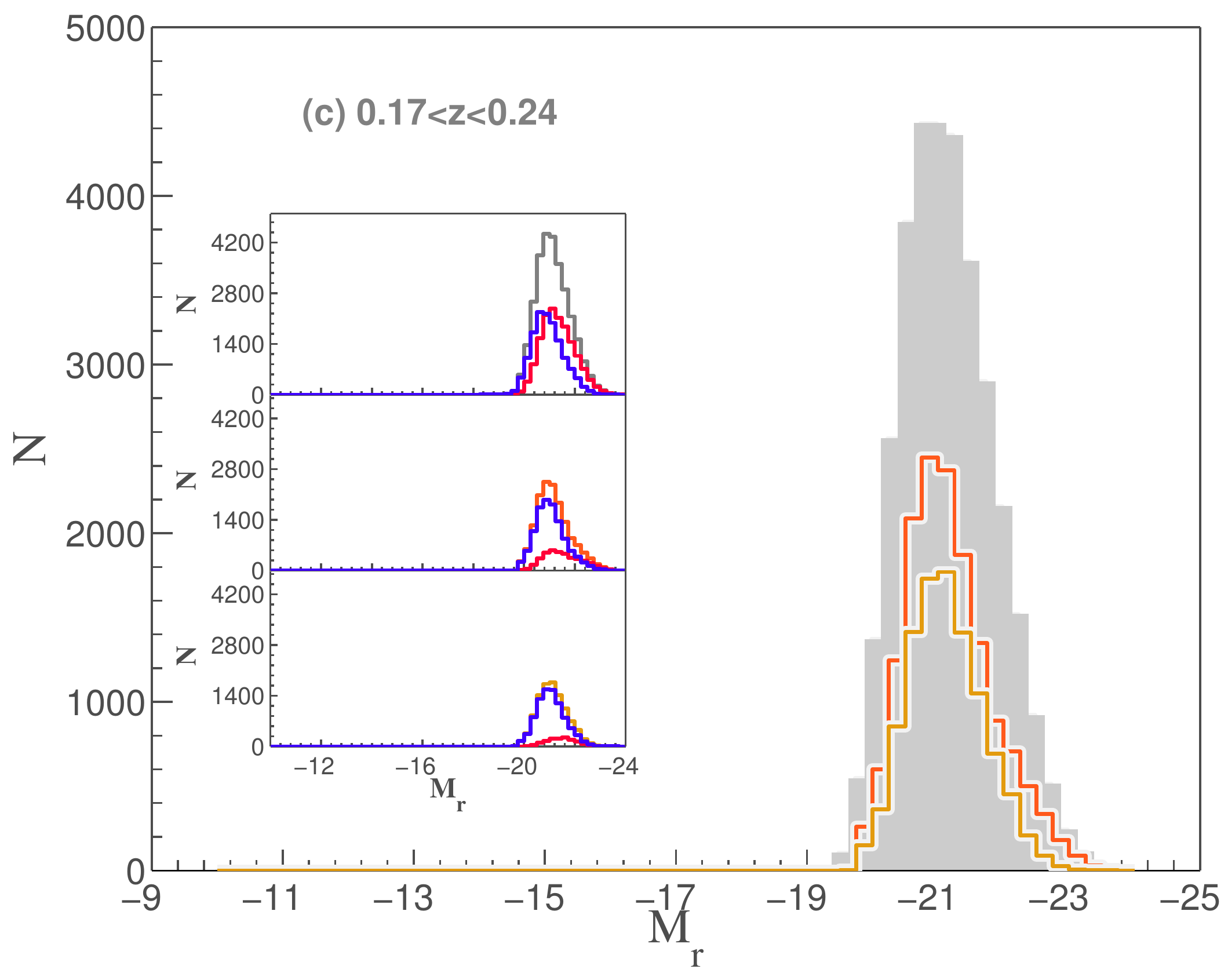}
		\includegraphics[scale=0.4]{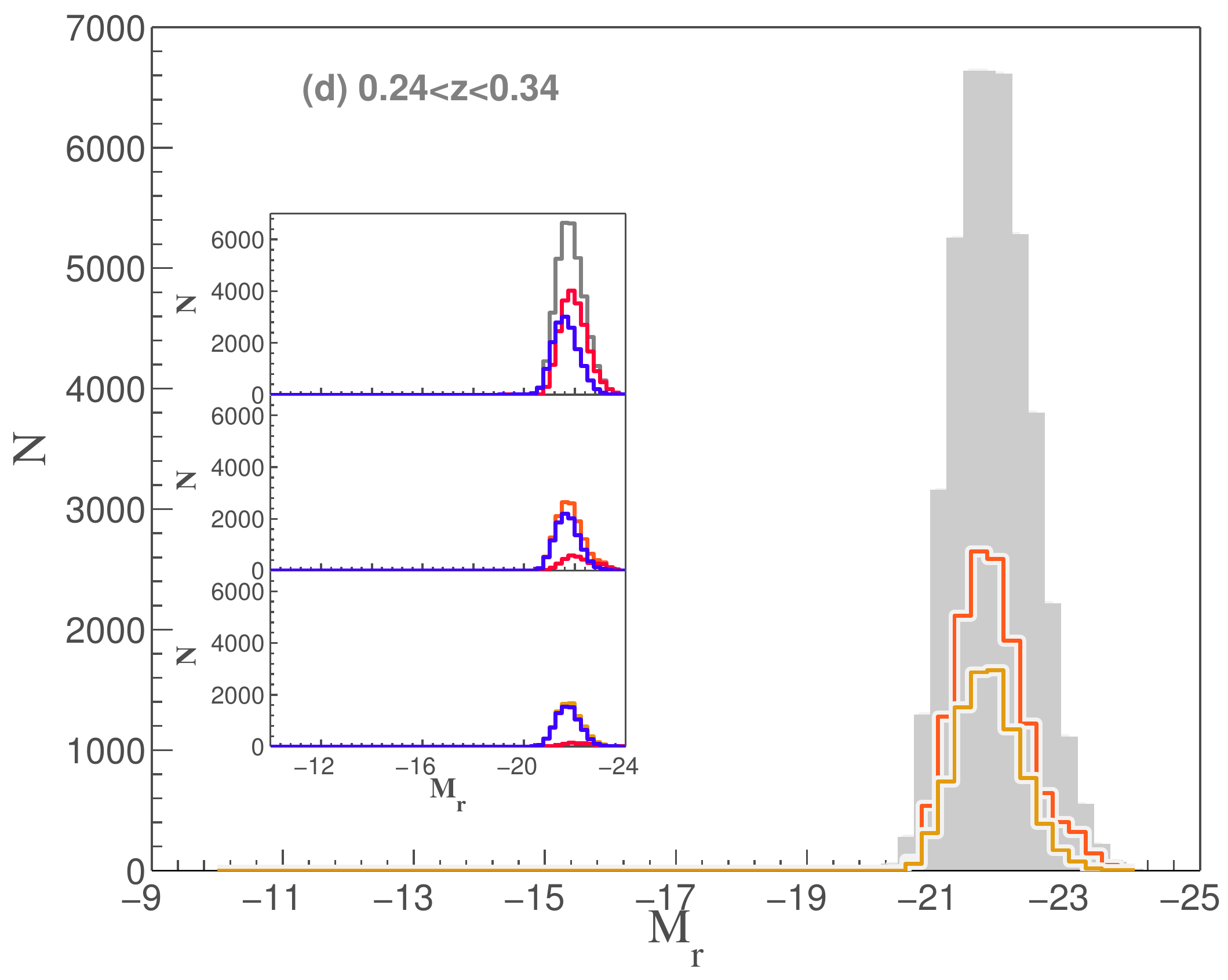}
		\caption{{The M$_r$ distribution of GAMA galaxies split in to four redshift bins. Each panel shows three distributions: all galaxies (grey histogram), those with observed H$\alpha$ flux greater than the detection limit of $25\times10^{-20}$W/m$^2$ (orange histogram), and those with H$\alpha$ fluxes $>1\times10^{-18}$W/m$^2$, i.e.\,the flux limit (yellow histogram). Each of the three histograms are further divided to indicate the distributions of galaxies classified as blue and red based on Eq.\,\ref{eq:color}. These are shown within the three insets in each main panel. The three insets top--to--bottom show the distributions of blue and red sub-populations corresponding to all galaxies (top), galaxies with H$\alpha$ fluxes $>25\times10^{-20}$W/m$^2$ (middle) and galaxies with H$\alpha$ fluxes $>1\times10^{-18}$W/m$^2$ (bottom).}}
		\label{fig:Halpha_in_Mr}
	\end{center}
\end{figure*}
\begin{figure*}
	\begin{center}
		\includegraphics[width=1\textwidth]{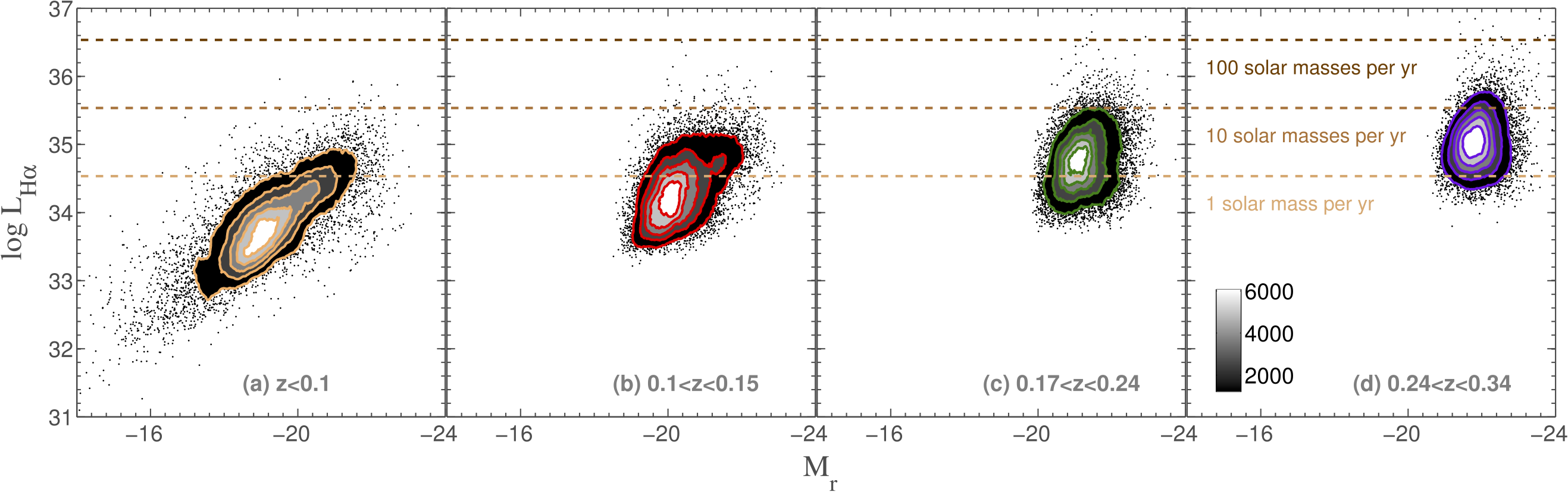}
		\caption{{The bivariate $\log$ L$_H\alpha$ (W) and M$_r$ distributions split in to four redshift bins. The dual $r$--band apparent magnitude and H$\alpha$ flux selection of our sample leads to an incompleteness in {optically faint star forming galaxies}. The $z<0.1$ sample probes the largest range in H$\alpha$ luminosity and M$_r$, therefore it is the most complete. The range probed by the higher redshift samples progressively drops with redshift.}}
		\label{fig:Halpha_in_Mr_distribution}
	\end{center}
\end{figure*}
The $z<0.1$ sample probes the largest range in both M$_r$ and L$_{H\alpha}$.
 The dual H$\alpha$--M$_r$ selection prevents the optically faint star forming galaxies from entering the sample at all redshifts, with its effects becoming more significant with increasing redshift. In order to assess the extent of this effect, we now consider the bivariate LFs.  
}

\vspace{-0.5cm}
{\color{black}\section{Constructing bivariate functions} \label{sec:construction}}

We use three different LF estimators (the classical 1/V$_{\rm max}$ method, the density corrected 1/V$_{\rm max}$ method and the bivariate step--wise maximum likelihood) to derive the bivariate LFs presented in \S\,\ref{sec:biLF}. The formulation of each of the three methods is described below. 

\subsection{The ``classical" method}\label{subsec:classical}

	The 1/V$_{\rm max}$ technique \citep{Schmidt68}, also referred to as the ``classical method", is widely used to estimate the co--moving space density as a function of luminosity. 
	
	The formulation of the 1/V$_{\rm max}$ method inclusive of incompleteness corrections {for univariate LFs} is described in \S\,4 of paper~I. The {bivariate} function derived using this definition is {
\begin{equation}
	\small
	\Phi[\log L_{H\alpha}, M_r] \times \Delta \log L\, \Delta M_r = \\ {\sum_i} \frac{1}{V_{i, max}}.
	\label{eq:vmax_def}
\end{equation}}
In this equation, V$_{i,max}$ represents the maximum volume out to which the $i^{th}$ object would be visible and still be part of the survey, $ \Delta \log\,L$ and {$\Delta M_r$} define the luminosity {and magnitude bin widths, respectively.}  
	
The GAMA-I sample used in this study is subjected to two different $r$--band magnitude limits \citep[$r<19.4$ for G09, 15 and $r<19.8$ for G12;][]{Driver11} and an emission--line selection. Given these constraints, the definition of V$_{i,\rm max}$ is 
\begin{equation}
	\small
	V_{i,max} = \min[(V_{i, max, H\alpha}), (V_{i, max, r}), (V_{i, z_{lim}})] \times c_i,
	\label{eq:vmax_def2}
\end{equation}
{where V$_{i,\rm max}$ is the minimum volume that a galaxy $i$ would have given the maximum volumes for that galaxy computed using an H$\alpha$ flux limit of $1\times10^{-18}$\,Wm$^{-2}$ (V$_{i, {\rm max}, H\alpha}$) and the magnitude limit of the GAMA survey (V$_{i, {\rm max}, r}$), and the volume (V$_{i, z_{lim}}$) of the redshift slice ($z_{lim}$) that $i^{th}$ galaxy resides in. The completeness correction, $c_i$, takes into account both the imaging and spectroscopic incompletenesses. The estimation of $c_i$ is described in detail in paper~I and in \cite{Loveday12}. }

{The classical method has the advantages of simplicity and it gives simultaneously both the shape and the absolute normalisation of the LF, however, it can be susceptive to cosmic (sample) variance \citep{Efstathiou88, Willmer97, Baldry12} when the survey volume is small}. For this reason, large datasets covering a substantial portion of the sky  are generally required to avoid the shape of the LF being distorted due to large scale structure. 

\begin{figure*}
	\begin{center}
	        \includegraphics[width=0.6\textwidth]{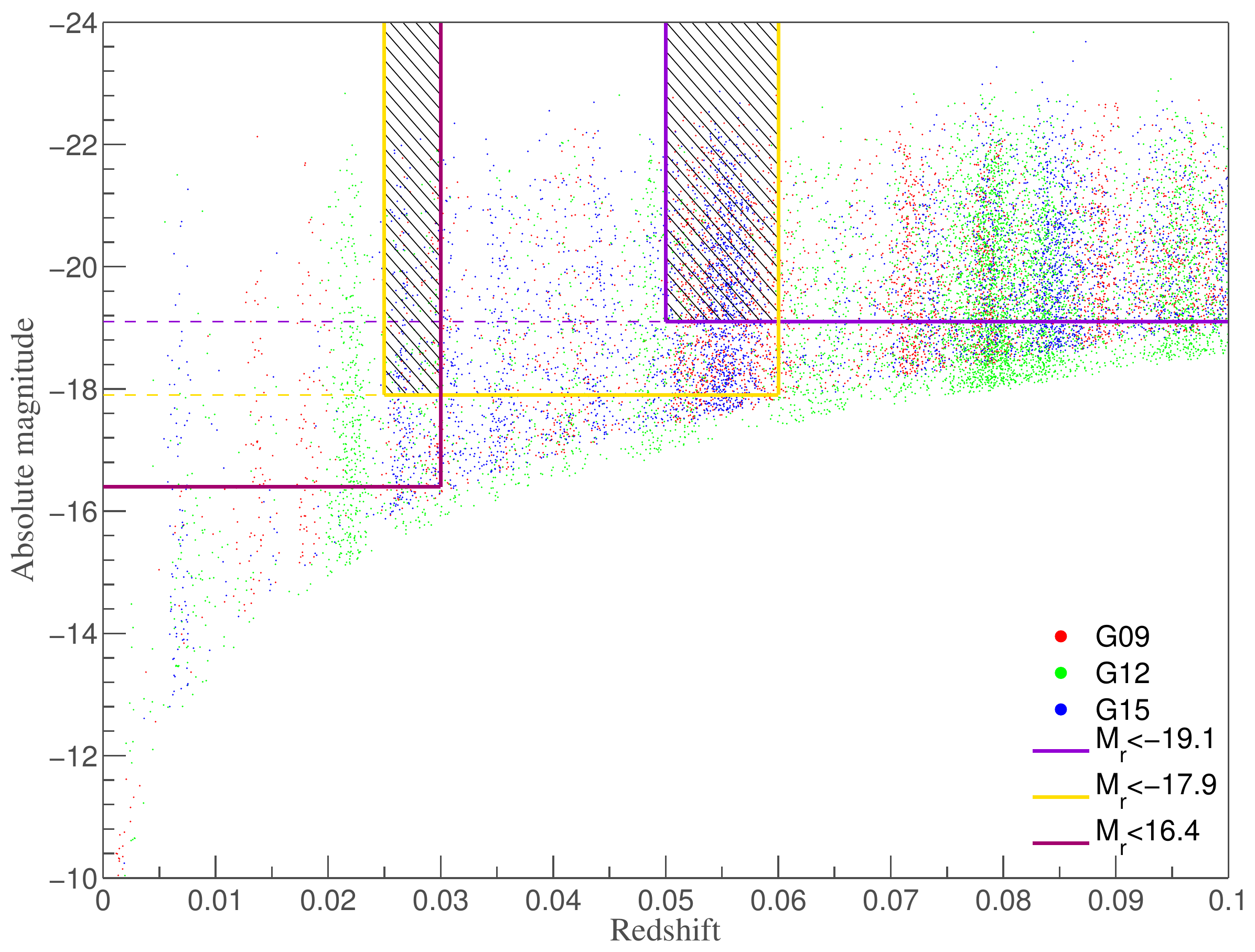}
		\includegraphics[width=0.6\textwidth]{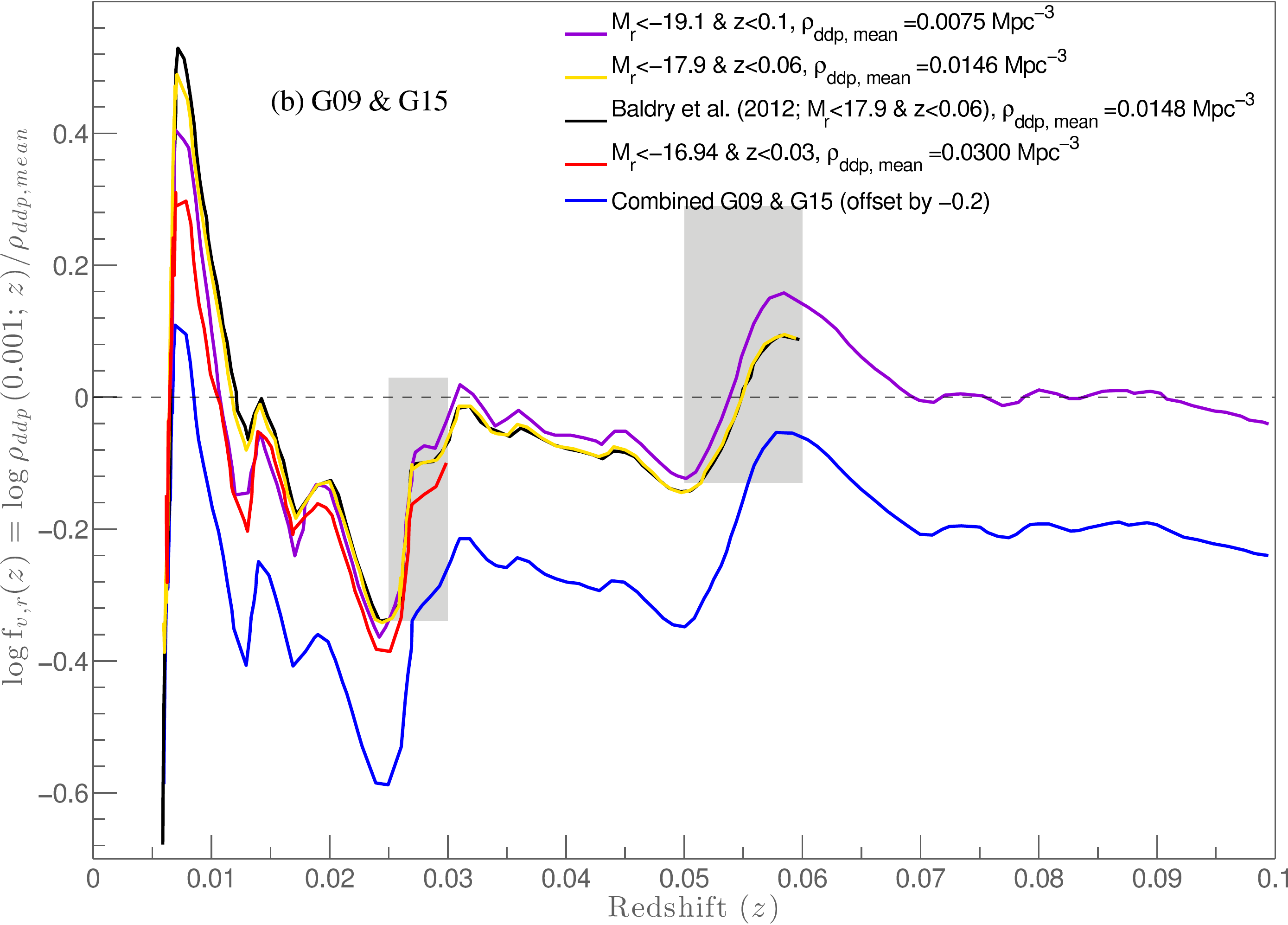}
		\includegraphics[width=0.6\textwidth]{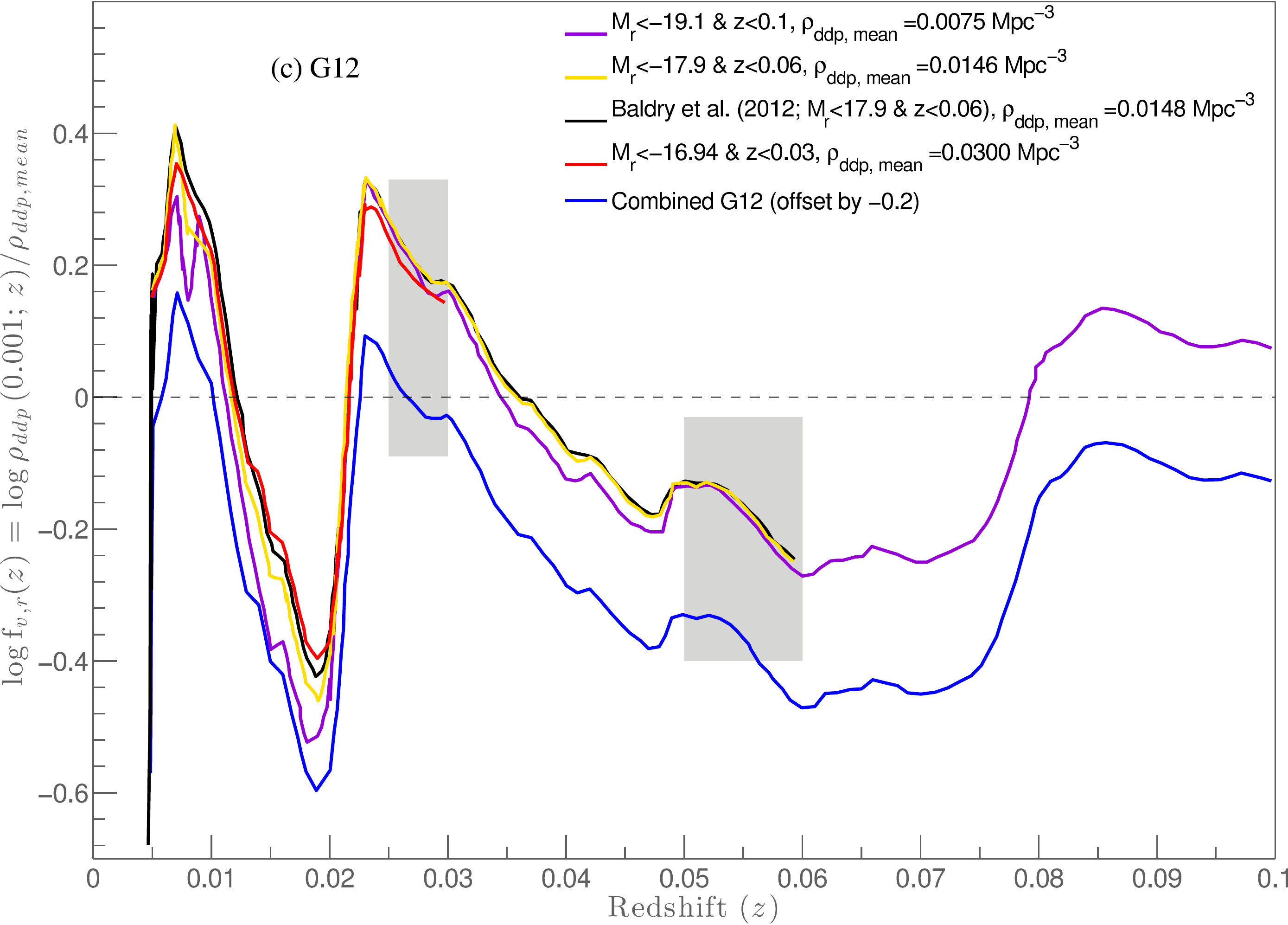}
		\caption{{(a) The distribution of absolute $r$--band magnitudes in redshift for \textit{all} $z<0.1$ galaxies colour coded by their GAMA region identifier. Using this distribution we define three overlapping volume limited samples, which are then used in the derivation of  $\log$ $f_{v, r}$ as a function of redshift for each volume limited sample (b and c). As the three GAMA fields have two different magnitude limits, $f_{v, r}$ is derived separately for G09 and G15 ($r<19.4$), and G12 ($r<19.8$). 
		The blue line, offset from the rest for legibility, indicates the variation in final $f_{v, r}$ used for the analysis. This line traces the $f_{v, r}$ versus redshift relation estimated from the lowest redshift volume sample (i.e.\,red line) until $z\sim0.03$ (i.e.\,first shaded region corresponding to the hatched region in the top panel) before smoothly transitioning to the relation derived from the second volume-limited sample (yellow line) and so on. The $f_{V,r}$ versus redshift relations derived in \citet{Baldry12} for $z<0.06$ galaxies are shown for comparison.}}
		\label{fig:lowz_ddp}
	\end{center}
\end{figure*}

\begin{figure*}
	\begin{center}
		\includegraphics[width=0.56\textwidth]{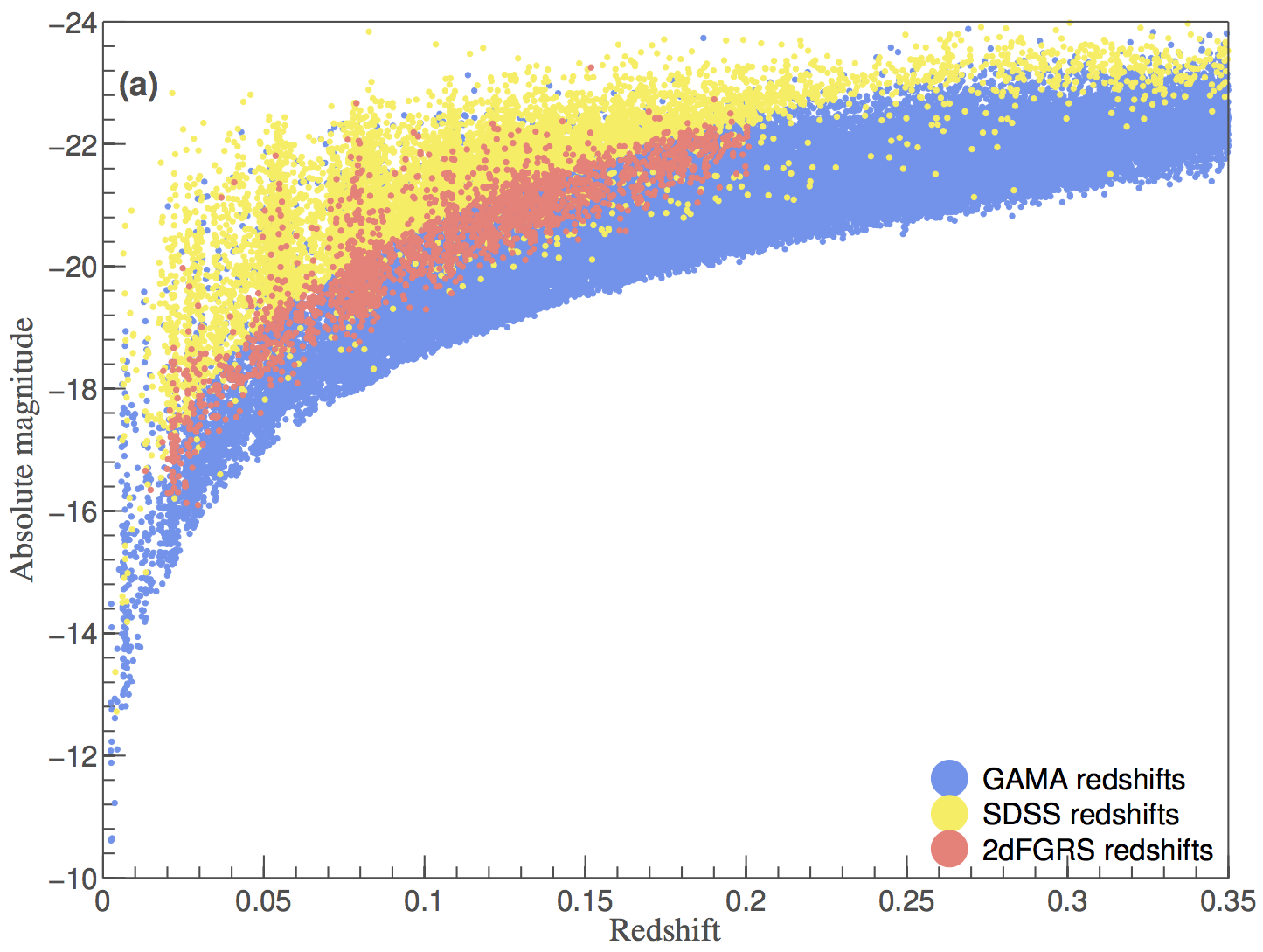}
		\includegraphics[width=0.56\textwidth]{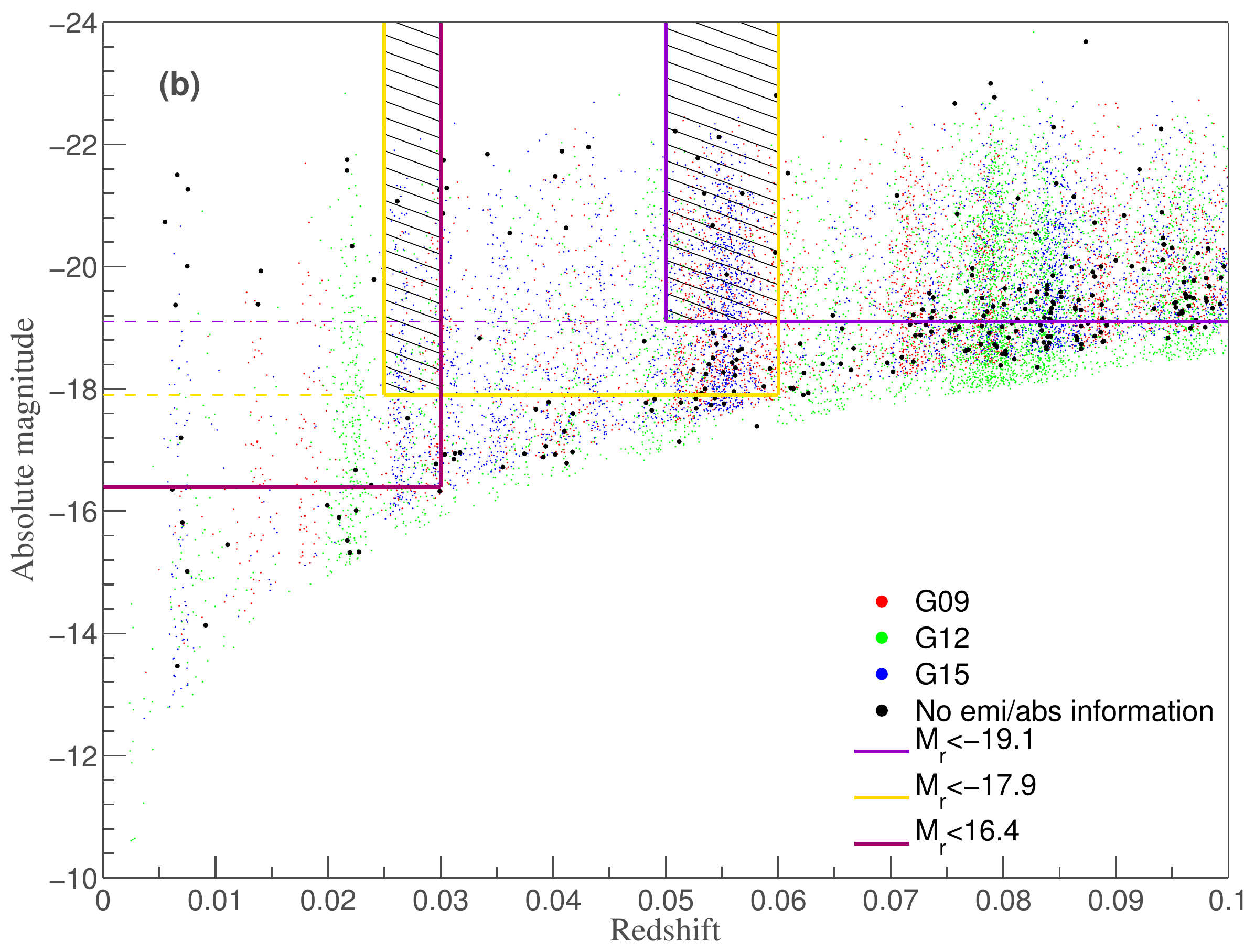}
		\includegraphics[width=0.56\textwidth]{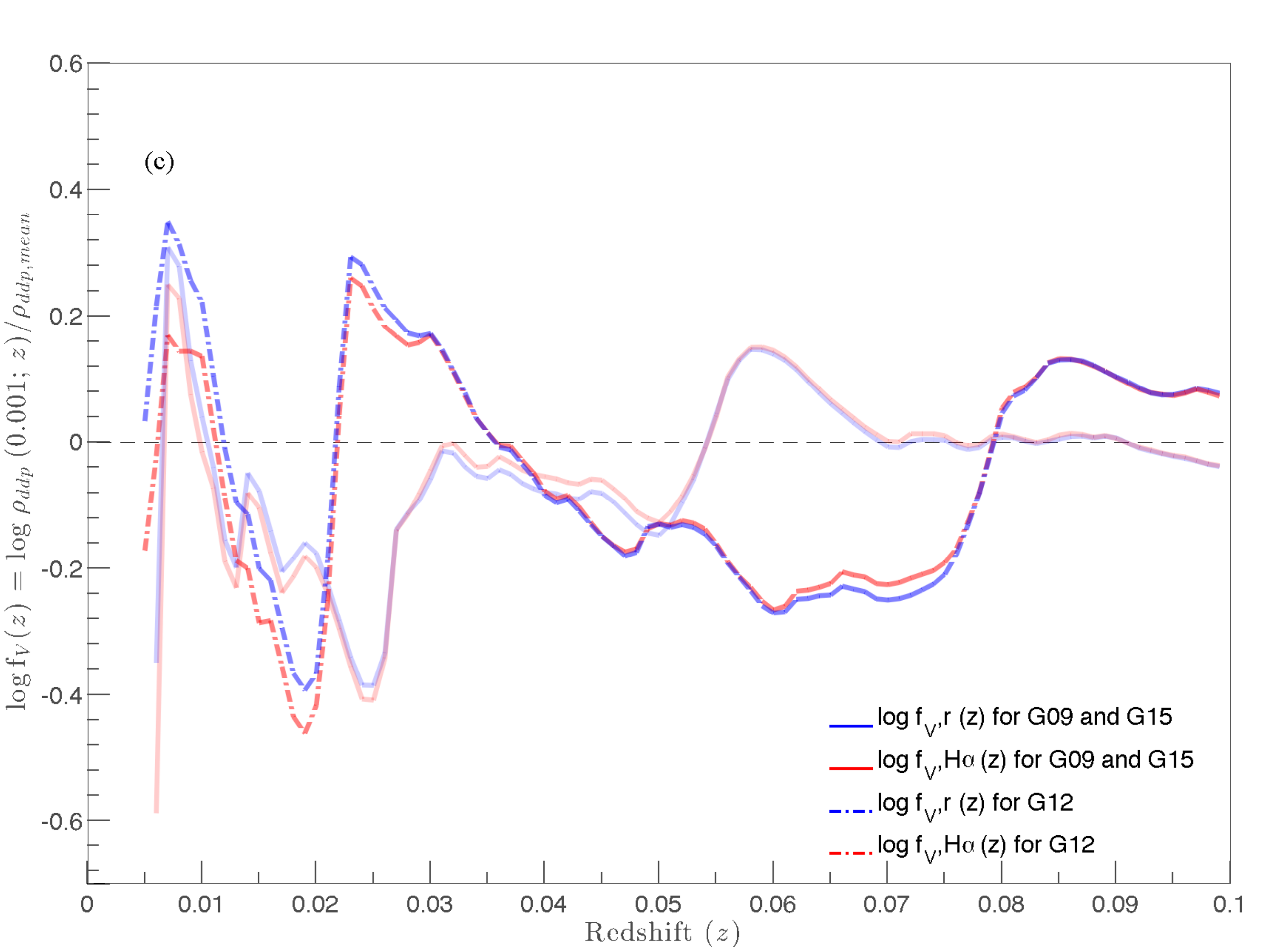}
		\caption{{(a) The distribution of absolute $r$--band magnitudes of H$\alpha$ SFR galaxies in redshift, colour coded by the survey from which the spectra are taken. (b) Same as Figure\,\ref{fig:lowz_ddp}, but for SF galaxies. The black points indicate the absolute magnitudes of galaxies for which we do not have any spectra. (c) The $f_{v, H\alpha}$ (red) and $f_{v, r}$ (blue) versus redshift relations for G09 and G15 (solid line), and G12 (dashed line). The lack of spectral line information makes the derived $f_{v, H\alpha}$ a lower limit, however, we demonstrate that the final $f_{v, H\alpha}$ versus redshift relation is close to the true relation (i.e.\,the true relation here is that of $f_{v, r}$) as there are only a few galaxies in a given redshift slice without this information. Note that as we use all galaxies to derive $f_{v, r}$, the lack of spectral line measurements does not affect $f_{v,r}$ estimates.}}
		\label{fig:lowz_HAvolumeLimited}
	\end{center}
\end{figure*}
	
	There are more sophisticated methods of estimating LFs to account for these disadvantages. Two such methods are the density corrected 1/V$_{\rm max}$ corrections \citep{Baldry06, Mahtessian11, Cole11, Baldry12} and the stepwise maximum likelihood method \citep[SWML; ][]{Efstathiou88}, a variant of the method proposed by \cite{Sandage79}.  
				
{Finally the univariate LF is obtained by integrating Eq.\,\ref{eq:vmax_def} over one of the variables.} 

\subsection{Density corrected maximum volume corrections} \label{subsec:ddp_vmax}

	\cite{Baldry06} \citep[but also][]{Mahtessian11, Baldry12}  describe a modification to the 1/V$_{\rm max}$ technique that takes into account the radial variation in the large--scale structure. They define a maximum volume weighted by density, V$_{\rm max}'$, 
\begin{equation}
	V_{i, {\rm max}}'= \frac{\rho_{ddp}(z_{min};z_{{\rm max}, i})}{\rho_{ddp, mean}(z_1; z_2)}\times V_{i, {\rm max}} = f_V \times V_{i, {\rm max}}, 
	\label{eq:density_vmax}
\end{equation}
where $\rho_{ddp}(z_1; z_2)$ is the number density of a density defining population (DDP) between redshifts $z_1$ and $z_2$. DDP is a volume limited sample and  $z_1$, $z_2$ are the minimum and maximum redshifts of that volume limited sample \citep{Baldry06, Baldry12}. 
$\rho_{ddp}(z_1; z_2)$ is estimated separately for each GAMA field, and the average between the three fields is taken to be $\rho_{ddp, mean}$. 

As we investigate the evolution of the bivariate L$_H\alpha$--M$_r$ function over a moderate range in redshift, the density weights ($f_v$) for V$_{{\rm max}, r-band}$ and V$_{{\rm max}, H\alpha}$ are estimated separately using several overlapping volume limited samples to improve the accuracy of V$_{{\rm max}, i}'$ \citep{Mahtessian11}. {The GAMA sample selection criteria detailed in \cite{Baldry12}, not restricted to their redshift range, is adopted to estimate density weights for V$_{{\rm max}, r-band}$. We use $f_{v, r}$ to denote the density weights estimated using this sample, which has a univariate $r$--band apparent magnitude selection. Figure\,\ref{fig:lowz_ddp}a shows the volume limited sample definitions used in the calculation of $f_{v,r}$ for the $z<0.1$ sample. The variation in $f_{v,r}$ is quantified separately for the {G09 and G15 (Figure\,\ref{fig:lowz_ddp}b), and G12 (Figure\,\ref{fig:lowz_ddp}c)} fields because of their different magnitude limits. In the regions where two volume limited samples are allowed to overlap, shown as hatched and shaded regions in {Figure\,\ref{fig:lowz_ddp}}, the $f_{v, r}$ corrections are combined such that the transition from one volume limited sample to the other is smooth. The blue line in Figure\,\ref{fig:lowz_ddp}b and \ref{fig:lowz_ddp}c show the final $f_{v,r}$ versus redshift relation used in the calculation of V$_{{\rm max, r}}'$ offset by $-0.2$ in log space for legibility. Also shown for comparison is the $f_{v,r}$ relation from \cite{Baldry12} for $z<0.06$ GAMA galaxies (black line). 
The negligible difference between the blue and black lines is a result of the different $r$--band magnitude types used as inputs to \textsc{kcorrect V4\_2} \citep{Blanton07}.
}

{
As the galaxy sample used for this study has a dual $r$--band magnitude and H$\alpha$ flux selection}, we also explored the impact of estimating the density corrections using {only the H$\alpha$ detections in a $r$--band selected sample. We use $f_{v, H\alpha}$ to denote this correction}. This serves, in principle, as a different correction, $f_{V,H\alpha}$, for V$_{{\rm max}, H\alpha}$, although it should be similar in practice. {To estimate $f_{V,H\alpha}$, we add 2dFGRS observations\footnote{\url{www2.aao.gov.au/2dfgrs/}} with $\eta$, a measure of the average absorption/emission line strength of a galaxy that strongly correlates with H$\alpha$ equivalent width, greater than $-1.2$ \citep{Madgwick02} to the SF galaxy sample used for this study (\S\ref{sec:data}). The AGNs are removed as described above using the 2dFGRS spectral line catalogue \citep{Lewis02}}. 

{The distribution of $r$--band absolute magnitudes in redshift for the selected GAMA H$\alpha$ emission line galaxies with spectra originating from GAMA, SDSS and 2dFGRS surveys is shown in Figure\,\ref{fig:lowz_HAvolumeLimited}a, with the GAMA survey providing the deepest spectroscopic observations followed by 2dFGRS and SDSS surveys. The same volume limited sample definitions introduced in Figure\,\ref{fig:lowz_ddp} are used to calculate $f_{V, H\alpha}$, but from a $r$--band magnitude versus redshift distribution comprising only H$\alpha$ SFR galaxies (Figure\,\ref{fig:lowz_HAvolumeLimited}b). The black points in Figure\,\ref{fig:lowz_HAvolumeLimited}b indicate galaxies with spectra originating from a survey other than GAMA, 2dFGRS or SDSS for which we currently do not have spectral line information, and are not included in the sample used to determine $f_{V, H\alpha}$. The spectroscopic incompleteness arising from the exclusion of these objects leads to a small discrepancy between  $f_{v,r}$ and $f_{v, H\alpha}$ as demonstrated in Figure\,\ref{fig:lowz_HAvolumeLimited}c. Even though this discrepancy is largest at the lowest redshifts, as might be expected due to the relatively low number of galaxies sampled by the lowest redshift volume--limited sample in comparison to other volume--limited samples, it has a negligible effect on the density corrected bivariate functions.} 

{An alternative approach that does not require binning was developed by \cite{Sandage79} (STY).  The STY maximum likelihood LF estimator, although not biased by the presence of the large scale structure, requires the assumption of a parametric form for the LF. A ``nonparametric" variant of STY method called the stepwise maximum likelihood (SWML) estimator was introduced by \cite{Efstathiou88}, mainly to overcome the inconvenience of not being able to adequately establish whether the chosen parameterisation represents a good fit to the data \citep{Efstathiou88, Willmer97}. While this technique is insensitive to density fluctuations \citep{Sandage79, Efstathiou88, Willmer97}, the luminosity bins in SWML methods are highly correlated such that any issue that occurs in a given bin may affect the whole luminosity function. In appendix\,\ref{sec:swml_lf}, we describe the formulation of the SWML estimator for bivariate functions. A comparison between the lowest redshift bivariate L$_{H\alpha}$--M$_r$ function constructed using SWML method and that constructed using the density corrected V$_{\rm max}$ method is presented in \S\,\ref{subsec:comparison}.}

\subsection{A comparison of bivariate LF estimators}\label{subsec:comparison}

The formulation of the classical 1/V$_{\rm max}$ and the density corrected 1/V$_{\rm max}$ is described in \S\,\ref{subsec:classical} and \S\,\ref{subsec:ddp_vmax}. 
\begin{figure}
	\includegraphics[width=0.5\textwidth]{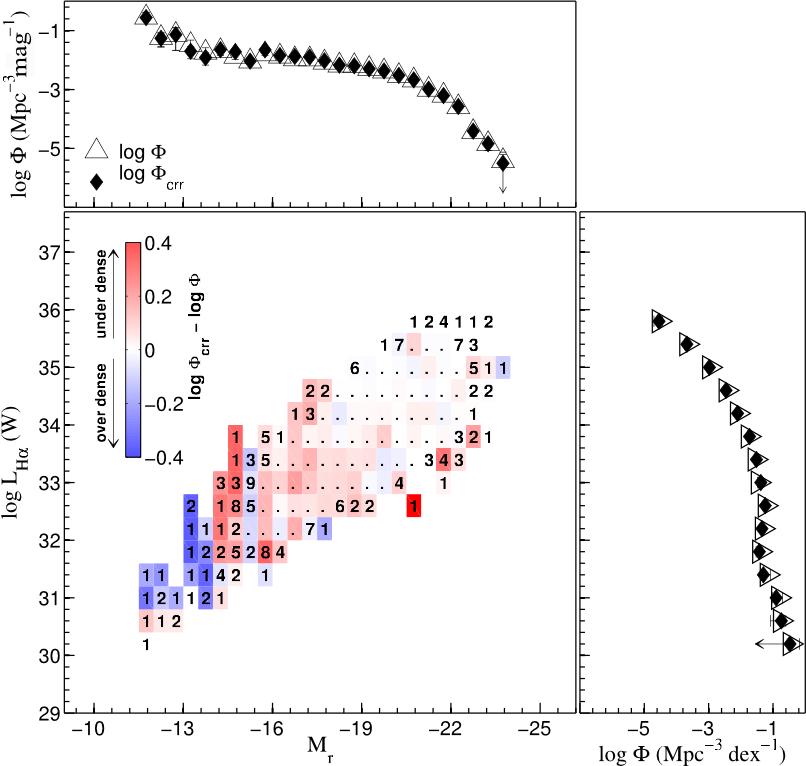}
	\caption{The residual map showing the difference in number densities derived using the classical ($\Phi$) and the density corrected ($\Phi_{crr}$) V$_{\rm max}$ methods for the $z<0.1$ bivariate functions. {The colours indicate the difference between $\Phi_{crr}$ and $\Phi$. A positive difference (redder colours) implies that $f_v$ corrects V$_{\rm max}$  for an under--density and negative (bluer colours) means the correction is for an over--density}. Top and right panels show the M$_r$ and H$\alpha$ univariate LFs.}
	\label{fig:res_map1}
\end{figure}
We present the residual between the two {bivariate functions}. {We focus here on the $z<0.1$ slice as this is the one which is most likely to differ between the two methods due to  the size of the volume surveyed.}

Figure\,\ref{fig:res_map1} shows the residual maps obtained by subtracting the bivariate L$_{H\alpha}$--M$_r$ function derived using the density corrected 1/V$_{\rm max}$ method from that derived using the classical  1/V$_{\rm max}$ method. 

\begin{figure*}
	\includegraphics[scale=0.35]{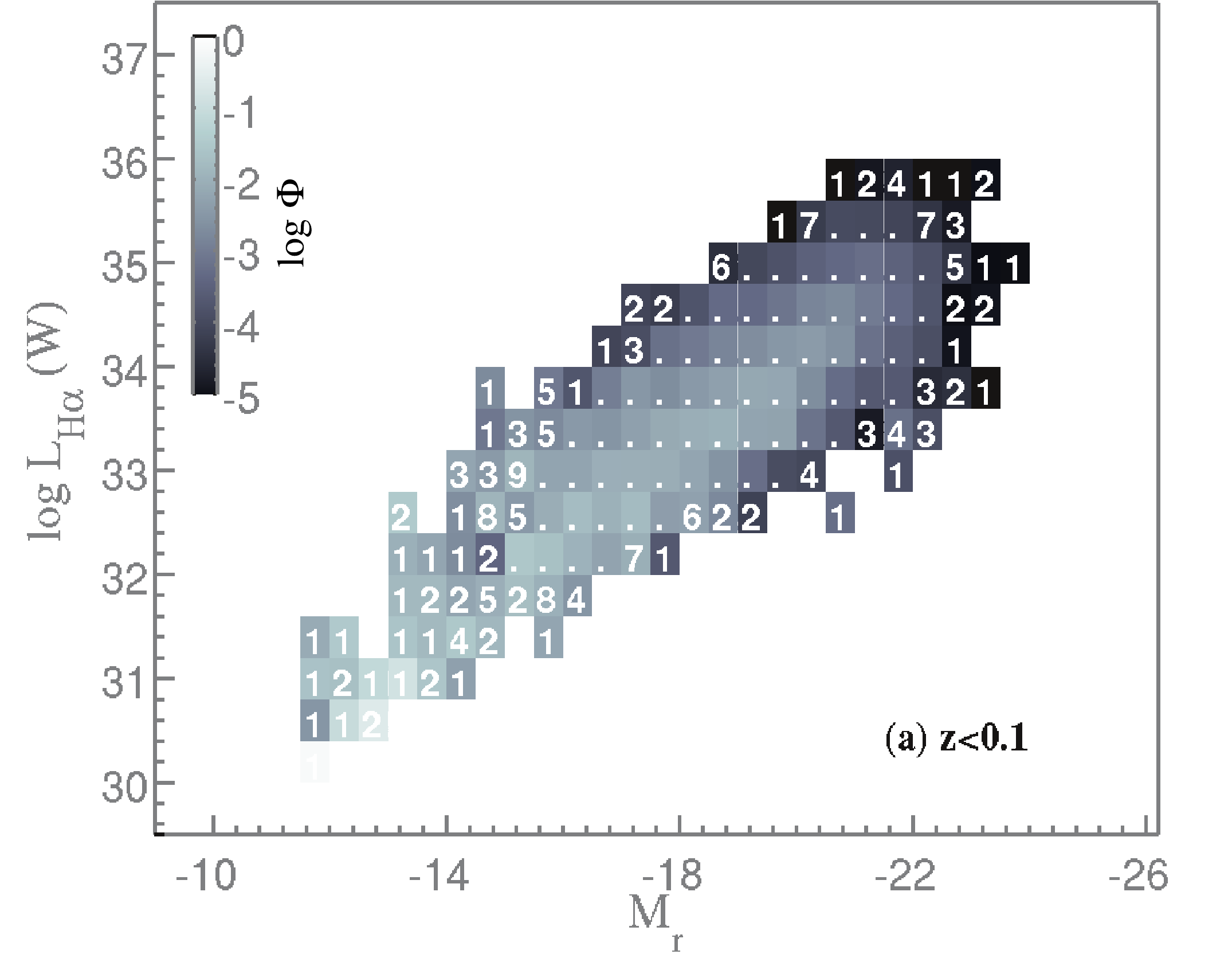}
	\includegraphics[scale=0.35]{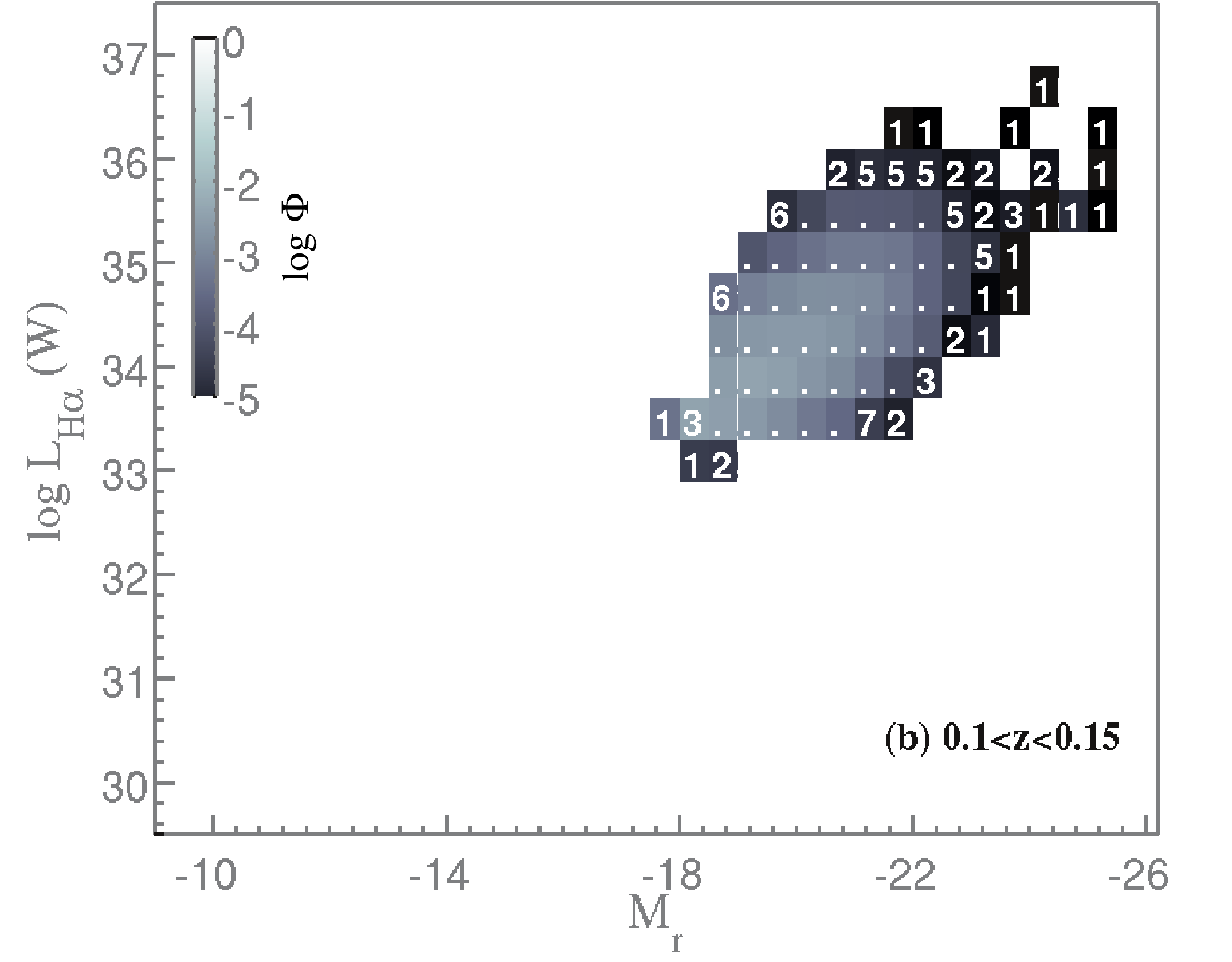}\\
	\includegraphics[scale=0.35]{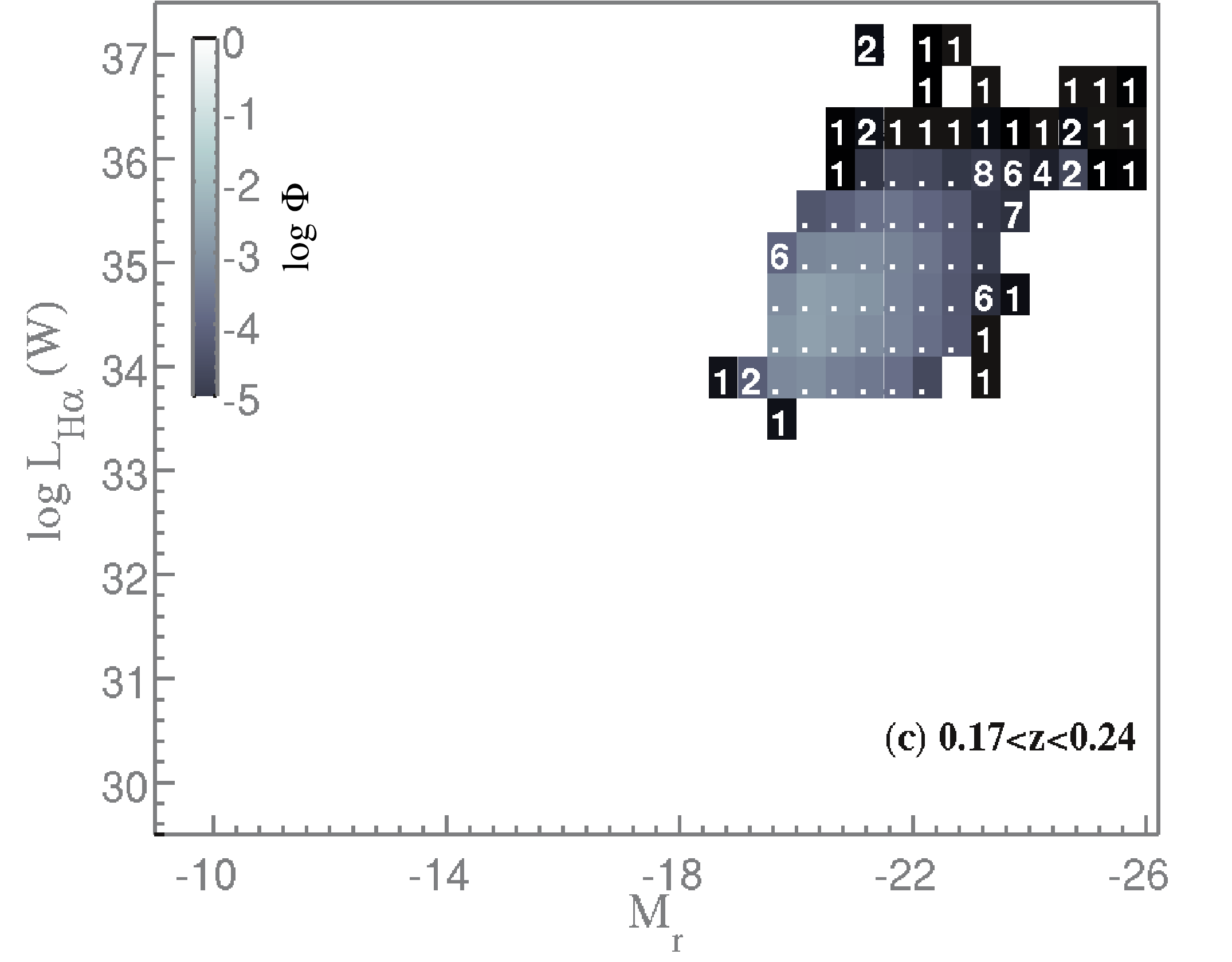}
	\includegraphics[scale=0.35]{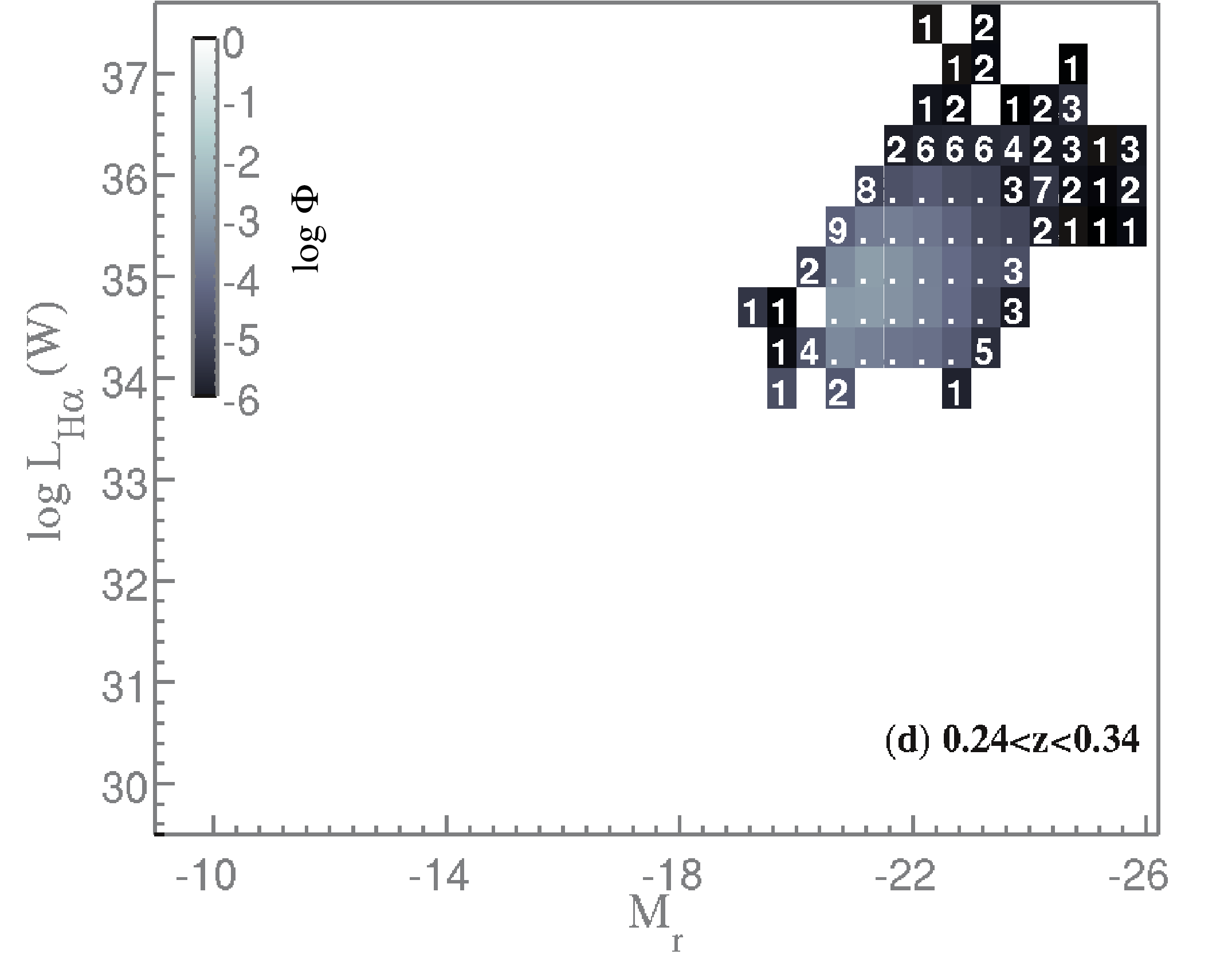}
	\caption{Bivariate L$_{H\alpha}$--M$_r$ functions of the H$\alpha$ SF galaxies computed using the classical 1/V$_{\rm max}$ method split into four redshift bins with redshift increasing across. The L$_{H\alpha}$--M$_r$ bins with less than 10 galaxies in them have their number of galaxies quoted, while the dots indicate bins with 10 or more galaxies. The gray scale corresponds to the number density in the unit of Mpc$^{-3}$ dex$^{-1}$ mag$^{-1}$.}
	\label{fig:vmax_bilf}
\end{figure*}

\begin{figure*}
	\includegraphics[scale=0.35]{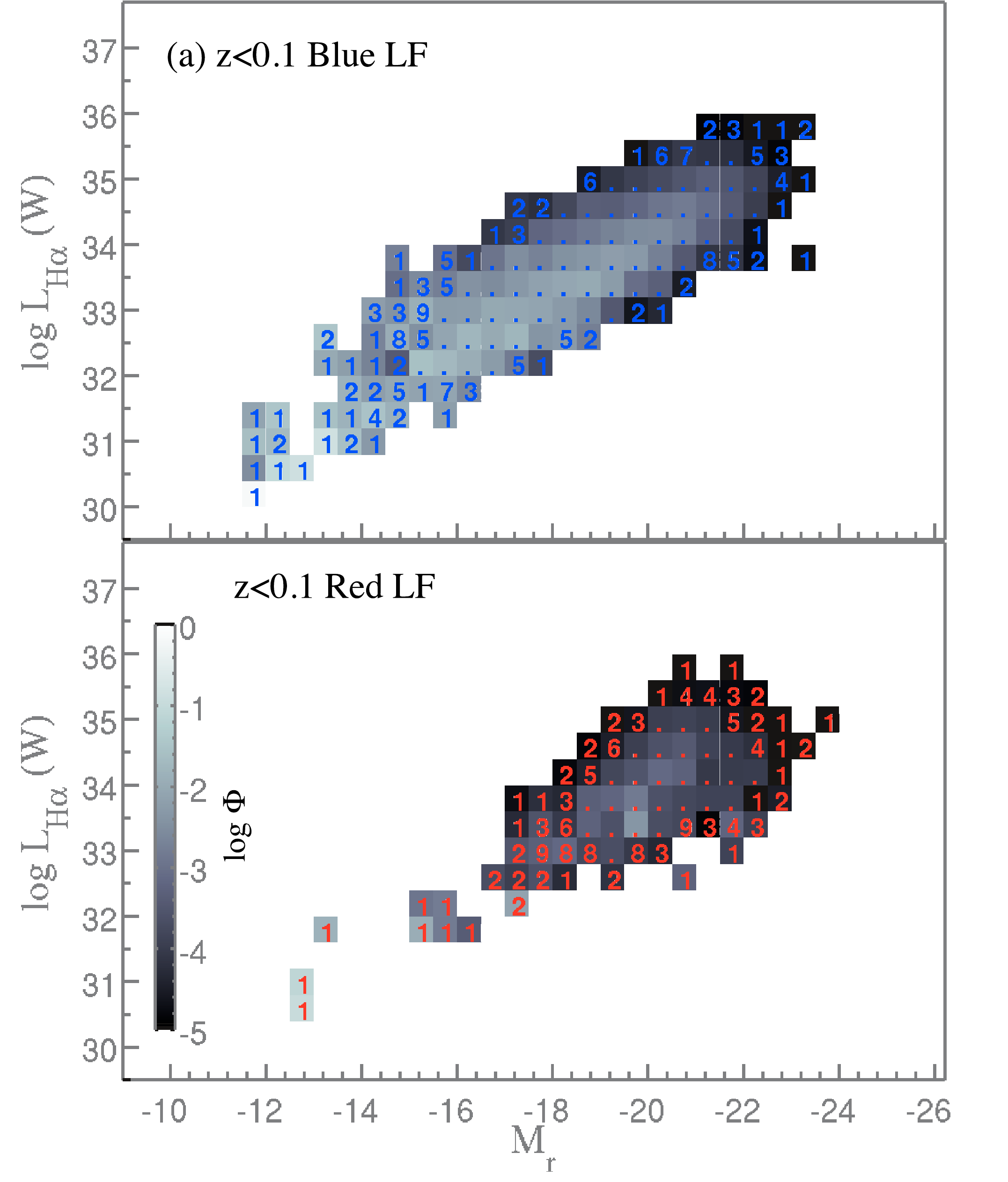}
	\includegraphics[scale=0.35]{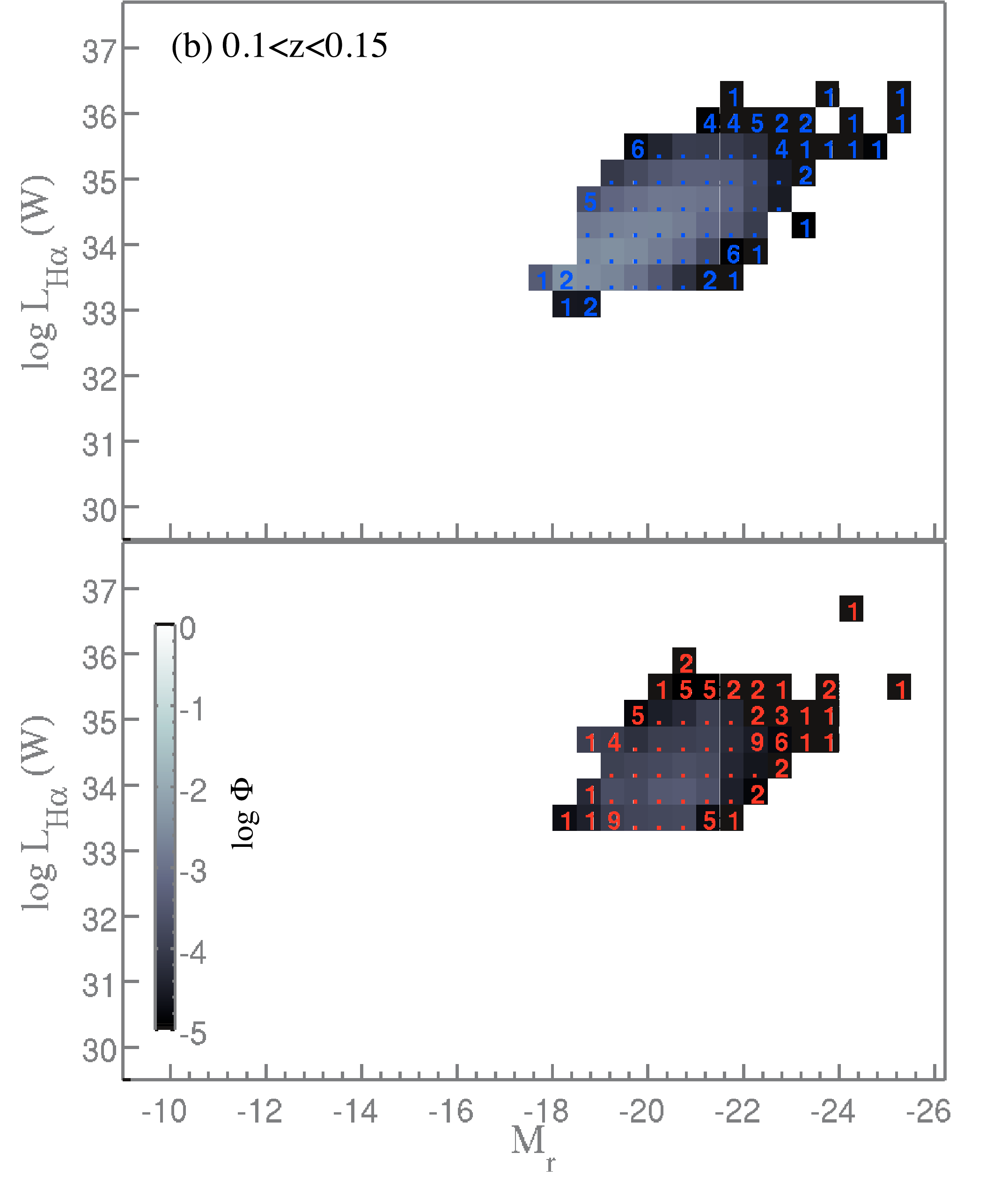}\\
	\includegraphics[scale=0.35]{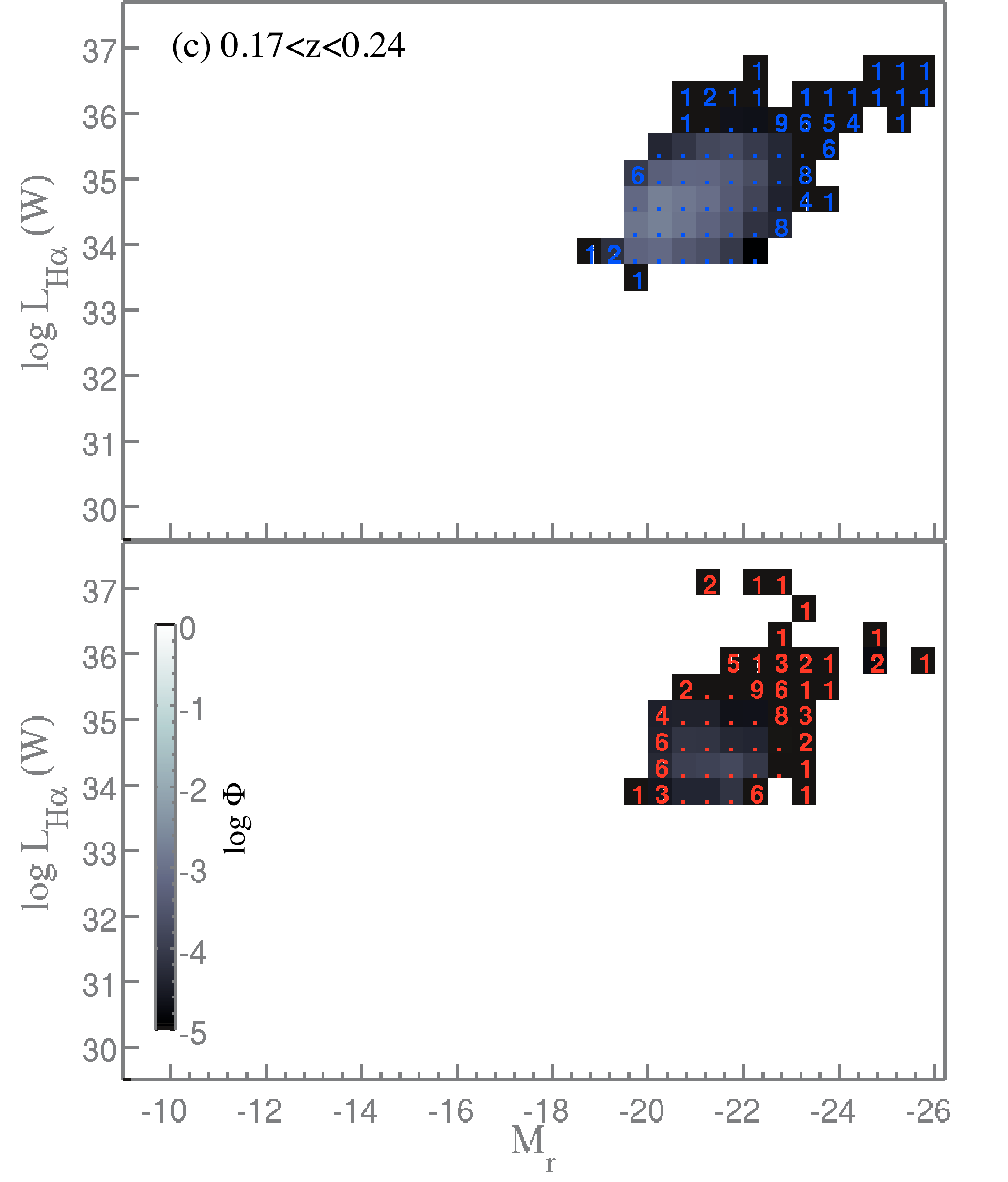}
	\includegraphics[scale=0.35]{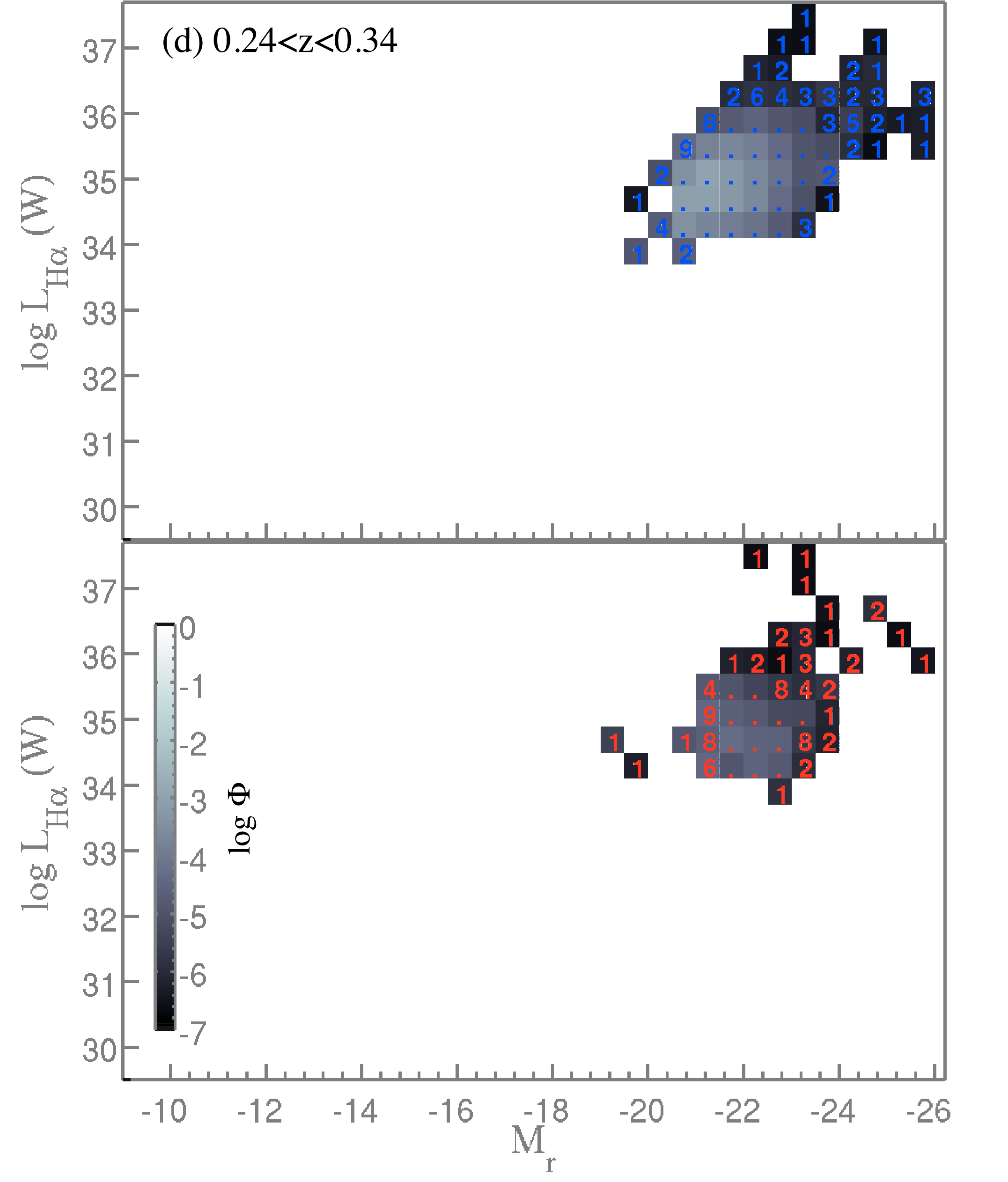}
	\caption{Bivariate L$_{H\alpha}$--M$_r$ functions of blue and red H$\alpha$ SF sub--populations constructed using the classical V$_{\rm max}$ method split in four redshift bins. The gray scale shown along side each red bivariate LF indicates the number densities ($\Phi$) in the unit of Mpc$^{-3}$ dex$^{-1}$ mag$^{-1}$ for both blue and red functions of the respective redshift range. Like in Figure\,\ref{fig:vmax_bilf}, the  L$_{H\alpha}$--M$_r$ bins with less than 10 galaxies in have their number of galaxies indicated.}
	\label{fig:vmax_bilf_photom}
\end{figure*}

As mentioned above, the {bivariate functions} based on the classical 1/V$_{\rm max}$ method, especially the $\Phi$ estimates at faint M$_r$ and L$_{H\alpha}$, are affected by large scale structure. The density corrected 1/V$_{\rm max}$ technique is designed to account for radial variations through $\rho_{ddp}$ defined in \S\,\ref{subsec:ddp_vmax}. {This is demonstrated in Figure\,\ref{fig:res_map1}, where the redder colour indicates that the applied density weight corrects the classical 1/V$_{\rm max}$ for an under--density and vice versa, with the colour intensity showing the significance of that correction for each L$_{H\alpha}$--M$_r$ bin.} As expected the density correction becomes progressively more important towards fainter M$_r$ and L$_{H\alpha}$ values {(i.e.\,towards increasingly smaller volumes)}.  {The top and right panels of Figure\,\ref{fig:res_map1} show the M$_r$ and H$\alpha$ LFs obtained from integrating the bivariate L$_{H\alpha}$--M$_r$ functions based on the classical (open symbols) and density corrected (filled symbols) 1/V$_{\rm max}$ methods in L$_{H\alpha}$ and M$_r$ directions, respectively. Clearly,} the density corrections to the bivariate functions have a small effect, and are limited to the faintest end of the bivariate function. This is not surprising given the low volume being probed combined with the small numbers of galaxies contributing to each bin. {Higher--$z$ residual maps are devoid of significant differences in this comparison.} We note that we see an almost identical result if we use the $f_{V,r}$ rather than $f_{V, H\alpha}$ in making the density--dependent correction to the V$_{\rm max, H\alpha}$. {In summary, we see a marginal difference in the bivariate functions between 1/V$_{\rm max}$ and the density corrected 1/V$_{\rm max}$, while no statistical difference is observed in the univariate LFs.

A similar conclusion is reached from the comparison between SWML with the density corrected V$_{\rm max}$ (Appendix\,\ref{sec:swml_lf}).}

{
\section{Bivariate L$_{H\alpha}$--M$_{\MakeLowercase{r}}$ functions}\label{sec:biLF}}

The GAMA bivariate L$_{H\alpha}$--M$_r$ functions {in redshift bins} constructed using the classical V$_{\rm max}$ method are shown in Figure\,\ref{fig:vmax_bilf}. Due to the magnitude limited nature of the survey, the range in M$_r$ and L$_{H\alpha}$ probed by the bivariate functions progressively decreases with increasing redshift. Particularly, the lowest--$z$ bivariate LF width with respect to M$_r$ and L$_{H\alpha}$ (i.e.\,the horizontal and vertical lengths of the bivariate distribution as a function of M$_r$ or L$_{H\alpha}$ respectively) indicates an overall decrease towards fainter L$_{H\alpha}$--M$_r$. This is supported by the reduction in number towards fainter M$_r$ and L$_{H\alpha}$. The decrease in horizontal width with increasing L$_{H\alpha}$ is likely to be a result of our sample being biased against red star formers ({\S\,\ref{sec:data}}). {The range in L$_{H\alpha}$--M$_r$ probed by the three higher--$z$ bivariate functions become more limited and faint L$_{H\alpha}$--M$_r$ bins become more incomplete with increasing redshift.} The L$_{H\alpha}$--M$_r$ bins with low galaxy number statistics are indicated in Figure\,\ref{fig:vmax_bilf}. Note that the errors in $\log\,\Phi$ for L$_{H\alpha}$--M$_r$ bins with small numbers of galaxies (e.g.\,numbers $\lesssim3$) is large.

We further assign a photometric blue or red class to {each star forming galaxy} in our sample. {For the analysis presented in this section, the blue--red classification is determined using the $g-r$ and M$^{0.1}_r$ colour--magnitude cut defined in \cite{Zehavi11} and used by \cite{Loveday12}.}
\begin{equation}
	(g-r)_{model}^{0.1} = 0.15 - 0.03M_r^{0.1}.
	\label{eq:color}
\end{equation}
There are a number of blue--red galaxy classification schemes in the literature \citep[e.g.][ Taylor et al.\, submitted]{Bell03, Baldry04, Peng10}. \cite{Baldry04, Baldry12}, for example, advocate a non--linear cut in ($u-r$) rest--frame colour and M$_r$. Since the observed bivariate data distribution in colour--magnitude plane is non--linear \citep{Baldry04}, a non--linear cut in ($u-r$)--M$_r$ space would improve our blue--red selection. {As our goal in this part of the analysis is to compare the univariate functions computed from the bivariate functions with the univariate functions of \cite{Loveday12}, the same colour--magnitude selection employed by \cite{Loveday12} is used.} A comprehensive discussion of different colour--magnitude classifications and their implications is presented in Taylor et al., (submitted).

\begin{figure*}
	\begin{center}
		\includegraphics[width=0.49\textwidth]{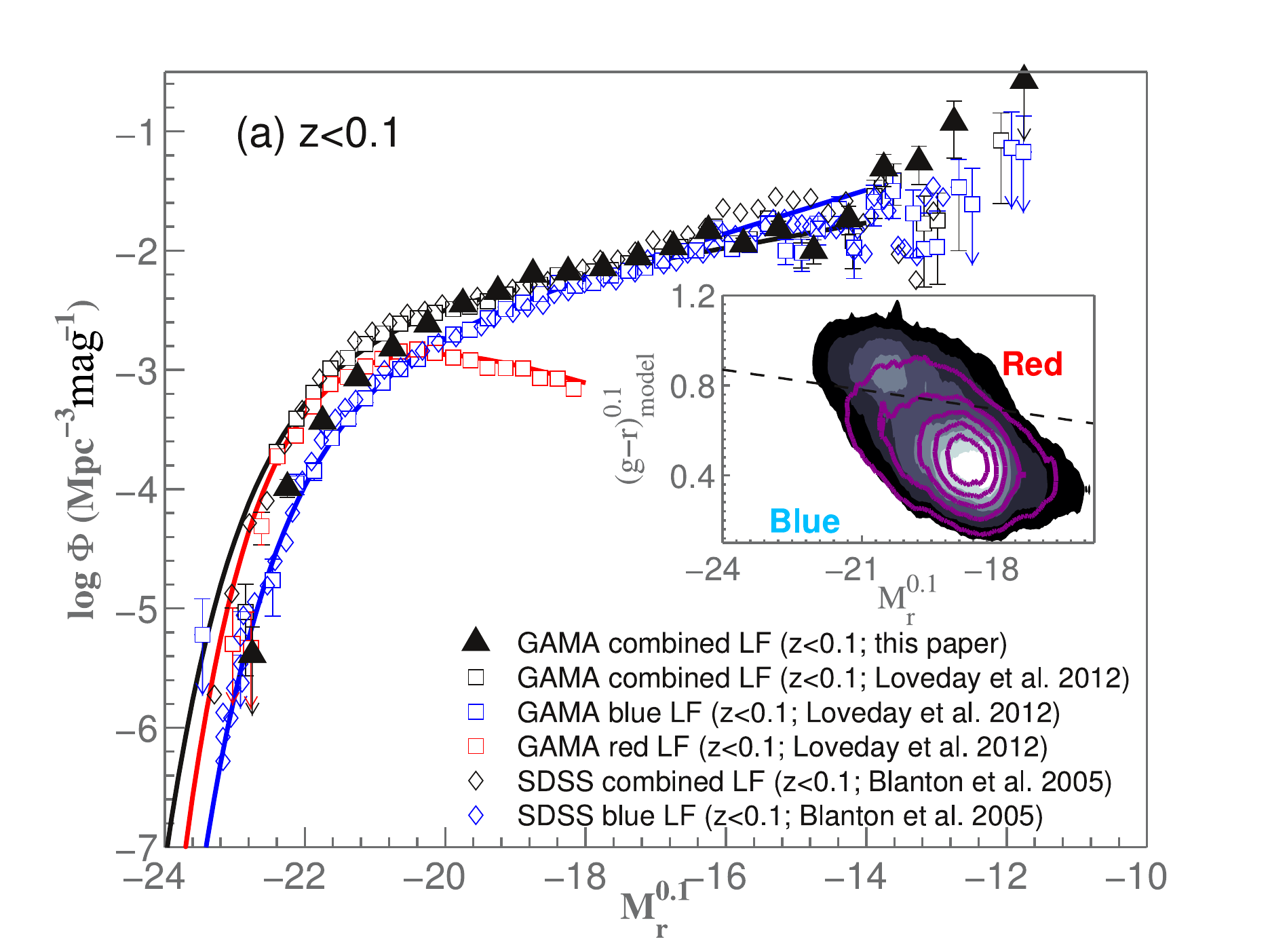}
		\includegraphics[width=0.49\textwidth]{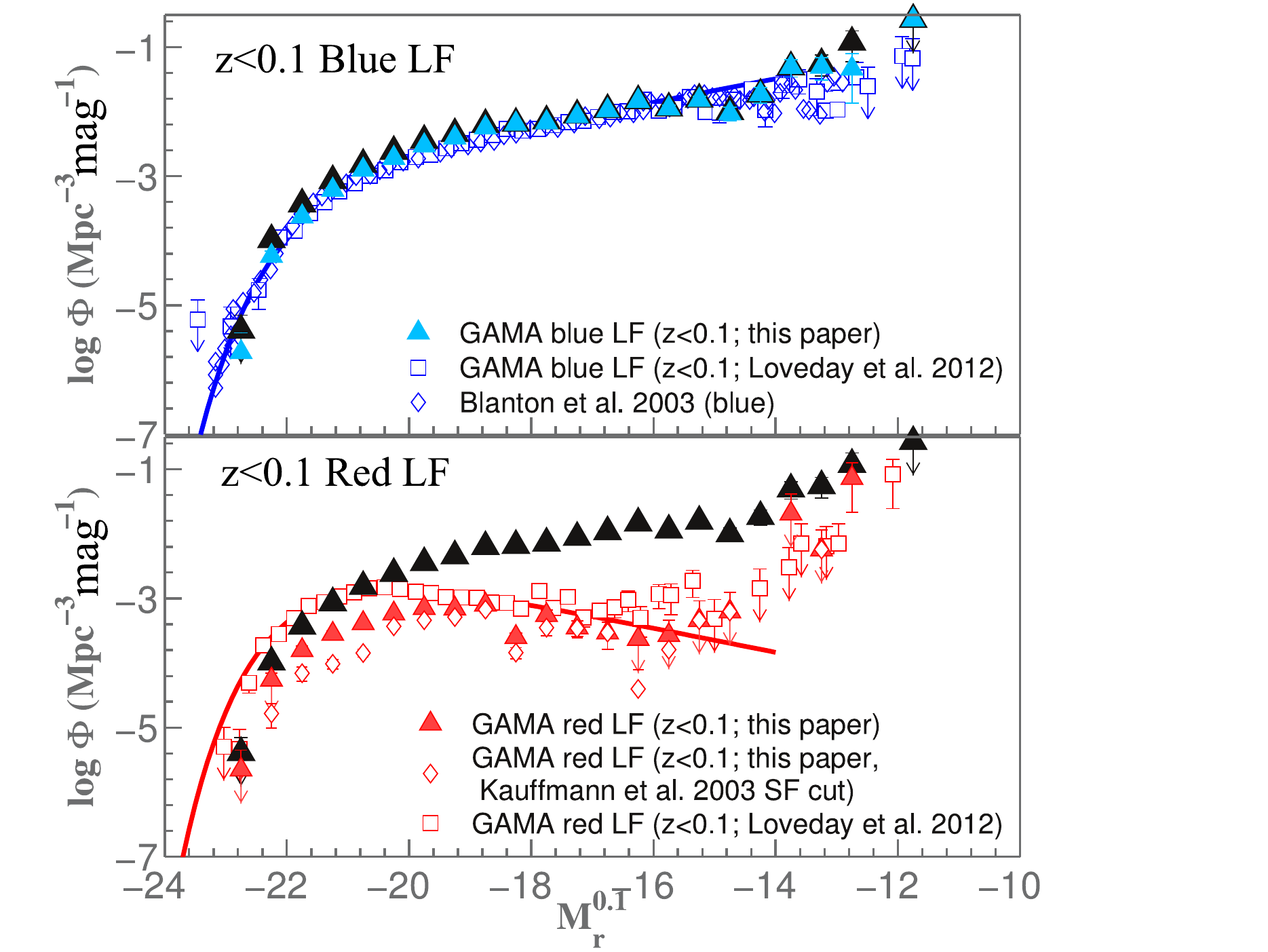}			
		\caption{GAMA M$_r$ total (left panel), blue (right top) and red (right bottom) LFs of star forming galaxies derived from integrating the $z<0.1$ bivariate L$_{H\alpha}$--M$_r$ functions. These LFs are compared against the M$_r$ LFs presented in \citet{Loveday12} for all GAMA galaxies (open squares) and \citet{Blanton05} for all SDSS galaxies (open diamonds). The solid lines show the best fitting Schechter functions to $z<0.1$ LFs from \citet{Loveday12}. The left panel inset shows the data density distributions in linear space of all $z<0.1$ galaxies regardless star formation (filled contours) and star forming galaxies with F$_{H\alpha}>1\times10^{-18}$\,W/m$^2$ (solid contours) in $(g-r)_{model}^{0.1}$ and M$_r^{0.1}$ plane. The colour--magnitude cut (Eq.\,\ref{eq:color}) used to split blue and red galaxies is shown as a dashed line. The open red diamonds in right bottom panel indicate the red SF M$_r^{0.1}$ LF constructed using the (pure) SF galaxies selected according to \citet{Kauffmann03} SF/AGN prescription instead of that of \citet{Kewley02}.}
		\label{fig:comprison1}
	\end{center}
\end{figure*}
\begin{figure*}
	\begin{center}		
		\includegraphics[trim=.7cm .1cm 1.5cm 1.0cm, clip=true, width=0.48\textwidth]{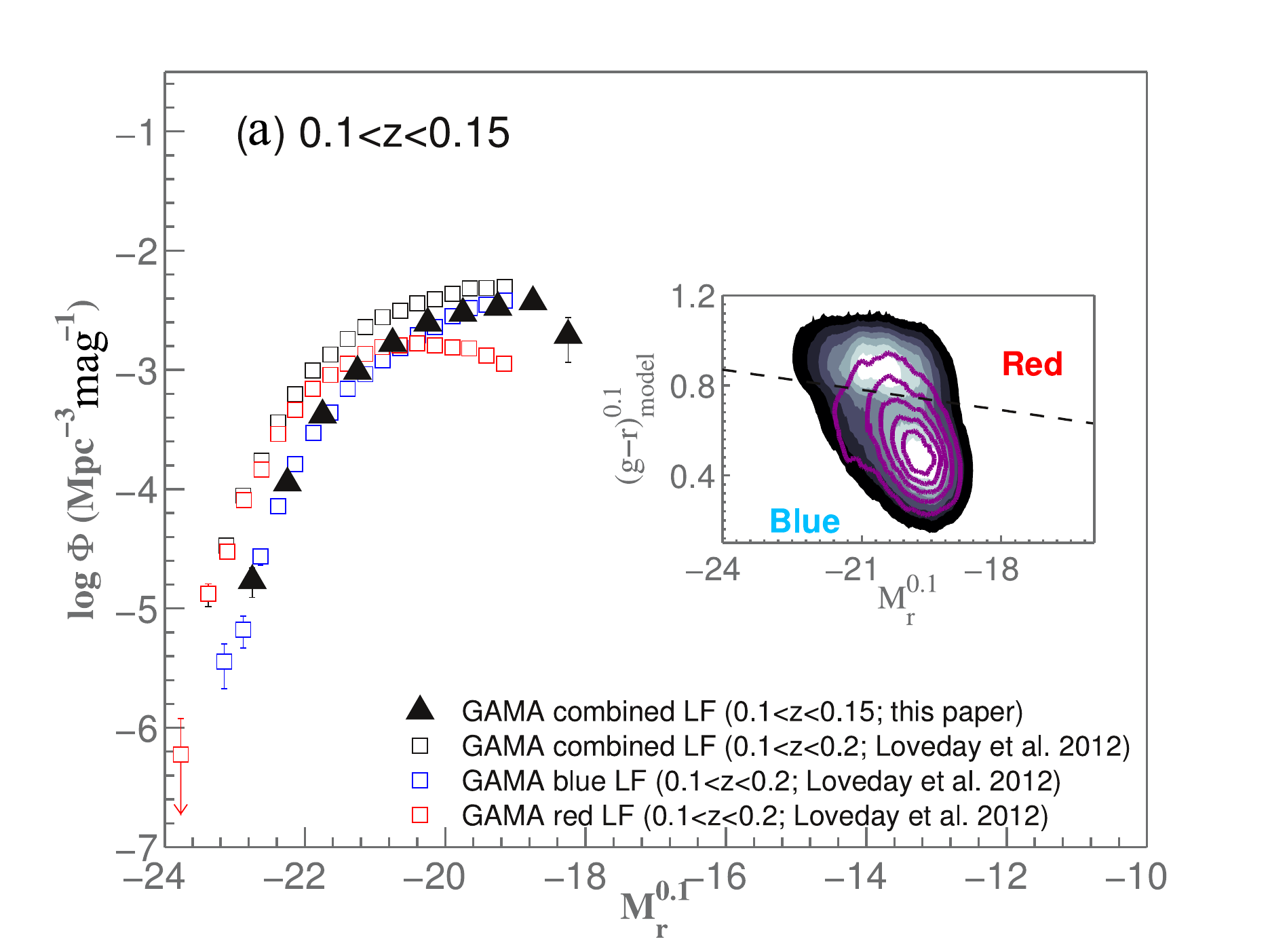}
		\includegraphics[trim=.7cm .1cm 1.5cm 1.0cm, clip=true, width=0.48\textwidth]{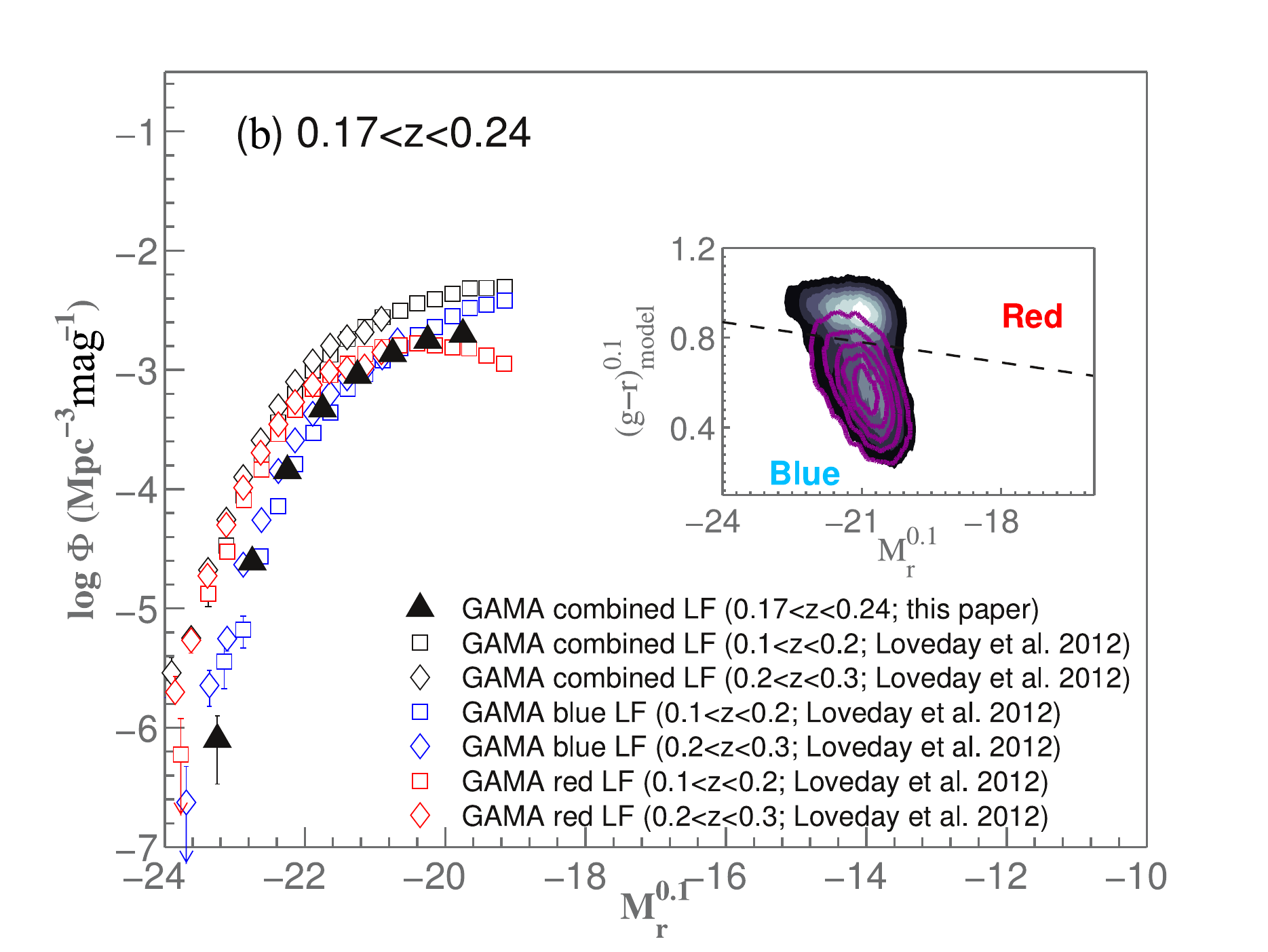}
		\includegraphics[trim=.7cm .1cm 1.5cm 1.0cm, clip=true, width=0.48\textwidth]{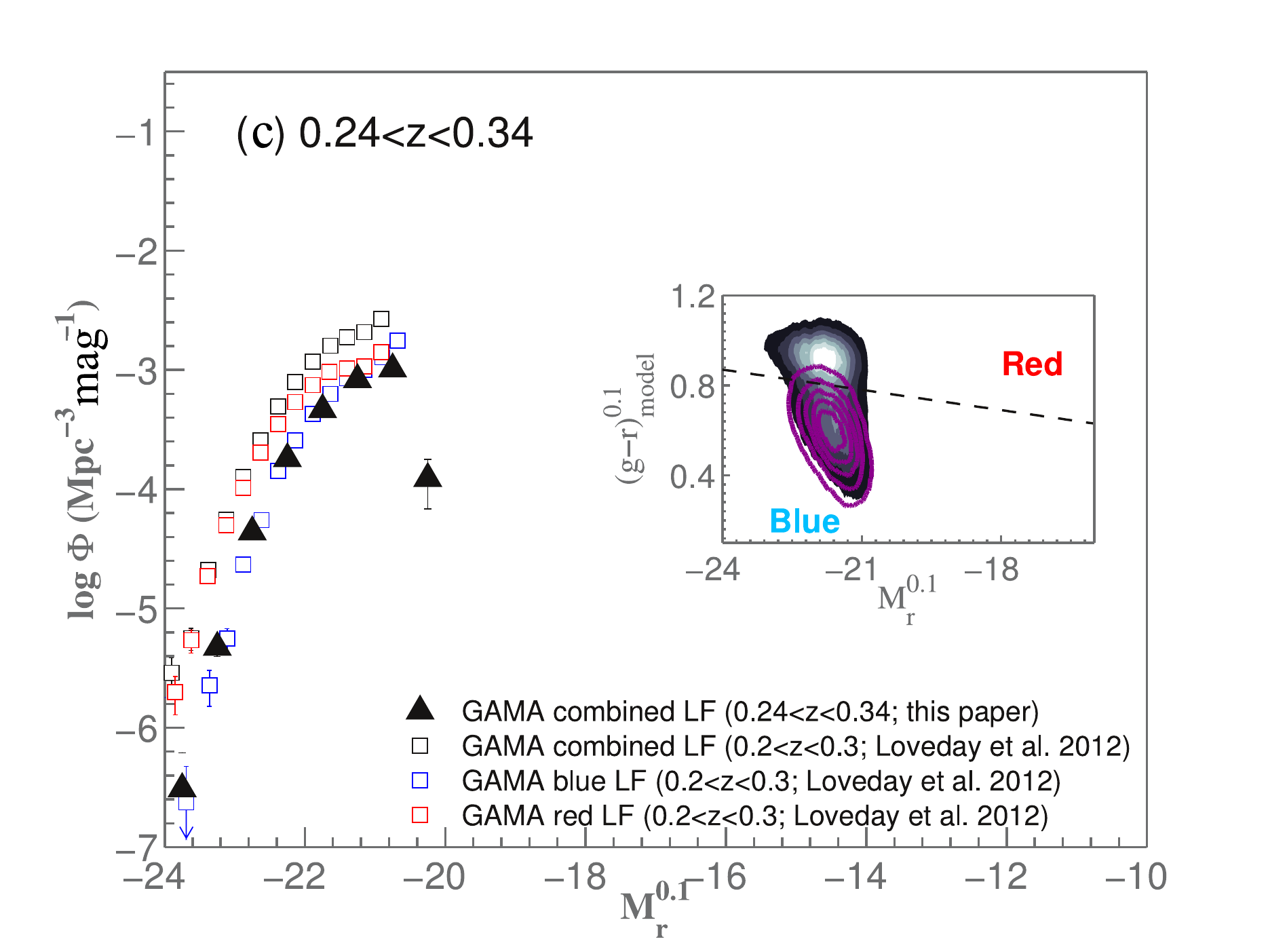}
		\caption{GAMA higher--$z$ M$_r^{0.1}$ LFs of all SF galaxies computed by integrating the bivariate L$_{H\alpha}$--M$_r$ functions. The open symbols in each panel show the GAMA all, blue and red M$_r^{0.1}$ LFs of \citet{Loveday12} over similar redshift ranges. The insets show the data density distributions in $(g-r)_{model}^{0.1}$ and M$_r^{0.1}$ of all galaxies regardless of star formation (grey shading) and SF galaxies with F$_{H\alpha}>1\times10^{-18}$\,W/m$^2$ (magenta contours). These distributions correspond to the redshift ranges given in the plot key. The colour--magnitude cut (Eq.\,\ref{eq:color}) is shown as a dashed line.}
		\label{fig:comprison2}
	\end{center}
\end{figure*}

{The bivariate L$_{H\alpha}$--M$_r$ functions of blue and red H$\alpha$ SF sub--populations split in four redshift bins are shown in Figure\,\ref{fig:vmax_bilf_photom}.} The range in M$_r$ and L$_{H\alpha}$ probed by the blue  L$_{H\alpha}$--M$_r$ functions {and their number densities are similar to that of the total H$\alpha$ SF LFs (Figure\,\ref{fig:vmax_bilf}) at all redshifts}, implying that the star forming bivariate LF, while drawn from both photometrically classified blue and red sub--populations, consists mostly of blue galaxies. 

\subsection{{$r$--band galaxy LFs of H$\alpha$ star formers}} \label{subsec:rLFs}

By integrating the bivariate L$_{H\alpha}$--M$_r$ functions over the L$_{H\alpha}$ axis, the M$_r$ LFs of the total H$\alpha$ SF sample and photometrically classified blue and red H$\alpha$ SF sub--samples can be recovered. We explore the evolution of luminosity densities (LDs) in \S\,\ref{sebsec:rlumdensity}.

{Figure\,\ref{fig:comprison1} shows the $z<0.1$ $r$--band galaxy LFs of all H$\alpha$ SF (left panel), blue H$\alpha$ SF (right top panel) and red H$\alpha$ SF (right bottom) galaxies computed from integrating the corresponding bivariate functions}\footnote{The bivariate LFs shown in Figures\,\ref{fig:vmax_bilf} and \ref{fig:vmax_bilf_photom} use $r$--band absolute magnitude $k$--corrected to $z=0$. To obtain the M$^{0.1}_r$ LFs shown in Figure\,\ref{fig:comprison1}, we calculate the bivariate LFs using $r$--band absolute magnitude $k$--corrected to $z=0.1$ to compare our results directly with \cite{Loveday12}.}. {These LFs are compared with the GAMA $r$--band LFs from \cite{Loveday12} and SDSS $r$--band LFs from \cite{Blanton05}, which are are shown as open squares and diamonds, respectively, in Figure\,\ref{fig:comprison1}}. 

The $z<0.1$ M$^{0.1}_r$ LF of all H$\alpha$ SF galaxies {(left panel of Figure\,\ref{fig:comprison1})} closely follows the M$^{0.1}_r$ LF of all GAMA galaxies from \cite{Loveday12}. {In fact, the bright end of the all H$\alpha$ SF M$^{0.1}_r$ LF very closely tracks the GAMA red M$^{0.1}_r$ LF \citep[][]{Loveday12}}, demonstrating that a significant fraction of the $z<0.1$ galaxies classified as red have detected H$\alpha$ emission at all M$^{0.1}_r$ values. 
This is further corroborated by the colour--magnitude distributions (Figure\,\ref{fig:comprison1} inset) of all $z<0.1$ galaxies regardless of star formation (filled contours) and $z<0.1$ H$\alpha$ SF galaxies (purple contours), where the purple contours extend beyond the blue population. {The small discrepancy evident between the blue M$^{0.1}_r$ LF of H$\alpha$ SF galaxies and that from \cite{Loveday12} is likely a result of the differences in the formulation of the 1/V$_{\rm max}$ technique. By taking into account the minimum of V$_{\rm max,r}$ and V$_{\rm max, H\alpha}$ (paper~I) rather than only V$_{\rm max,r}$ as done by \cite{Loveday12}, where we take into account the maximum volume corrections based on both $r$--band magnitude and H$\alpha$ flux limits.}

{Furthermore, we use the \cite{Kauffmann03} criteria on discriminating pure--SF and SF--AGN composites to further remove likely SF--AGN composites from the red SF sample. The resultant M$^{0.1}_r$ LF is shown as open red diamonds in the right bottom panel of Figure\,\ref{fig:comprison1}. The open symbols are slightly displaced from  the M$^{0.1}_r$ LF constructed using SF galaxies selected based on the \cite{Kewley02} criteria (filled symbols) at the bright end of the LF, and overlap with the filled symbols at the faint--end, implying that only a small  fraction of red galaxies are removed from the original red SF sample by the \cite{Kauffmann03} cut.} {We have also used the method of \cite{Cid11} for differentiating SF galaxies from AGNs. Even with the AGN/SF cuts recommended by \cite{Cid11}, more than $50\%$ of the red galaxy sample at a given redshift retain their star forming status. So that a fraction of galaxies classified as red at all M$^{0.1}_r$ values are in fact currently forming stars.} Moreover, the analysis of \cite{Lopez13} exploring the properties of SDSS and GAMA galaxies find that a large fraction of galaxies detected in GAMA are SF in comparison to SDSS as GAMA is deeper than SDSS, and therefore more sensitive to low mass galaxies at low redshift, which are mostly dominated by star formation. 

{The higher--$z$ ($z$ up to $0.34$) M$^{0.1}_r$ LFs of all H$\alpha$ SF galaxies {(Figure\,\ref{fig:comprison_app})} overlap with the GAMA blue LFs of \cite{Loveday12} over similar redshift ranges, while the fractional contribution from red H$\alpha$ SF galaxies to the GAMA red  M$^{0.1}_r$ LFs \citep{Loveday12} progressively drops with increasing redshift. The contours in the insets of Figure\,\ref{fig:comprison2} depicting the distribution of all galaxies (filled contours) and SF galaxies (magenta contours) at different redshifts show that the contours of SF galaxies become more restricted to the blue sub--population with increasing redshift, collaborating the drop in red SF number densities seen in Figure\,\ref{fig:comprison_app}. This is most likely due to the flux limit biasing our sample against these systems (Figure\,\ref{fig:Halpha_in_Mr}), and possibly also due to the difficulty in measuring spectral lines in more obscured, weak emission line systems at higher redshifts.} 

\section{Modelling of bivariate L$_{H\alpha}$ dependent functions and implications for luminosity and SFR densities } \label{sec:funcfits}

LFs, emission--line and photometric alike, are traditionally parameterised with a \cite{Schechter76} function, which is then integrated to obtain a luminosity or SFR density. 
The linear form of the Schechter function{\footnote{The logarithmic form of the Schechter function is expressed as, \begin{equation}
	\begin{aligned}
	\Phi(\log L) d\,\log L & = \ln(10) \Phi^* 10^{(\log{L}-\log{L^{*}})(\alpha+1)} \\
					& \exp[-10^{(\log{L}-\log{L^{*}})}] d\,\log{L}. 
	\nonumber
	\end{aligned}
\end{equation}
}}, 
\begin{equation}
	\Phi(L) dL = \Phi^* \left(\frac{L}{L^{*}}\right)^{\alpha} \exp \left(-\frac{L}{L^{*}}\right) d \left(\frac{L}{L^{*}}\right), 
	\label{eq:schechter}
\end{equation}
behaves as a power--law with a slope $\alpha$ for luminosities ($L$) less than the characteristic luminosity ($L^*$) and as an exponential for $L>L^*$, with the normalisation given by $\Phi^*$. From this, the predicted luminosity density is given by,
{
\begin{equation}
	\rho_{_{Lfit}} = \int\limits_0^\infty L\,\Phi(L)\,dL = \Phi^*L^*\Gamma(\alpha+2),   
	\label{eq:schechter_LD}
\end{equation}
where $\Gamma$ is the Gamma function, and a conservative limit is given by, 
\begin{equation}
	\rho_{_{Lfit, lim}} = \int\limits_0^{L_min} L\,\Phi(L)\,dL = \Phi^*L^*\Gamma(\alpha+2, L_{min}/L^*).    
	\label{eq:schechter_LD}
\end{equation}
where $\Gamma$ is now the incomplete Gamma function. 
}

Galaxy broadband LFs are, generally, well described by a Schechter function \citep[e.g.][]{Hill10, Loveday12}. Several studies \citep[e.g.][]{Blanton05}, however, find that a double Schechter function is best suited at capturing the whole shape of the galaxy LF, especially if the range in magnitude probed is relatively large. 

While the galaxy broadband LFs show an exponential--like fall in $\Phi$ with increasing brightness, the star--forming LFs (e.g.\,H$\alpha$, far infrared, radio) show a less steep fall in $\Phi$ \citep[][paper~I]{Saunders90, Salim12}. For this reason, to characterise the star forming LFs we adopt a \cite{Saunders90} {function. The linear form of the Saunders function,} 
\begin{equation}
	\Phi(L) dL =  \Phi^{*} \left(\frac{L}{L^{*}}\right)^{\alpha} \exp \left[-\frac{1}{2\sigma^2}\log^2\left(1+\frac{L}{L^{*}}\right)\right] d \left(\frac{L}{L^{*}}\right),
	\label{eq:saunders}
\end{equation}
behaves in a similar fashion to a Schechter function for $L<L^*$ and as a Gaussian in log luminosity with a width ($\sigma$) for $L>L^*$.

\begin{figure*}
	\begin{center}
		\includegraphics[trim=.5cm .1cm 1.5cm .5cm, clip=true, width=0.49\textwidth]{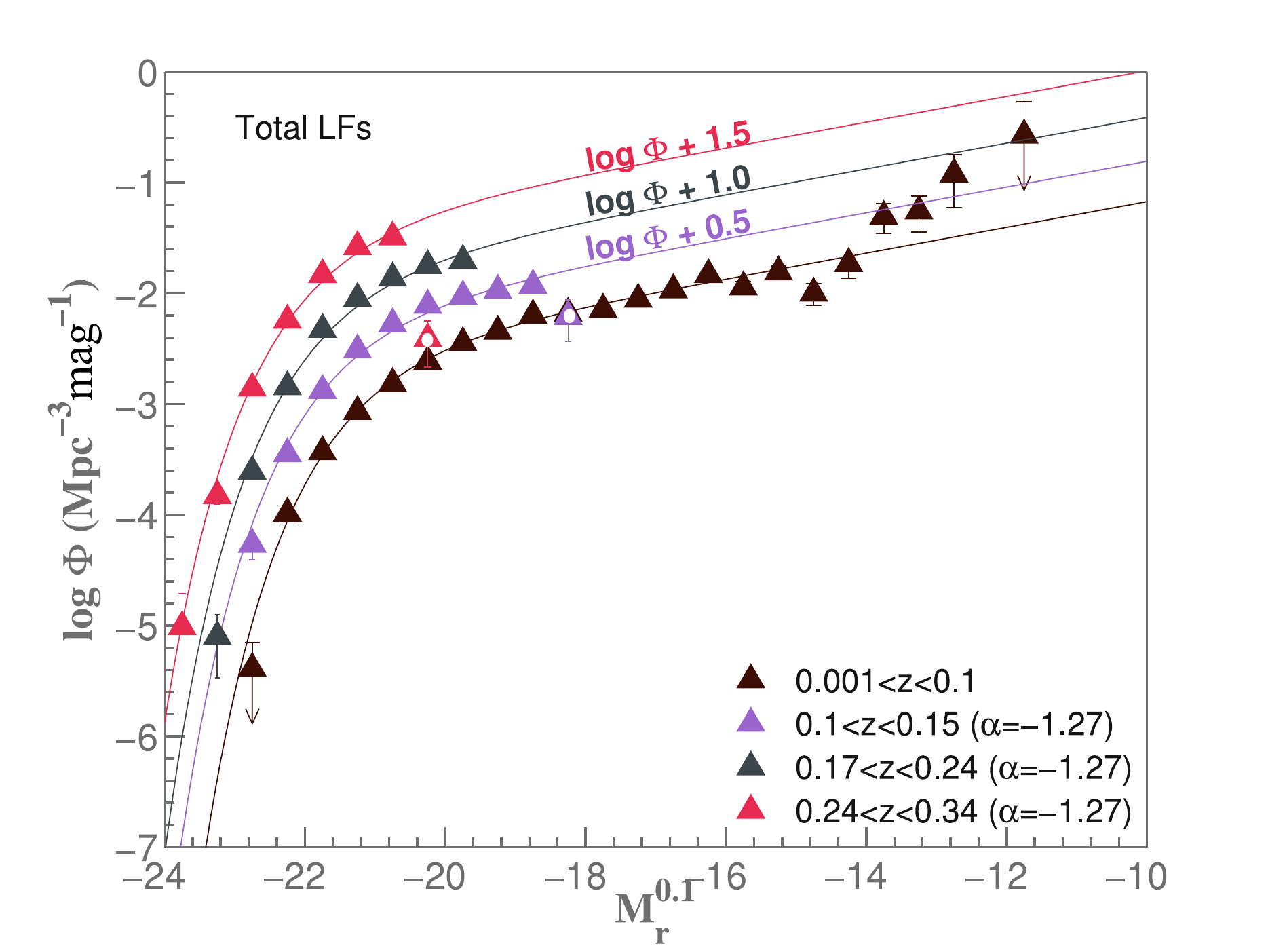}
		\includegraphics[trim=.5cm .01cm 1.5cm .1cm, clip=true, width=0.49\textwidth]{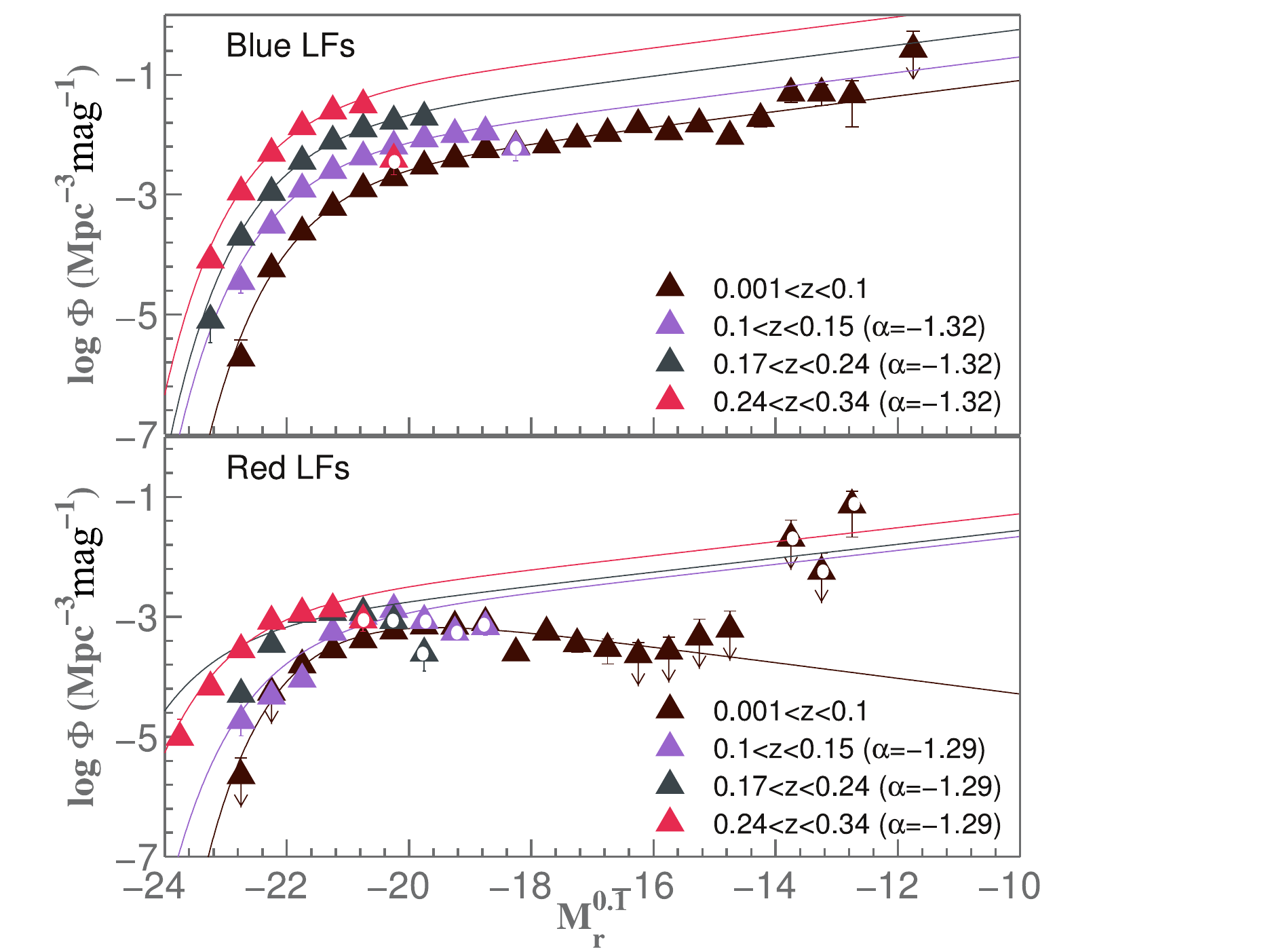}
		\caption{The best fitting Schechter functions to the M$^{0.1}_r$ LFs shown in Figures\,\ref{fig:comprison1}, \ref{fig:comprison2} and \ref{fig:comprison_app}. The faint--end slope, $\alpha$, is fixed at $-1.29$ for the higher--$z$ combined (left panel) and red SF (right bottom) LFs. It is fixed at $-1.32$ for the higher--$z$ blue LFs (Right top). Each higher--$z$ LF is scaled up by the factors stated in the left panel to aid legibility.}
		\label{fig:MrLF_fits}
	\end{center}
\end{figure*}

{We fit Schechter functions to the M$^{0.1}_r$ LFs of SF galaxies shown in Figures\,\ref{fig:comprison1} and \ref{fig:comprison_app} using a Levenberg--Marquardt routine to find the minimum $\chi^2$ to the binned LF data points. The resultant Schechter functional fits and their best fitting parameters are presented in Figure\,\ref{fig:MrLF_fits} and Table\,\ref{table:LFfits}. Due to the lack of faint SF galaxies at higher--$z$, a consequence of the survey magnitude selection, the best fitting slopes for the $0.1<z<0.15$, $0.17<z<0.24$ and $0.24<z<0.34$ LFs cannot be measured reliably from the observed LF. Instead, we constrain the faint--end slopes of higher--$z$ LFs to be equal to the best fitting slopes of their respective $z<0.1$ LFs.} {The low--$z$ red SF galaxy LF has a poorly constrained $\alpha$ due to a lack of bright galaxies necessary to disentangle the degeneracy between $\alpha$ and M$^*$, therefore we assume $\alpha=-1.29$ (i.e.\,$\alpha$ estimated from the $z<0.1$ M$^{0.1}_r$ LF all H$\alpha$ SF galaxies) for the higher--$z$ red LFs.} 

\begin{table*}
		\begin{minipage}{\textwidth}
			\begin{center}
			\caption{The best fitting Schechter function parameters corresponding to the functional fits shown in Figure\,\ref{fig:MrLF_fits}, and the luminosity densities computed from integrating the fits.}
			\begin{tabular}{l|cccc}
\hline
         $z$ 	 & 	M$^*_r$ 	&   $\log$ $\Phi^*$ 	&  $\alpha$            &  $\log\,\rho_{Lfit}$ \\
                 & 			&   (Mpc$^{-3}$)	 & 		     	     &  (L$_{\odot}\,$Mpc$^{-3}$)\\
                               
\hline
\hline

Combined 		& 			&   				& 		     	      & 		\\
$z<0.10$ 	 						
& $-20.93\pm0.15$ 		
& $-2.41\pm0.14$		
& $-1.29\pm0.06$ 		
& $7.97\pm0.05$   					\\ % L density 

$0.10<z<0.15$ 	 						
& $-21.21\pm0.06$ 		
& $-2.54\pm0.05$		
& $-1.29$\footnote{$\alpha$ is fixed at $-1.29$, the best fitting slope for the $z<0.1$ blue and red combined M$^{0.1}_r$ LF of H$\alpha$ SF galaxies.} 		
& $7.96\pm0.03$   					\\ % L density 

$0.17<z<0.24$ 	 						
& $-21.34\pm0.13$ 		
& $-2.64\pm0.15$		
& $-1.29^{a}$ 		
& $7.91\pm0.09$   					\\ % L density 

$0.24<z<0.34$ 	 						
& $-21.52\pm0.02$ 		
& $-2.72\pm0.04$		
& $-1.29^{a}$ 		
& $7.90\pm0.03$   					\\ % L density 

Blue 		& 			&   				& 		     	      &			\\
$z<0.10$ 	 						
& $-20.80\pm0.07$ 		
& $-2.46\pm0.07$		
& $-1.32\pm0.05$ 		
& $7.89\pm0.01$   					\\ % L density 

$0.10<z<0.15$ 	 						
& $-21.21\pm0.08$ 		
& $-2.62\pm0.06$		
& $-1.32$\footnote{$\alpha$ is fixed at $-1.32$, the best fitting slope for the $z<0.1$ blue M$^{0.1}_r$ LF of H$\alpha$ SF galaxies.} 	 		
& $7.89\pm0.06$    					\\ % L density 

$0.17<z<0.24$ 	 						
& $-21.25\pm0.07$ 		
& $-2.66\pm0.11$		
& $-1.32^b$	 		
& $7.86\pm0.08$    					\\ % L density 

$0.24<z<0.34$ 	 						
& $-21.40\pm0.06$ 		
& $-2.71\pm0.08$		
& $-1.32^b$	 		
& $7.89\pm0.05$    					\\ % L density 

Red 		& 			&   				& 		     	      &			\\
$z<0.10$ 	 						
& $-20.76\pm0.09$ 		
& $-2.85\pm0.08$		
& $-0.67\pm0.07$ 		
&  $7.31\pm0.04$    					\\ % L density 

$0.10<z<0.15$ 	 						
& $-21.44\pm0.25$ 		
& $-3.45\pm0.17$		
& $-1.29^{a}$		
& $7.14\pm0.06$    					\\ % L density 

$0.17<z<0.24$ 	 						
& $-22.74\pm1.07$ 		
& $-4.00\pm0.27$		
& $-1.29^{a}$ 		
& $7.12\pm0.16$    					\\ % L density 

$0.24<z<0.34$ 	 						
& $-22.16\pm0.17$ 		
& $-4.16\pm0.22$		
& $-1.29^{a}$ 		
& $6.82\pm0.14$    					\\ % L density 

\hline
			\end{tabular}
		\label{table:LFfits}
			\end{center}
		\end{minipage}
\end{table*}

{\subsection{Functional fits to bivariate functions}\label{subsubsec:simple_model}}

In this section, {we explore a simple analytic approach} to modelling the bivariate L$_{H\alpha}$--M$_r$ function. {Such a} parameterisation of bivariate functions is useful in comparing the distributions drawn from differently selected samples and for studying the redshift evolution inclusive of selection biases.

\cite{Choloniewski85} and \cite{deJong00} developed a formalism to link the distribution of galaxy scale sizes, assumed to be Gaussian, to the luminosity parameterised by a Schechter function. {Their} analytic expression {as well as other} related functional forms have widely been used to model bivariate brightness profiles \citep[e.g.][]{deJong00, Cross01, Driver05, Ball06},  size--luminosity \citep[e.g.][]{deJong00, Huang13} and colour--luminosity relationships \citep[e.g.][]{Chapman03, Chapin09}. Similar functional forms have also been formulated by \cite{Yang05} and \cite{Cooray05} in the context of conditional LFs that specify the average number of galaxies with luminosities in the range $L\pm\Delta L/2$ that reside in a given halo mass. 

Our motivation for modelling the GAMA bivariate functions is to correct for the apparent lack of evolution in H$\alpha$ SFR densities with redshift reported in paper~I. The low redshift bivariate model can be used as reference to gain an understanding of the extent to which the higher--$z$ ($0.1\lesssim z\lesssim0.34$) samples are affected by incompleteness. 

{Below we describe the construction of a simple analytic model}, denoted $\Psi(M, L_{H\alpha})$, to describe the  bivariate functions presented in this paper, assuming that the bivariate functions can be written as a product of two functions \citep{Choloniewski85, Corbelli91, Hopkins98}. Naturally, as a Schechter function best describes the M$_r$ LFs presented in \S\,\ref{subsec:rLFs} and a Saunders function best describes the GAMA H$\alpha$ LFs (paper~I), these functions are adopted to represent the bivariate distributions.  

To link these functional forms in a bivariate analytic relation, we begin by establishing how {L$^*_{H\alpha}$} in the H$\alpha$ LF varies as a function of M$_r$. {The relationship between L$^*_{H\alpha}$ and M$_r$ is clearly evident in Figure\,\ref{fig:normalised_LFs}, where we divide the lowest--$z$ L$_{H\alpha}$--M$_r$ function by L$_{H\alpha}$ LF at a given M$_r$.} The symbols indicate the best fitting  L$^*_{H\alpha}$ in different magnitude ranges, estimated by fitting a Saunders function (Eq.\,\ref{eq:saunders}) with a non--varying faint--end slope (blue) to the data and fitting a Gaussian (green),
{
\begin{equation}
	\Phi(L) dL =  \frac{\Phi^*}{\sigma\sqrt{2\pi}}\exp\left[-\frac{1}{2\sigma^2}\log^2\left(\frac{L}{L^*}\right)\right] d \left(\frac{L}{L^{*}}\right),
	\label{eq:gaussian}
\end{equation}
with a Gaussian width, $\sigma$, that is allowed to vary to the data.}

{The relationship between $L^*_{H\alpha}$ and M$_r$ seen in Figure\,\ref{fig:normalised_LFs} can be approximated as 
\begin{equation}
	L^{*}_{H\alpha} =  L^{*}_{H\alpha}(M^0_r) 10^{-0.4\beta(M_r-M^0_r)}. 
	\label{eq:connect}
\end{equation}
}
Using Eq.\,\ref{eq:connect} to connect Eq.\,\ref{eq:schechter} and \ref{eq:saunders}, and expressing the distribution in terms of M$_r$, we arrive at the full bivariate expression: {
\begin{equation}
	\begin{aligned} 
		 & \Psi(M_r, L_{H\alpha})  =  \Phi(M_r) \times \Phi(M_r, L_{H\alpha}) = 0.4\,\ln(10) \\
		                    & \Psi^{^*}10^{-0.4(\alpha_{_M}+1)(M_{_r}-M^{^*}_{_r})} \\
				   & \times \exp[-10^{-0.4(M_{_r}-M^{^*}_{_r})}] \\
				   & \times \left\{ \left[\frac{L_{_{H\alpha}}}{L^{^*}_{_{H\alpha}}(M^{^0}_{_r})}\right]\left[10^{0.4\beta(M_{_r}-M^{^0}_{_r})}\right] \right\}^{\alpha_{_L}} \\
				   & \times \exp\left\{-\frac{1}{2\sigma^2}\log^2\left[1+\left(\frac{L_{_{H\alpha}}}{L^{^*}_{_{H\alpha}}(M^{0}_{_r})}\right)\left(10^{0.4\beta(M_{_r}-M^{^0}_{_r})}\right)\right]\right\}.
	\end{aligned}
	\label{eq:bivariate}
\end{equation}
\begin{figure}
	\begin{center}
		\includegraphics[width=0.4\textwidth]{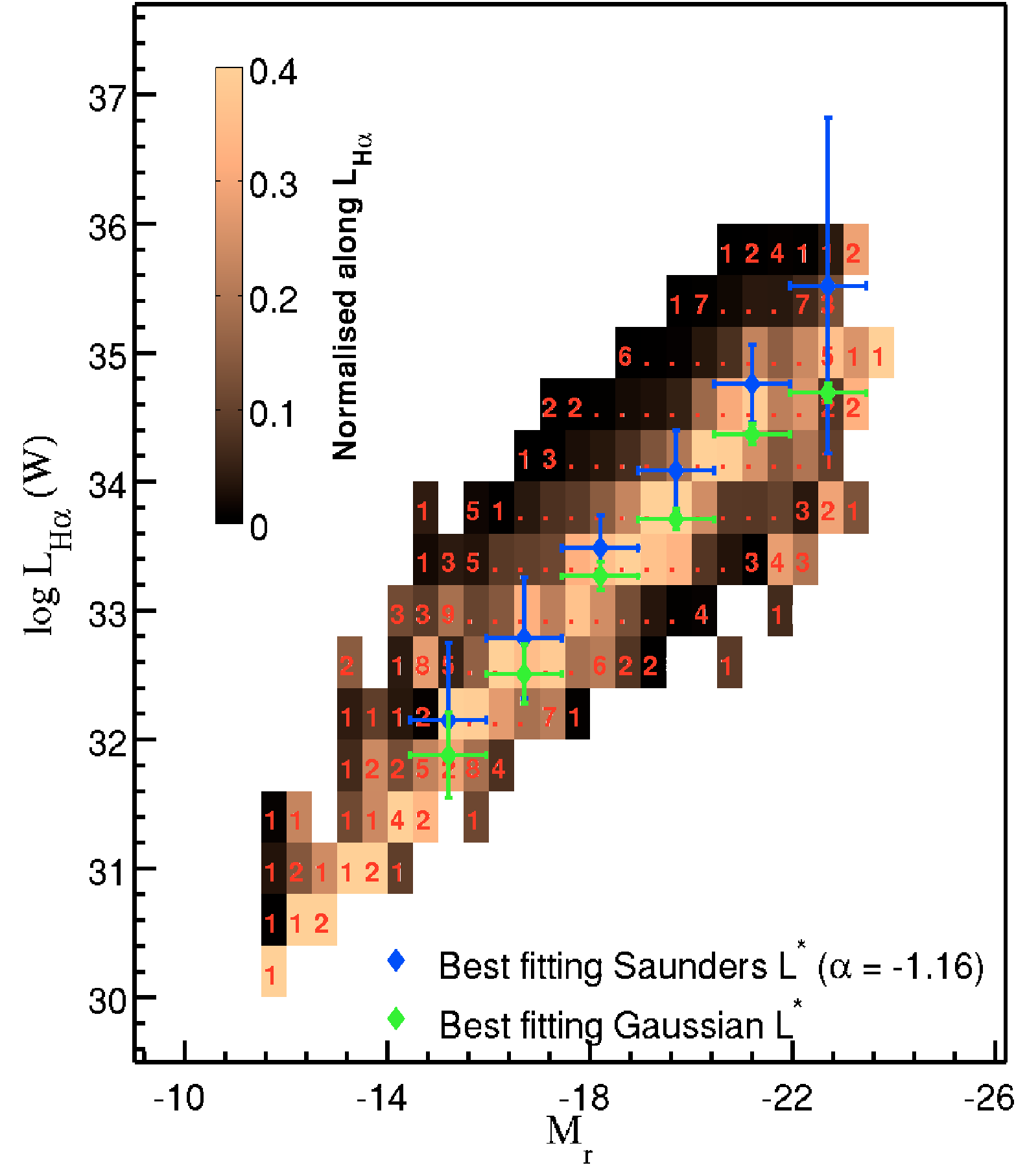}
		\caption{The low--$z$ bivariate function (Figure\,\ref{fig:vmax_bilf}a) normalised along the L$_{H\alpha}$ direction. The symbols indicate the variation in characteristic luminosity ($L^*_{H\alpha}$) with respect to M$_r$ for Saunders functional fits (blue) and Gaussian fits (green) to the H$\alpha$ LFs in different M$_r$ ranges. For the Saunders fits, the faint--end slope ($\alpha$) is fixed at $-1.16$, the best fitting $\alpha$ of the $z<0.1$ GAMA H$\alpha$ LF (paper~I). The horizontal errors indicate the M$_r$ range probed and the {vertical errors indicate the errors associated with best fitting L$_{H\alpha}^*$ values}.  In both cases, the variation in {$L^*_{H\alpha}$} can be approximated through a power--law (Eq.\,\ref{eq:connect}).}
		\label{fig:normalised_LFs}
	\end{center}
\end{figure}
\begin{figure}
	\begin{center}
		\includegraphics[trim=.1cm 0.05cm .001cm .05cm, clip=true, width=0.48\textwidth]{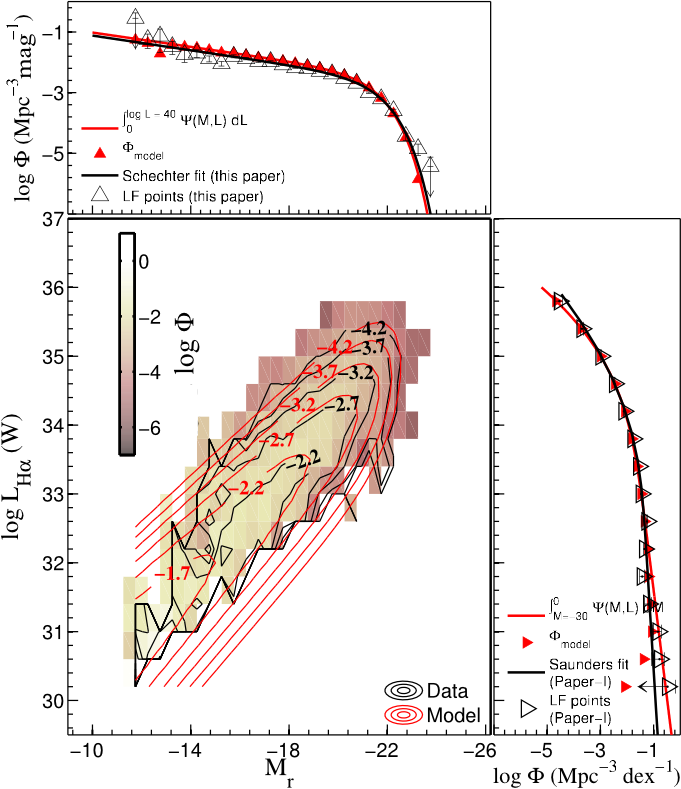}
		\caption{The $z<0.1$ bivariate function and the fitted model (Eq.\,\ref{eq:bivariate}). The model contours provide a reasonable description of the data (black lines). The top and right panels compare the univariate M$_r$ (Figure\,\ref{fig:comprison1}) and H$\alpha$ (Figure\,7 of paper~I) LFs (symbols) computed from integrating the $z<0.1$ bivariate M$_r$--H$\alpha$ function (Figure\,\ref{fig:vmax_bilf}a) with that predicted by the model. Also shown are the Schechter and Saunders functional fits to the univariate M$_r$ and H$\alpha$ LFs (Figure\,11 of paper~I). The colour scale indicates log number densities in the unit of Mpc$^{-3}$ dex$^{-1}$ mag$^{-1}$.}
		\label{fig:simple_model}
	\end{center}
\end{figure}
} We assume $M^0_r = -19.0$ and fit the, now, seven free parameter model to the $z<0.1$ L$_{H\alpha}$--M$_r$ function through a nonlinear $\chi^2$ minimisation routine based on Levenberg--Marquardt method using the Poisson errors of the bivariate function measurements. The resultant best fitting model is shown in Figure\,\ref{fig:simple_model} and the best fitting model parameters are given in Table\,\ref{table:model_param2}. 

{To determine the best fitting parameters for the higher--$z$ ($0.1\lesssim z\lesssim0.34$) bivariate functions, we fix $\alpha_M$ and $\alpha_L$ to be equal to the best--fitting $z<0.1$ model values. The best fitting values determined in this manner are given in Table\,\ref{table:model_param2}. Given the log--normal like shape of the $z<0.1$ normalised bivariate function at brighter magnitudes (Figure\,\ref{fig:normalised_LFs}), which is similar to that exhibited by size--magnitude or colour--magnitude distributions \citep{Choloniewski85, Chapman03, Huang13}, a Schechter--Gaussian model can also be fitted. 
\begin{equation}
	\begin{aligned} 
		 & \Psi(M_r, L_{H\alpha})  = 0.4\,\ln(10) \\
		                    & \Psi^{^*}10^{-0.4(\alpha_{_M}+1)(M_{_r}-M^{^*}_{_r})} \\
				   & \times \exp[-10^{-0.4(M_{_r}-M^{^*}_{_r})}] \\
				   & \times \exp\left\{-\frac{1}{2\sigma^2}\log^2\left[1+\left(\frac{L_{_{H\alpha}}}{L^{^*}_{_{H\alpha}}(M^{0}_{_r})}\right)\left(10^{0.4\beta(M_{_r}-M^{^0}_{_r})}\right)\right]\right\}.
	\end{aligned}
	\label{eq:bivariate_gauss}
\end{equation}
The number density contours produced by a Schechter--Gaussian model is identical to the $\Phi$ contours shown in Figure\,\ref{fig:simple_model}. Therefore in Table\,\ref{table:model_param2} we also provide the best--fitting Schechter--Gaussian model parameters for the bivariate functions shown in Figure\,\ref{fig:vmax_bilf}.}  
We note that in the two highest redshift bins, we have a limited sampling of galaxies fainter than broadband or emission line characteristic luminosity (Figure\,\ref{fig:vmax_bilf}). 

\begin{table*}
	\begin{center}
			\caption{The best fitting parameters of the fitted models (Eq.\,\ref{eq:bivariate}) to the four bivariate L$_{H\alpha}$--M$_r$ functions. The faint--end slopes, $\alpha_M$ and $\alpha_L$, of the higher--$z$ bivariate models are fixed to be equal to the $z<0.1$ model values.}
			\begin{tabular}{ >{\centering\arraybackslash}p{0.15\textwidth}  >{\centering\arraybackslash}p{0.13\textwidth}  >{\centering\arraybackslash}p{0.13\textwidth} >{\centering\arraybackslash}p{0.13\textwidth} >{\centering\arraybackslash}p{0.13\textwidth}}
\hline
&&&&\\
			\end{tabular}
			\begin{tabular}{ >{\raggedleft\arraybackslash}p{0.1\textwidth}  >{\centering\arraybackslash}p{0.6\textwidth}}
\vspace{-0.2cm}
	Parameter       		 &                       {\small Schechter--Saunders function}   		\\ 
			
			\end{tabular}
			\begin{tabular}{ >{\centering\arraybackslash}p{0.15\textwidth}  >{\centering\arraybackslash}p{0.13\textwidth}  >{\centering\arraybackslash}p{0.13\textwidth} >{\centering\arraybackslash}p{0.13\textwidth} >{\centering\arraybackslash}p{0.13\textwidth}}
				     &		 $0.001<z<0.1$	&  $0.1<z<0.15$              &  $0.17<z<0.24$ 	& $0.24<z<0.34$    		\\
\hline
\hline
         M$^*_r$ 	 	     &   $-21.12\pm0.10$		&  $-21.57\pm0.09$  & 	$-21.77\pm0.26$  	& 	$-22.62\pm0.14$	\\
         $\alpha_M$  	     &   $-1.29\pm0.06$ 		&       -  			 &       - 				&       -    				\\
         $\log$\,$\Psi^*$ (Mpc$^{-3}$)    &   $-4.17\pm1.54$		&  $-4.49\pm0.45$    & 	$-4.79\pm0.49$  	& 	$-4.40\pm0.93$	\\
         $\log$\,L$^*_{H\alpha}$\,(W)&  $32.30\pm0.65$ 		&  $32.52\pm0.33$   & 	$32.75\pm0.59$  	& 	$33.55\pm0.64$	\\
         $\beta$        	     &   $0.92\pm0.03$		&  $0.85\pm0.12$     & 	$0.59\pm0.30$  	& 	$0.46\pm0.25$		\\
         $\sigma$        	     &   $0.46\pm0.03$		&  $0.46\pm0.05$     & 	$0.48\pm0.04$  	& 	$0.38\pm0.10$		\\
         $\alpha_L$           &   $1.53\pm1.15$   	        &       -  			 &       - 				&       -    				\\	
\hline
			\end{tabular}
						\begin{tabular}{ >{\raggedleft\arraybackslash}p{0.1\textwidth}  >{\centering\arraybackslash}p{0.6\textwidth}}
\vspace{-0.2cm}
	Parameter       		 &                       {\small Schechter--Gaussian function}   		\\ 
			
			\end{tabular}
						\begin{tabular}{ >{\centering\arraybackslash}p{0.15\textwidth}  >{\centering\arraybackslash}p{0.13\textwidth}  >{\centering\arraybackslash}p{0.13\textwidth} >{\centering\arraybackslash}p{0.13\textwidth} >{\centering\arraybackslash}p{0.13\textwidth}}
				     &		 $0.001<z<0.1$ 	&  $0.1<z<0.15$              &  $0.17<z<0.24$ 	& $0.24<z<0.34$    		\\
\hline
\hline
         M$^*_r$ 	 	      &  $-21.12\pm0.11$	&  $-21.57\pm0.10$  & 	$-21.77\pm0.27$  	& 	$-22.63\pm0.14$	\\
         $\alpha_M$           &  $-1.29\pm0.06$  	&       -  			&       -  				&       -  				\\
         $\log$\,$\Psi^*$ (Mpc$^{-3}$)     &  $-2.67\pm0.14$	&  $-2.94\pm0.14$    & 	$-3.08\pm0.33$  	& 	$-3.53\pm0.31$	\\
         $\log$\,L$^*_{H\alpha}$\,(W)&  $33.55\pm0.04$	&  $33.79\pm0.14$   & 	$34.12\pm0.42$  	& 	$34.45\pm0.36$	\\
         $\beta$        	      &  $0.92\pm0.03$	&  $0.85\pm0.12$     & 	$0.59\pm0.30$  	& 	$0.44\pm0.24$		\\
         $\sigma$        	      &  $0.45\pm0.02$	&  $0.45\pm0.04$     & 	$0.47\pm0.04$  	& 	$0.38\pm0.07$		\\
\hline
			\end{tabular}
			\label{table:model_param2}
	\end{center}
\end{table*}
\begin{table*}
	\begin{center}
%		\begin{minipage}{14cm}
			\caption{The SFRDs estimated from integrating the best--fitting Schechter--Saunders models over three different L$_{H\alpha}$ and M$_r$ ranges. The limiting $\log$ L$_{H\alpha}=37$ (denoted L$_b$ in the table) and M$_r=-26$ (denoted M$_b$ in the table).  The second table column ($\log\dot{\rho}_{*(0,0)-(L_b, M_b)}$) reports the SFRDs computed from integrating the best--fitting higher--$z$ bivariate models from [L$_{H\alpha}$, M$_r$] = [$0$, $0$] to [L$_b$, M$_b$], the third provides those computed from integrating the bivariate models between the observed lowest $z<0.1$ $\log$ L$_{H\alpha}$ and M$_r$ values and [L$_b$, M$_b$], and the final column provides the SFRDs obtained from integrating each model between the lowest $\log$ L$_{H\alpha}$ and M$_r$ values at their respective redshift range and [L$_b$, M$_b$].}
			\begin{tabular}{llll}
\hline
	Redshift          
	&    $\log\dot{\rho}_{*(0,0)-(L_b, M_b)}$                       
	&    $\log\dot{\rho}_{*(L_{f,z1},M_{f,z1})-(L_b, M_b)}$          
	&    $\log\dot{\rho}_{*(L_{f,z_x},M_{f,z_x})-(L_b, M_b)}$)  \\
	
	range
	&    (M$\odot$yr$^{-1}$Mpc$^{-3}$)
	&    (M$\odot$yr$^{-1}$Mpc$^{-3}$)
	&    (M$\odot$yr$^{-1}$Mpc$^{-3}$)   \\
\hline
\hline
        $0.1<z<0.15$    
        & $-2.00$
        & $-2.00$
        & $-2.03$ (-14.25, 33.1)    \\          
        
        $0.17<z<0.24$    
        & $-1.75$
        & $-1.77$
        & $-2.03$ (-18.75, 33.5)    \\          
        
        $0.24<z<0.34$    
        & $-1.72$
        & $-1.77$
        & $-2.14$ (-19.25, 33.9)    \\          

\hline
			\end{tabular}
			\label{table:model_param3}
%		\end{minipage}
	\end{center}
\end{table*}

More complex fitting methods like that of \cite{deJong00} could be considered, but for the purposes of this analysis, however, we find the best--fit model shown in Figure\,\ref{fig:simple_model} to be sufficient as it provides a good qualitative and a quantitative description of  the low--$z$ bivariate L$_{H\alpha}$--M$_r$ function. 
Furthermore, the approximate Schechter and Saunders functional fit forms of GAMA M$_r$ and H$\alpha$ LFs can be recovered from numerically integrating $\Psi(M_r, L_{H\alpha})$ along L$_{H\alpha}$ and M$_r$ axes, respectively. Moreover, by integrating $\Psi(M_r, L_{H\alpha})$ with respect to both L$_{H\alpha}$ and M$_r$, {we recover} the $z<0.1$ H$\alpha$ SFR density reported in paper~I.  

The best fitting {$z<0.1$ M$^*_r$ and $\alpha_M$ bivariate model parameters  given in Table\,\ref{table:model_param2} agree within uncertainty with the best fitting parameters determined for the $z<0.1$ univariate M$_r$ LF (Table\,\ref{table:LFfits})}. The relationship between {$\log$ L$^*_{H\alpha}$ and M$^*_r$} is emphasised in Figure\,\ref{fig:normalised_LFs}. Note that $\alpha_L$ represents the faint--end slope, {and the positive slope implies a decrease in number density. This is not unexpected as the distribution of the normalised number densities with respect to log L$_{H\alpha}$ within a given magnitude bin has a log--normal shape than a power--law shape (Figure\,\ref{fig:normalised_LFs} ).}

In summary, both Schechter--Saunders and Schechter--gaussian models are able to provide a good representation of the lowest--$z$ bivariate L$_{H\alpha}$--M$_r$ function. As for the higher--$z$ ($0.1\lesssim z\lesssim0.34$) bivariate L$_{H\alpha}$--M$_r$ functions, the models provide a good description of the bright end of the bivariate functions, where the data exists. It becomes increasingly difficult to constrain the models to obtain a good description of the faint--end of the bivariate functions with increasing redshift as the range in L$_{H\alpha}$ and M$_r$ probed decreases. 

{Finally, an alternative method of modelling bivariate functions is proposed by \cite{Takeuchi10}. This approach relies on using a copula to connect two marginal distributions (e.g.\,H$\alpha$ and M$_r$ LFs) thereby constructing their bivariate distribution. \cite{Takeuchi10} and \cite{Takeuchi13} advocate the use of a Farlie--Gumbel--Morgenstern or a Gaussian copula, both of which are explicitly related to the linear correlation coefficient, to connect two given marginals. See \cite{Takeuchi10} and \cite{Sato11} for a rigorous mathematical definition of copula theory, dependence measures between two variables and how to estimate the bivariate distribution given two or more marginals.}

\subsection{Luminosity and density of SF galaxies} \label{sebsec:rlumdensity}

\begin{figure}
	\begin{center}
		\includegraphics[trim=.1cm 0.05cm 8.5cm .01cm, clip=true, width=0.35\textwidth]{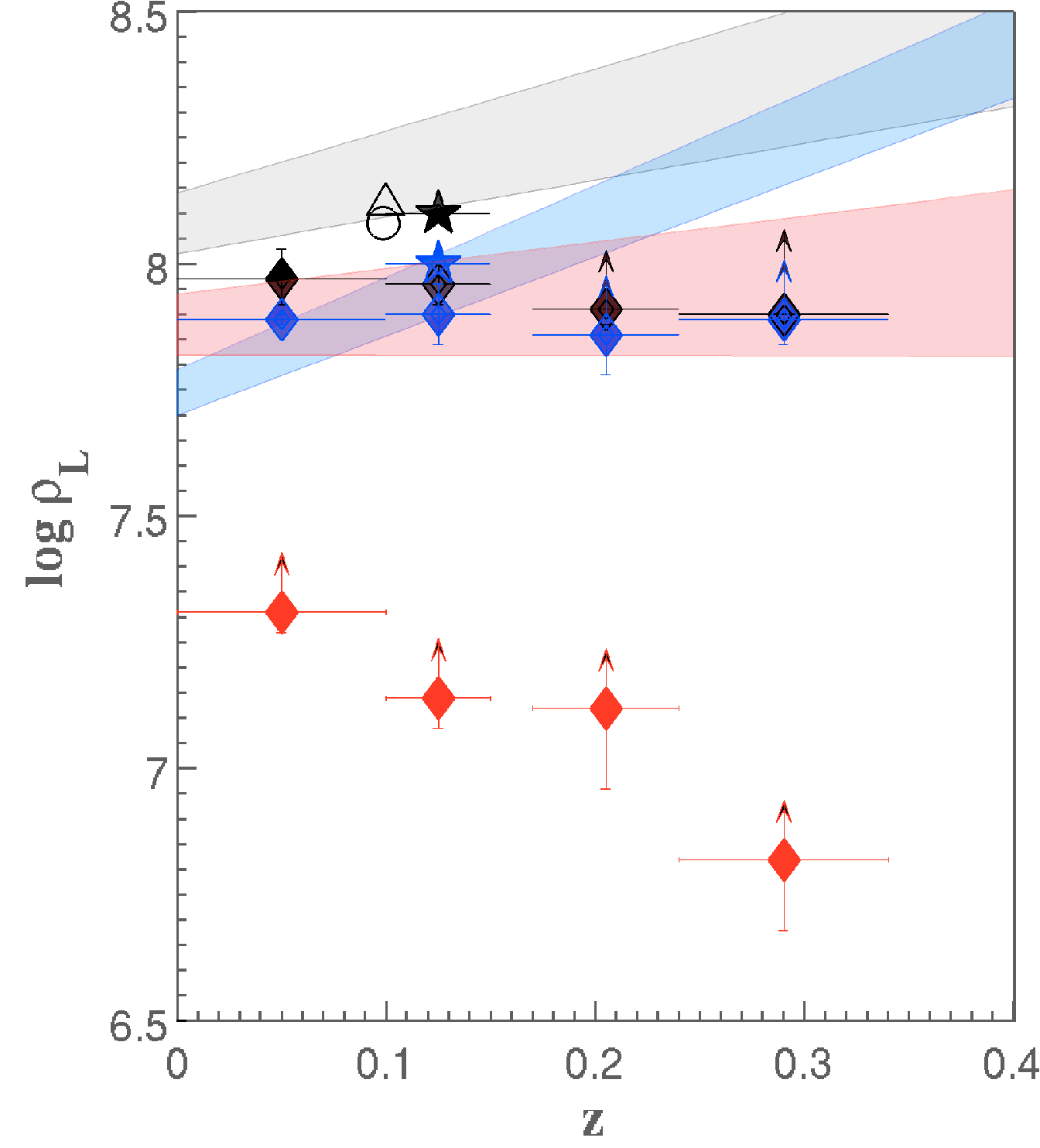}
		\caption{The evolution of {$r$--band} luminosity density ($\rho_L$) of all, and blue and red H$\alpha$ SF galaxies as a function of redshift (filled black, blue and red symbols respectively), compared to the evolution of $\rho_L$ of all, and blue and red GAMA galaxies as a function of redshift \citep[shaded regions;][] {Loveday12}. Also shown are the $\rho_L$ measurements at low--$z$ from \citet{Blanton03} (open circle) and \citet{Montero09} (open triangle). {The filled stars at $z\sim0.12$ indicate the value obtained from integrating the best--fitting bivariate analytical form for the $0.1<z<0.15$ bivariate L$_{H\alpha}$--M$_r$ function}.}
		\label{fig:rband_density}
	\end{center}
\end{figure}

{ 
In this section we present the $r$--band luminosity and H$\alpha$ SFR density of SF galaxies computed from integrating the bivariate L$_{H\alpha}$--M$_r$ functions.} 
The luminosity density ($\rho_L$) evolution observed in the GAMA H$\alpha$ SF population is shown in Figure\,\ref{fig:rband_density}. The filled diamonds indicate the GAMA $\rho_L$ derived from integrating the four Schechter functions shown in Figure\,\ref{fig:MrLF_fits}. 
{The filled stars indicate the density estimated from integrating the best--fitting Schechter--Saunders model to the  $0.1<z<0.15$ bivariate L$_{H\alpha}$--M$_r$ function. The $z<0.1$ estimate is not shown here as it is similar to that obtained from integrating the Schechter functional fit to the univariate $z<0.1$ M$_r$ LF}. We also show the GAMA $\rho_L$ confidence limits from \cite{Loveday12}, and low redshift density estimates from \cite{Blanton03} and \cite{Montero09}. We see a result here that is consistent with the SF {populations} of the broadband blue and red LFs shown in Figure\,\ref{fig:comprison_app}. {That is the $\rho_L$ estimated from the best--fitting Schechter functions do not show a evolution in $\rho_L$ with redshift, in contrast to \cite{Loveday12}}. This is a natural consequence of the SF populations comprising only a small fraction (10--20\%) of the total red galaxy population. The decrease in $\rho_L$ of  blue SF population at higher--$z$ ($z>0.17$) is likely due to the difficulty in measuring H$\alpha$ in higher--$z$ galaxy spectra {as higher--$z$ galaxies likely to have fainter optical magnitudes and lower spectral signal--to--noise than their low--$z$ counterparts}. Given our sample is already biased against red SF galaxies, mainly as a result of the H$\alpha$ flux limit, the drop in $\rho_L$ corresponding those galaxies is not unexpected.   
\begin{figure*}
	\begin{center}
		\includegraphics[scale=0.55]{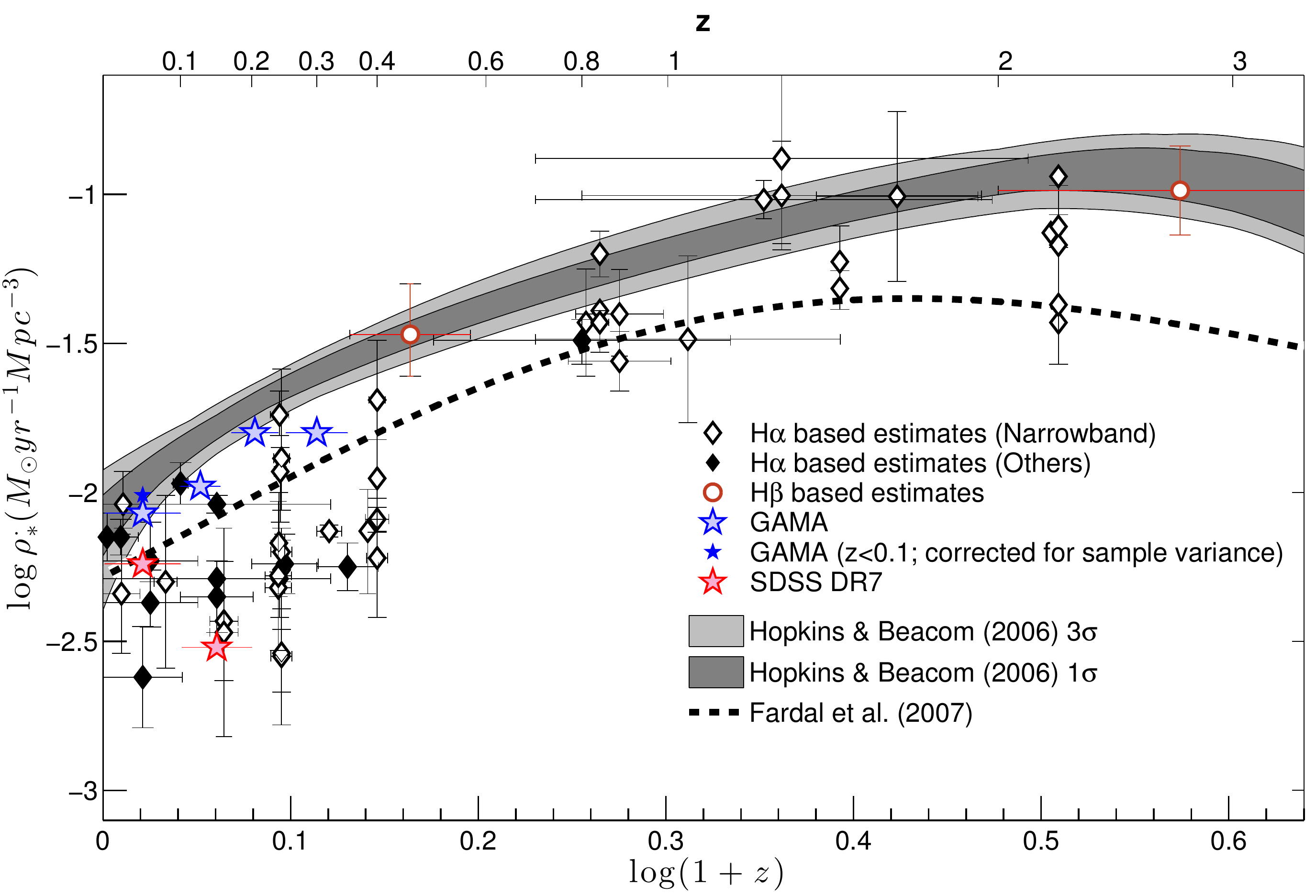}
		\caption{The cosmic SFR history, taken from paper~I, with our new measurements shown as blue stars. The GAMA SFRDs are based on integrating the analytic fits to the bivariate L$_{H\alpha}$--M$_r$ LFs (Figures\,\ref{fig:simple_model}). Published estimates based on narrowband surveys and slitless spectroscopy data are shown as open symbols, and those based on broadband surveys as filled symbols. The grey bands and the dashed line correspond to the best--fitting star formation histories of \citet{HB06} and \citet{Fardal07}, respectively. }
		\label{fig:SFRD_density}
	\end{center}
\end{figure*}

\begin{figure*}
		\includegraphics[scale=0.34]{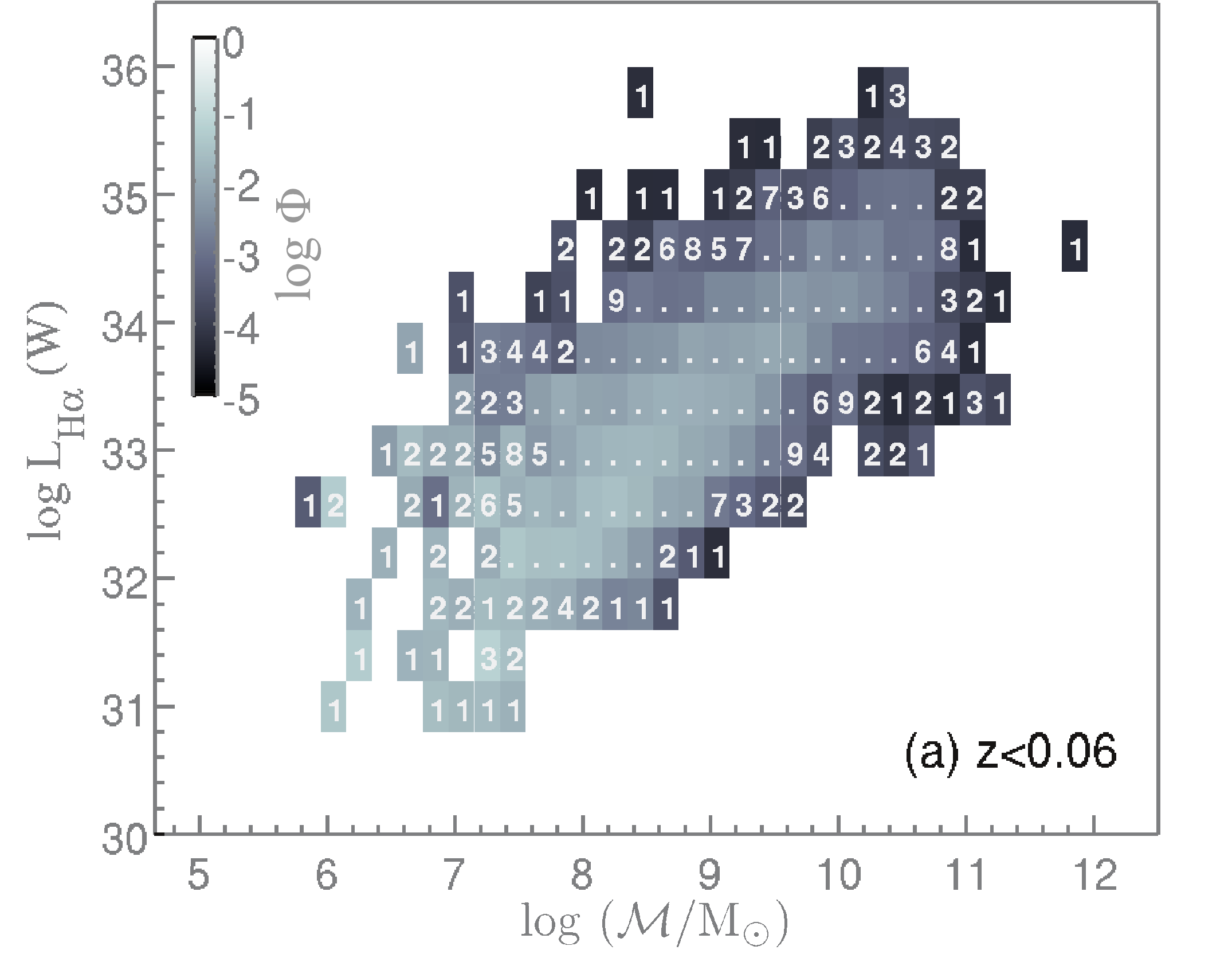}
		\includegraphics[scale=0.27]{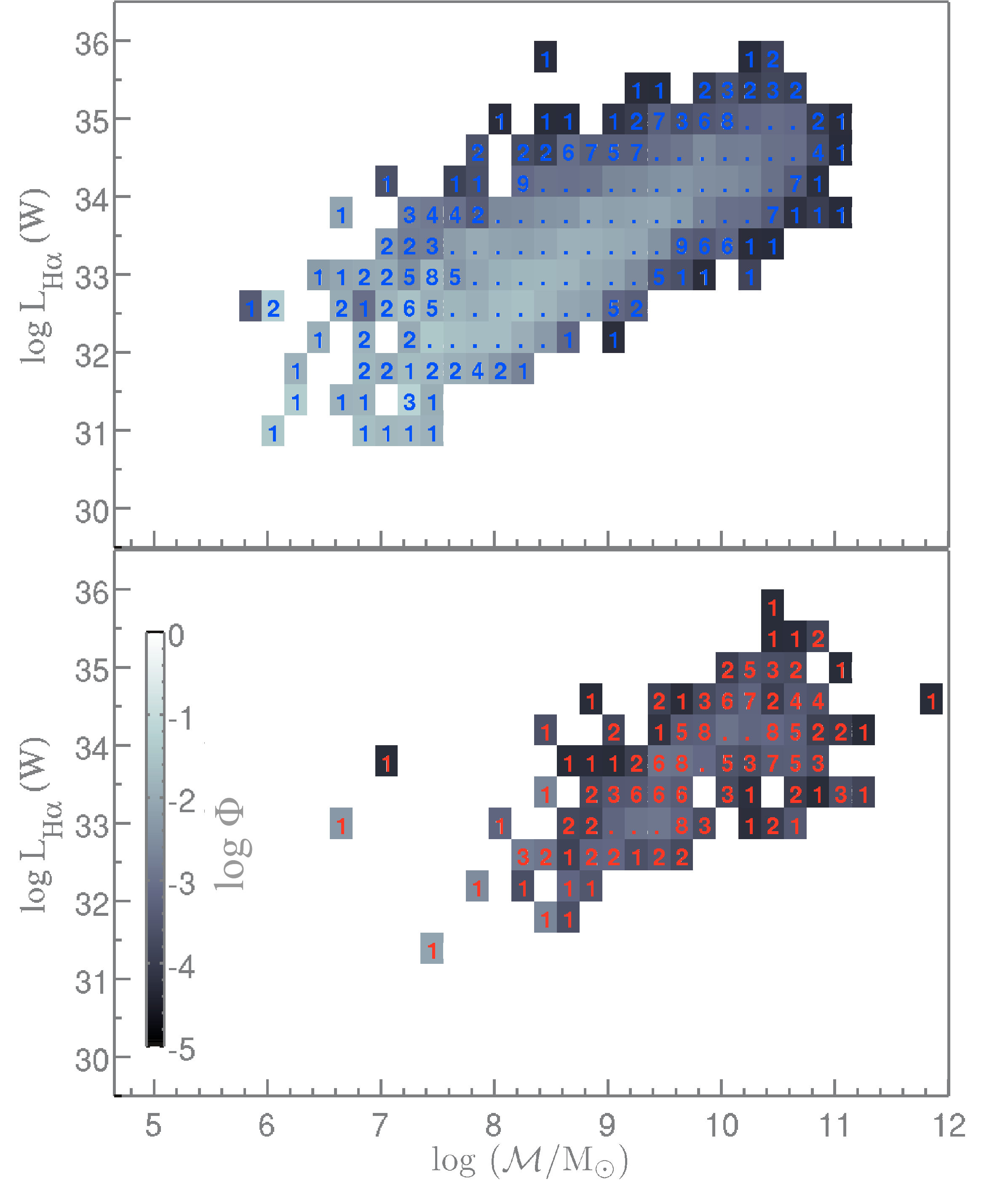}\\
		\includegraphics[scale=0.41]{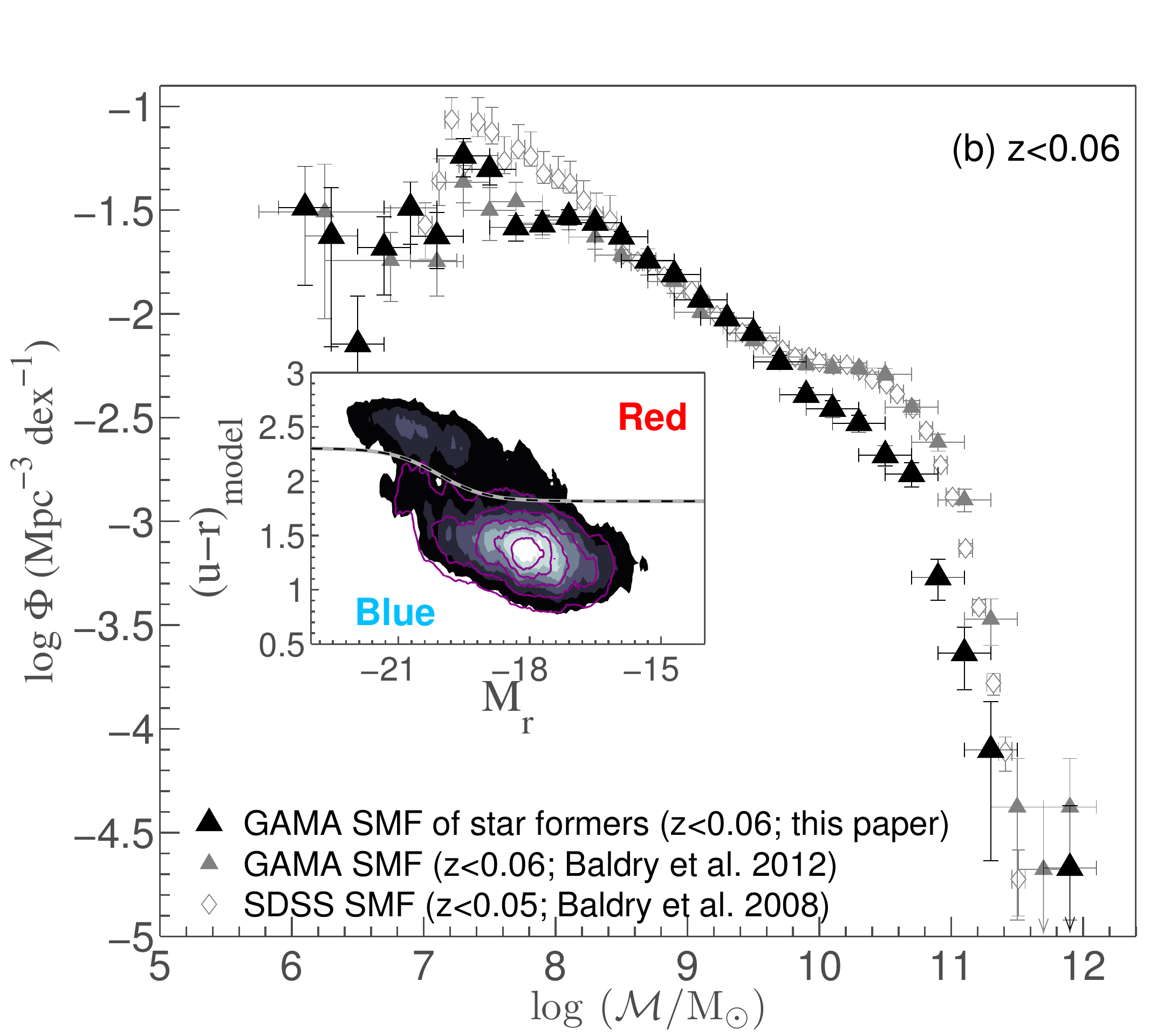}
		\includegraphics[scale=0.33]{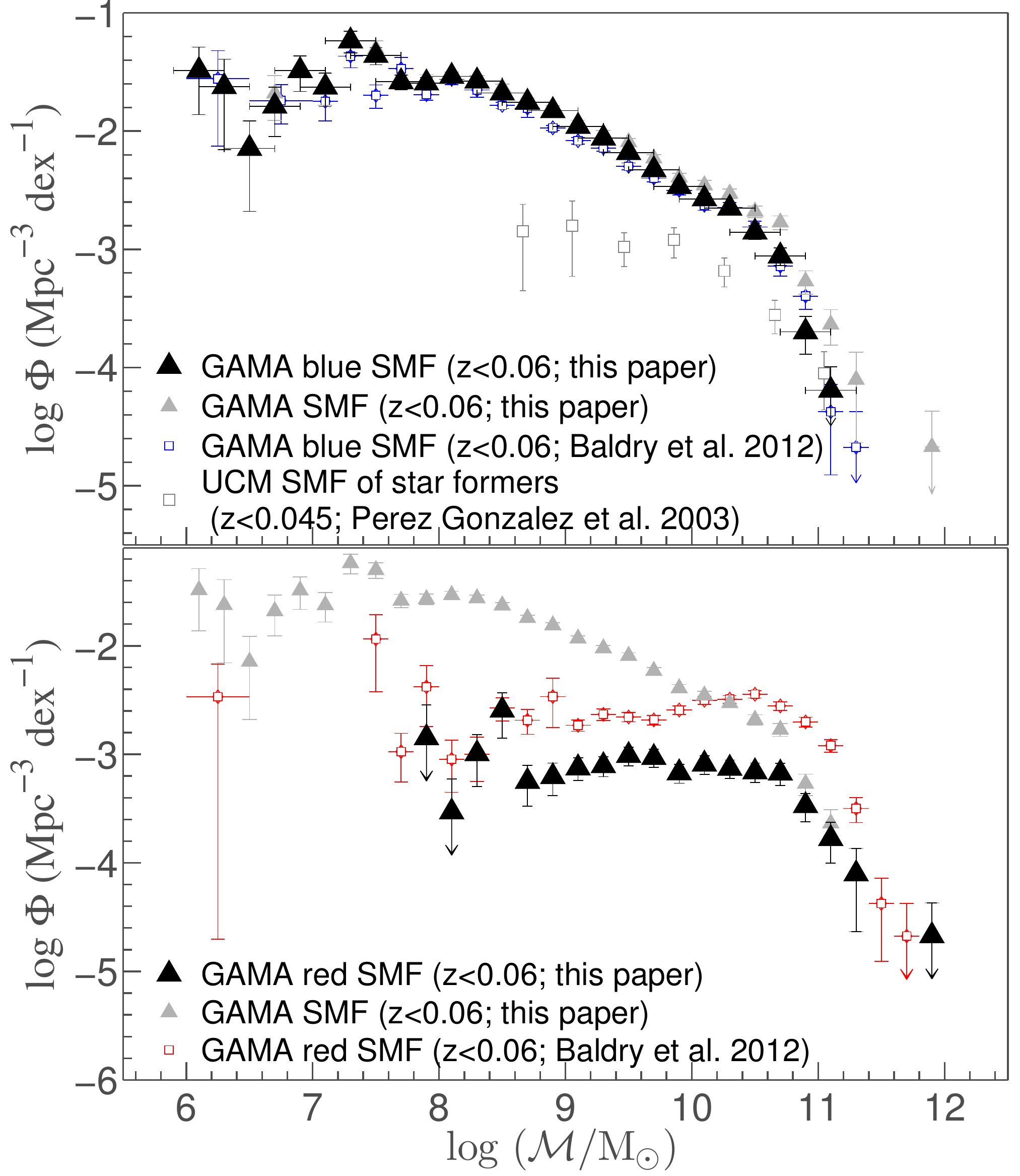}
		\caption{(a) The $z<0.06$ GAMA bivariate L$_{H\alpha}$--$\mathcal{M}$ functions of all, blue and red SF galaxies (left, top right and bottom right panel respectively), constructed using the density corrected 1/V$_{\rm max}$ method described in \S\,\ref{subsec:ddp_vmax}. The grey scale indicates the log of number densities ($\Phi$) in the unit of Mpc$^{-3}$ dex$^{-2}$. (b) The SMFs obtained from integrating the bivariate L$_{H\alpha}$--$\mathcal{M}$ functions over L$_{H\alpha}$ are compared to the $z<0.06$ GAMA SMFs of \citet{Baldry12} and $z<0.05$ SDSS SMF of \citet{Baldry08}.}
		\label{fig:Baldry_vs_ours_combined}
\end{figure*}

\subsection{Star Formation Rate Density}

As mentioned previously, our primary motivation behind modelling the bivariate L$_{H\alpha}$--M$_r$ functions is to overcome the bivariate sample effects introduced by the dual H$\alpha$ flux and magnitude selection imposed on our sample. {As a result of this effect, the higher--$z$ ($z>0.1$) SFRDs presented in paper~I are underestimates}. In Figure\,\ref{fig:SFRD_density} we show the SFRDs derived by integrating the bivariate analytic fit to L$_{H\alpha}$--M$_r$ function. By modelling the low--redshift L$_{H\alpha}$--M$_r$ distribution over $-26<$M$_r<-10$ and $30<$log L$_{H\alpha}<37$ {\color{black}(Figure\,\ref{fig:simple_model})}, and assuming the faint--end bivariate distribution is similar for the higher--$z$ ($0.1\lesssim z\lesssim0.34$) samples, we can estimate a correction for the missing optically faint star forming galaxies at those redshifts. In Figure\,\ref{fig:SFRD_density} it can be seen that the resulting SFRDs are much more consistent with that from other published measurements \citep[e.g.][]{HB06}, than the direct estimates from paper~I. 
Note that the two highest redshift ranges probed likely overestimated due to poorly constrained bivariate functions. 

{
\section{Importance of bivariate L$_{H\alpha}$--$\mathcal{M}$ functions} \label{sec:biLFs_SMF}}

\begin{figure*}
		\includegraphics[scale=0.35]{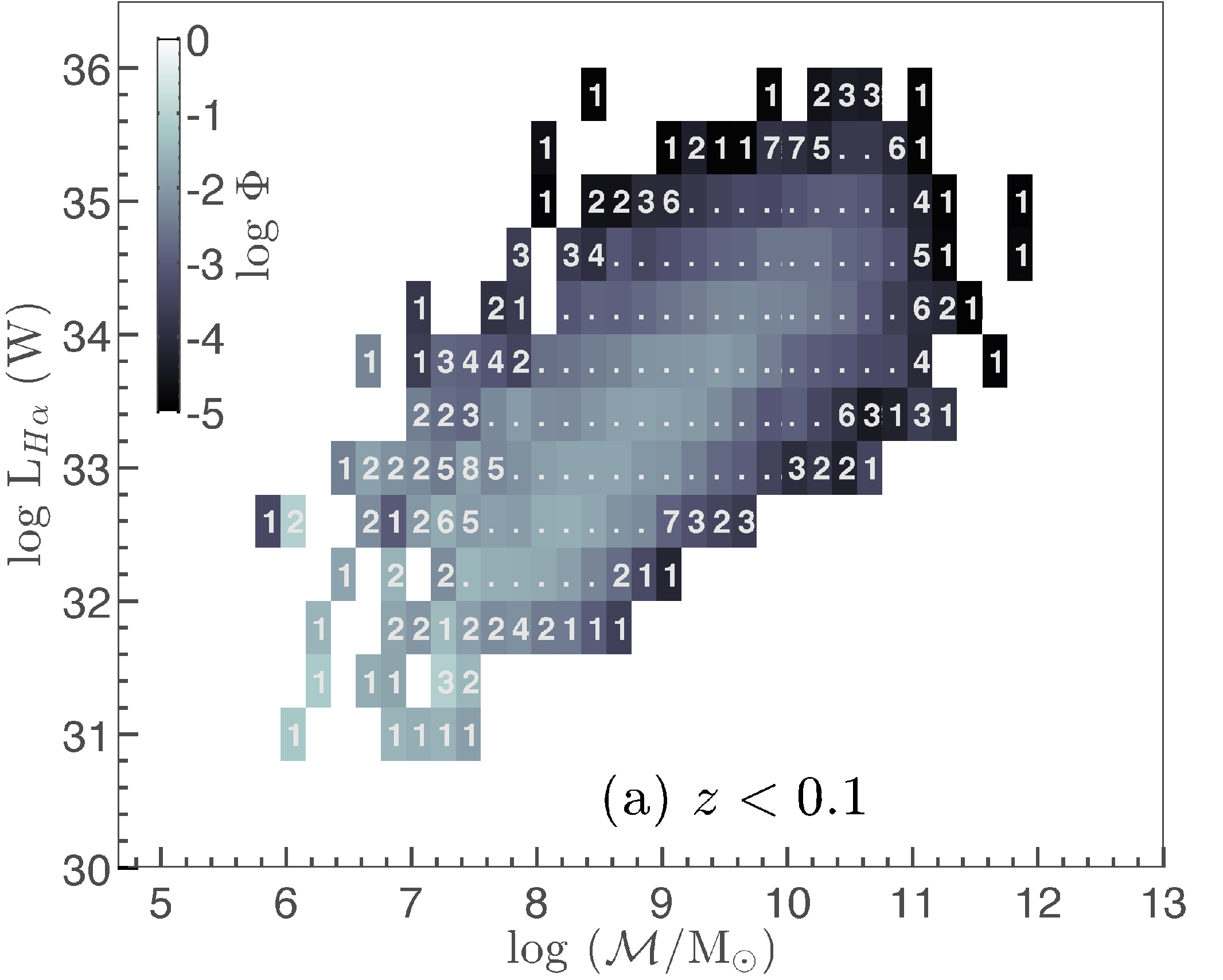}
		\includegraphics[scale=0.35]{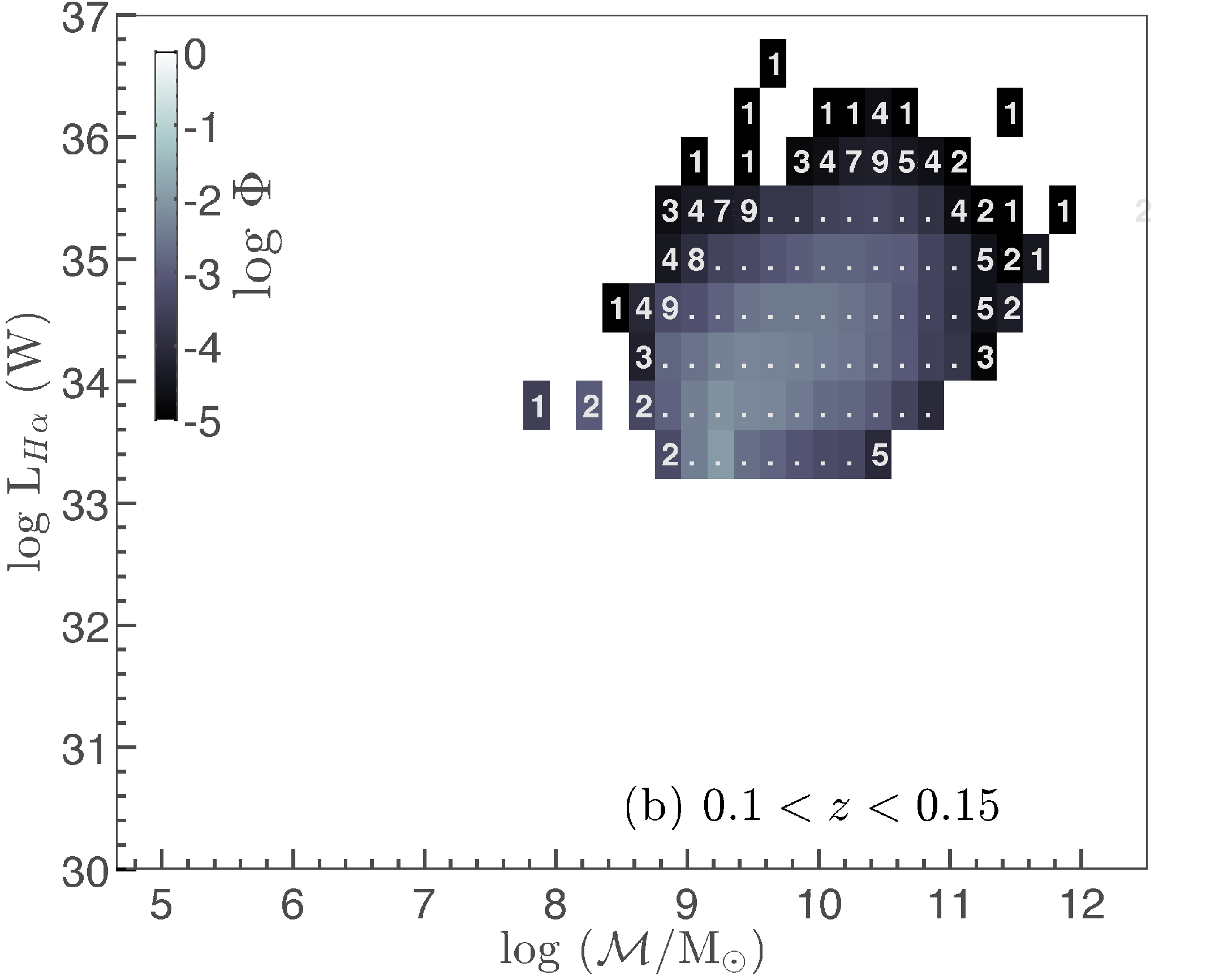}
		\includegraphics[scale=0.35]{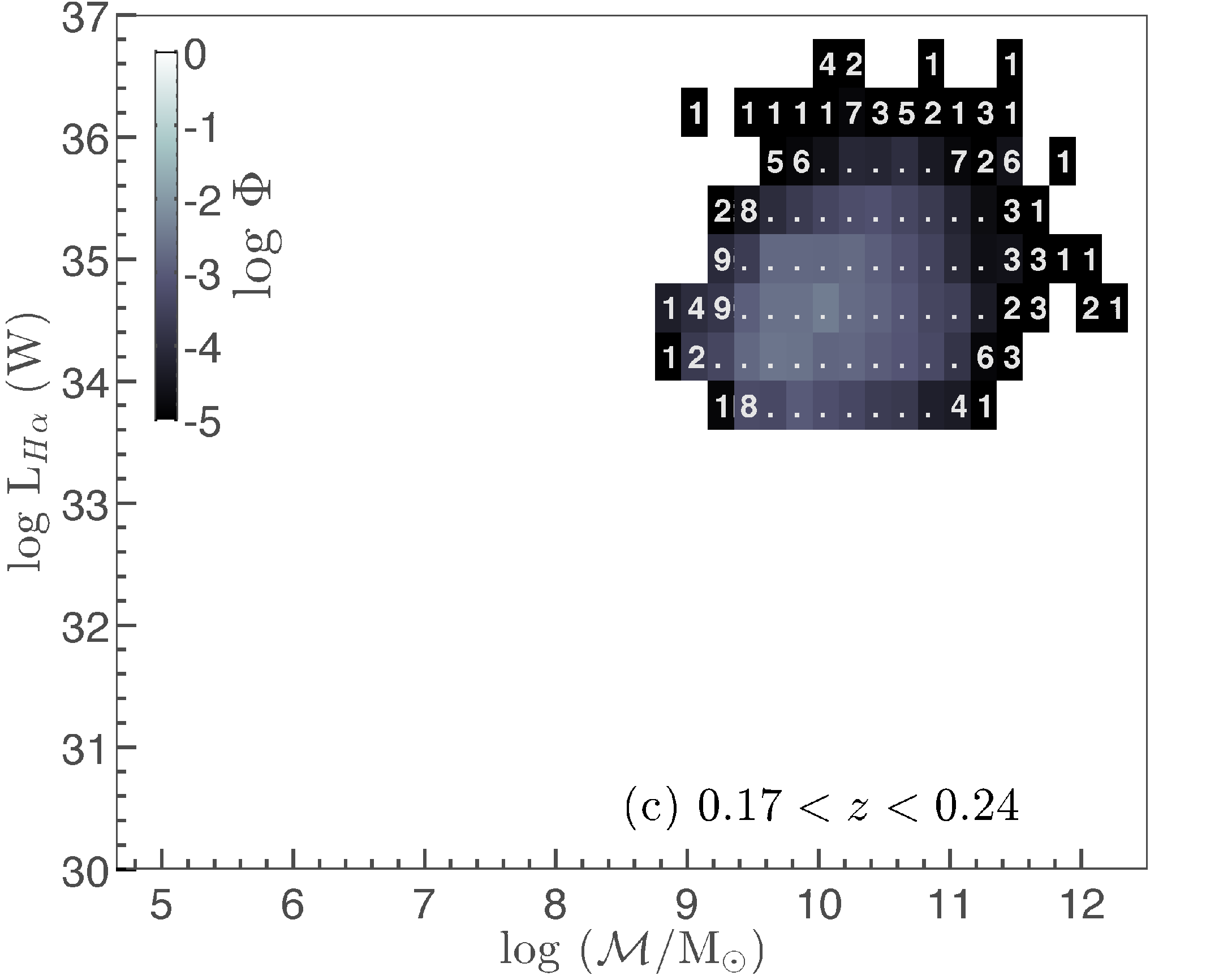}
		\includegraphics[scale=0.35]{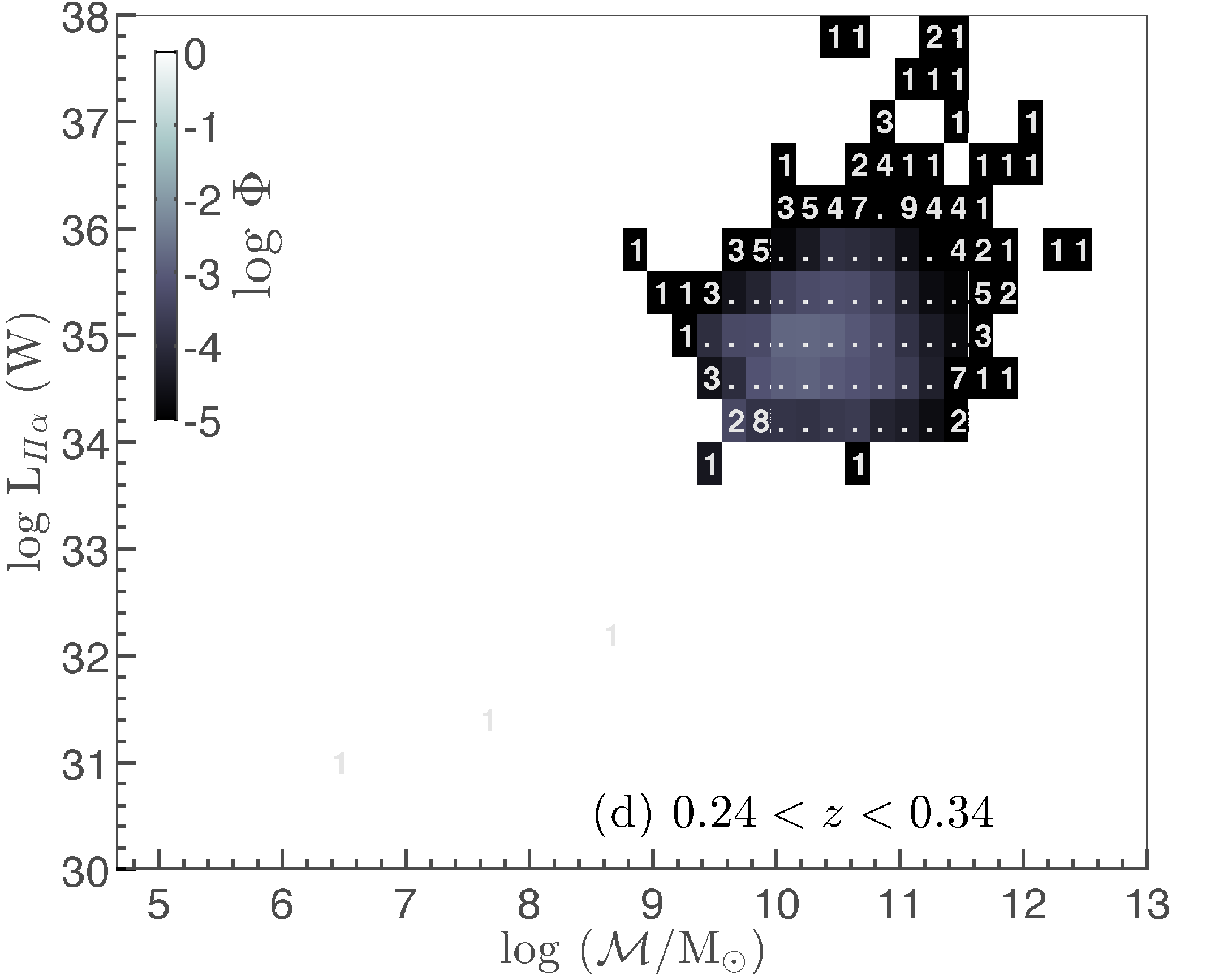}
		\caption{The bivariate L$_{H\alpha}$--$\mathcal{M}$ functions of all SF galaxies split in to four redshift bins. The grey scale indicates the log number densities ($\Phi$) in the unit of Mpc$^{-3}$ dex$^{-2}$. }
		\label{fig:SMFsinz_biLF}
\end{figure*}

In this section, we investigate the evolution of the bivariate L$_{H\alpha}$--$\mathcal{M}$ functions, and the SMFs that result from integrating the bivariate functions for the full sample and those split by colour. {The bivariate L$_{H\alpha}$--$\mathcal{M}$ functions of all SF galaxies in redshift bins are shown Figure\,\ref{fig:SMFsinz_biLF}. Again the lowest redshift function probes the largest range in both L$_{H\alpha}$ and $\mathcal{M}$, while the higher--$z$ ($0.1<z<0.34$) functions indicate a decrease in range with increasing redshift like seen in Figure\,\ref{fig:vmax_bilf} for bivariate L$_{H\alpha}$--M$_r$ functions.}.

The colour--magnitude relation used to obtain a photometric blue and red classification in this part of the analysis is different to that discussed in \S\,\ref{sec:biLF}.
To reiterate, our objective in constructing bivariate and univariate SMFs of star forming galaxies with blue and red photometric class is to compare our results with the existing GAMA results in a consistent manner \citep[e.g.][]{Baldry12}. For the stellar mass based analyses presented in the subsequent sections, as they are focussed on SMFs, we use the colour--magnitude relation introduced by \cite{Baldry04} (Eq.\,11 of their paper) and used by \cite{Baldry12} to construct the $z<0.06$ GAMA SMFs split by colour. 

\subsection{The Stellar Mass Functions of SF galaxies} \label{subsec:Baldryvsus}

\cite{Baldry12} present the $z<0.06$ GAMA galaxy SMFs determined using the density corrected 1/V$_{\rm max}$ method. They further show the GAMA SMFs of photometrically classified blue and red galaxies based on the colour ($u-r$)--magnitude (M$_r$) relationship given in \cite{Baldry04}. Using the density corrected 1/V$_{\rm max}$ method discussed in \S\,\ref{subsec:ddp_vmax}, we construct the $z<0.06$ bivariate L$_{H\alpha}$--$\mathcal{M}$ functions and the SMFs of H$\alpha$ SF galaxies. Our results compared to the GAMA SMFs from \cite{Baldry12} are shown in Figure\,\ref{fig:Baldry_vs_ours_combined}.

\begin{figure*}
		\includegraphics[width=0.49\textwidth]{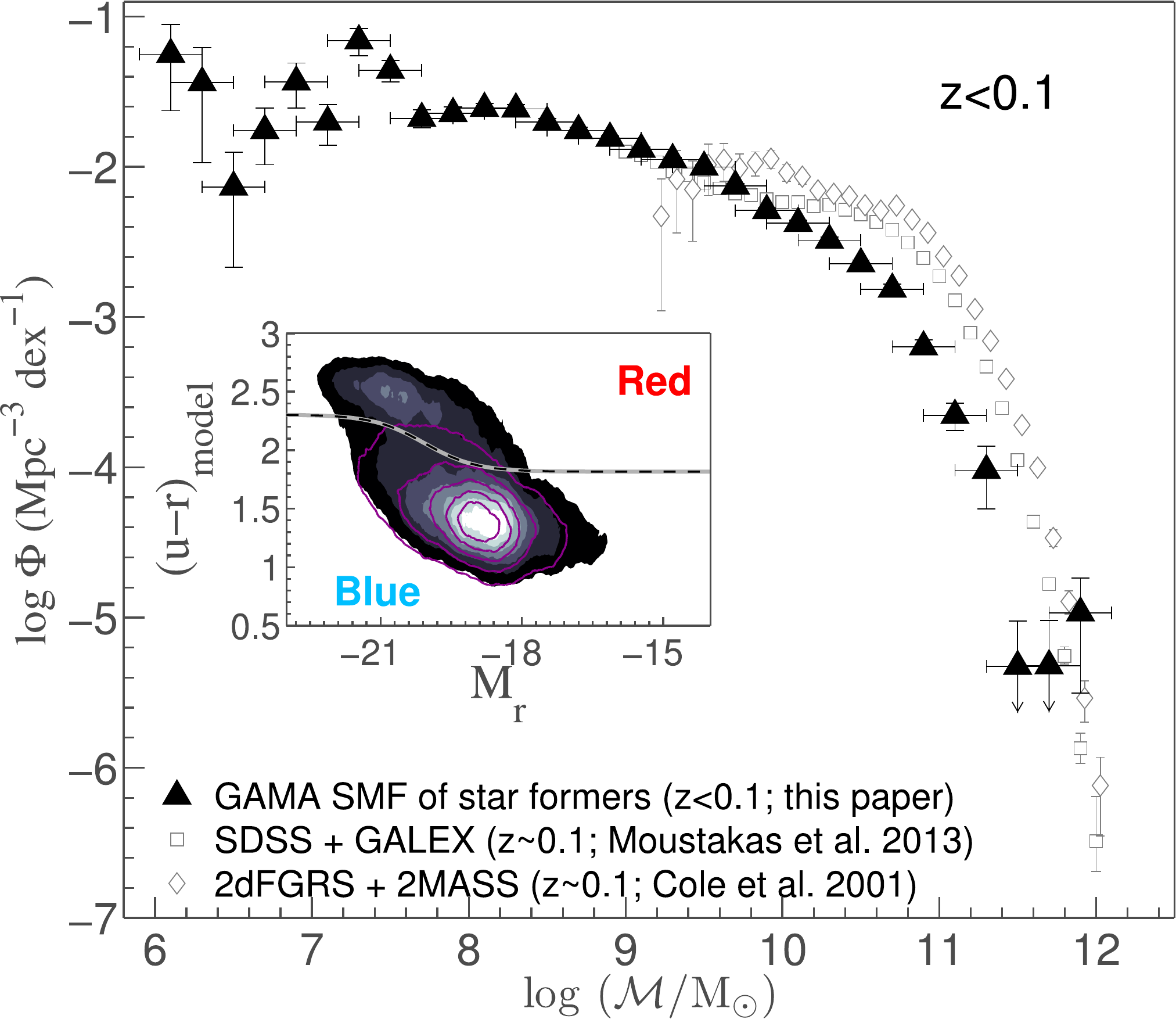}
		\includegraphics[width=0.49\textwidth]{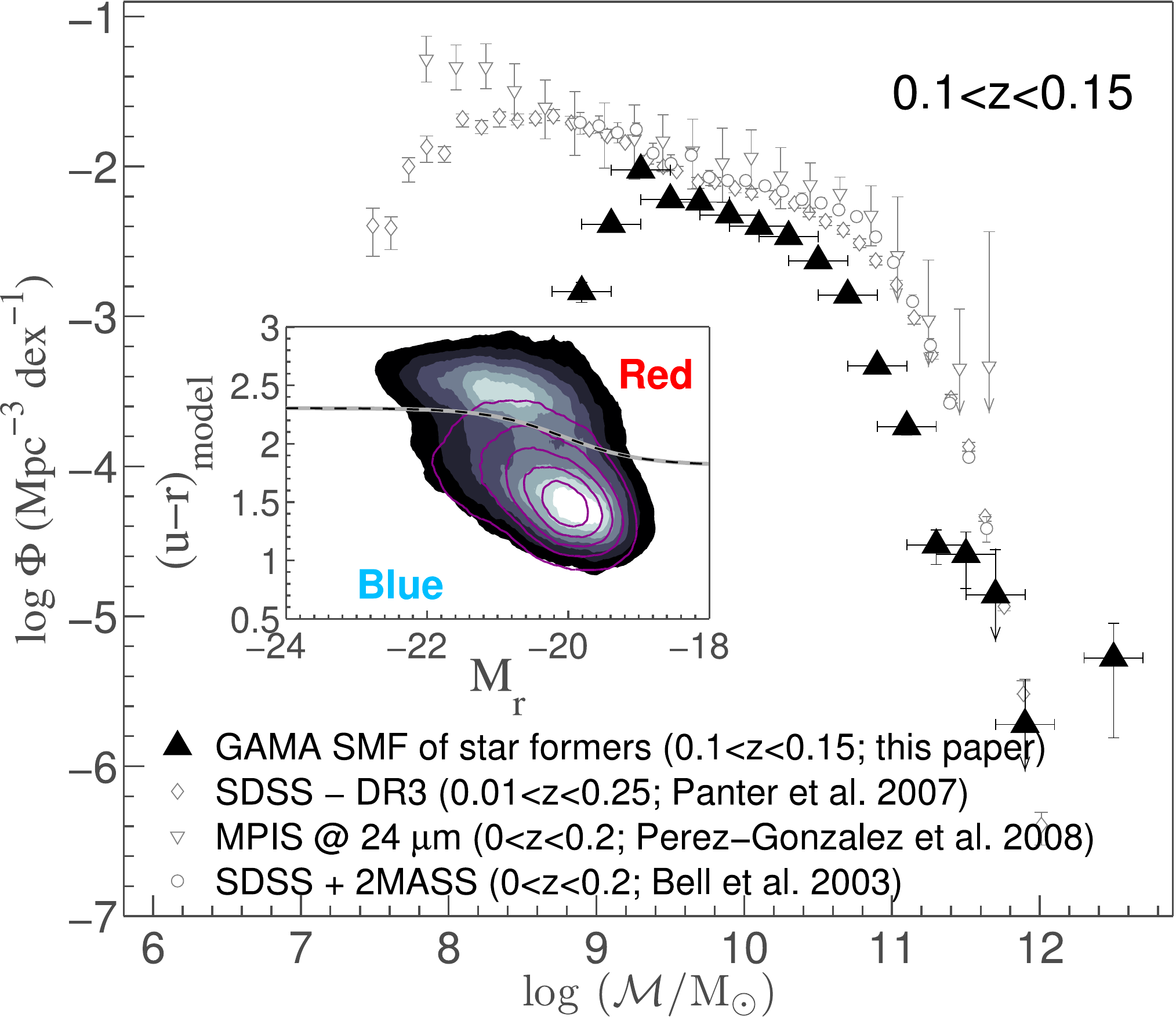}
		\includegraphics[width=0.49\textwidth]{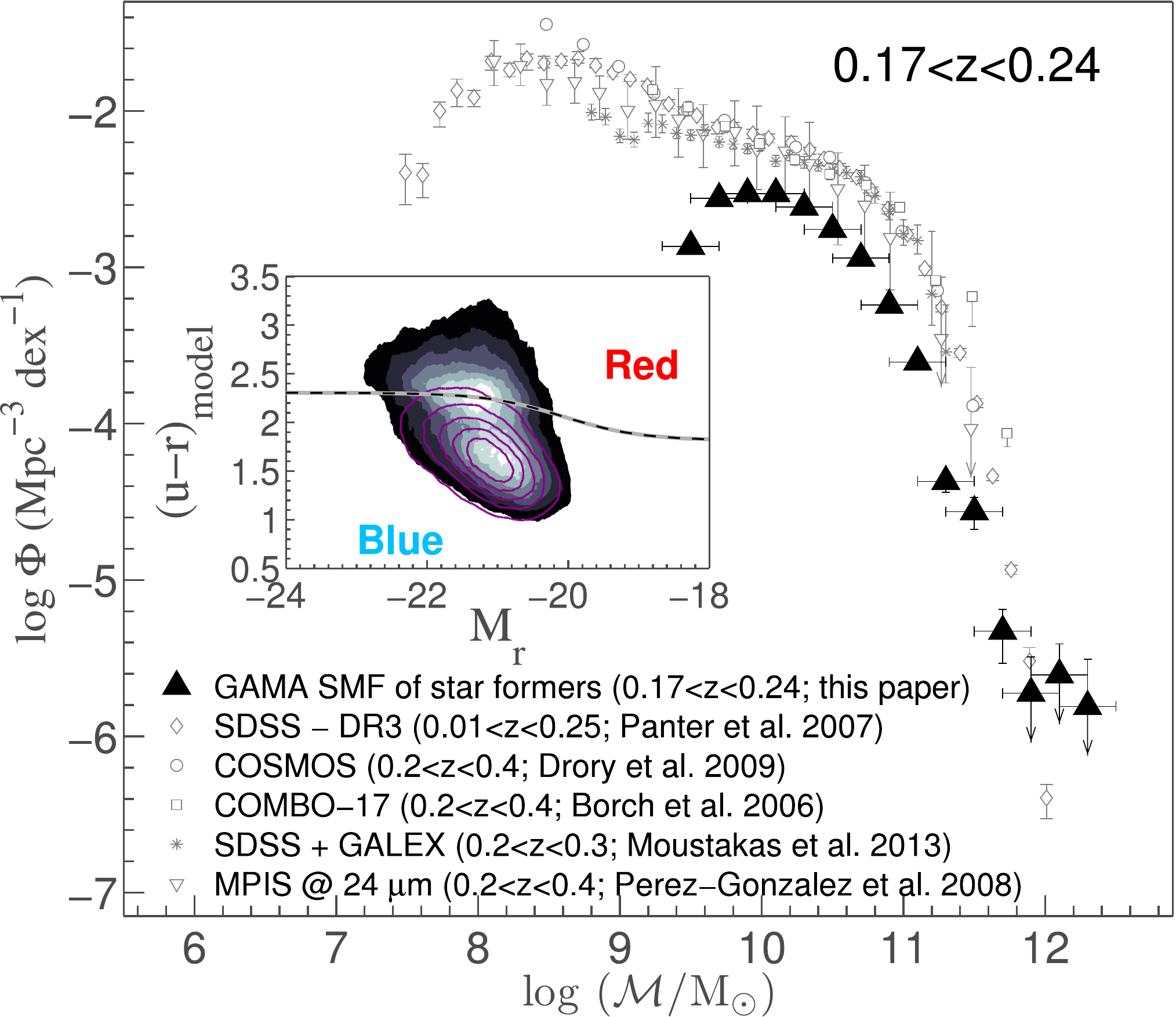}
		\includegraphics[width=0.49\textwidth]{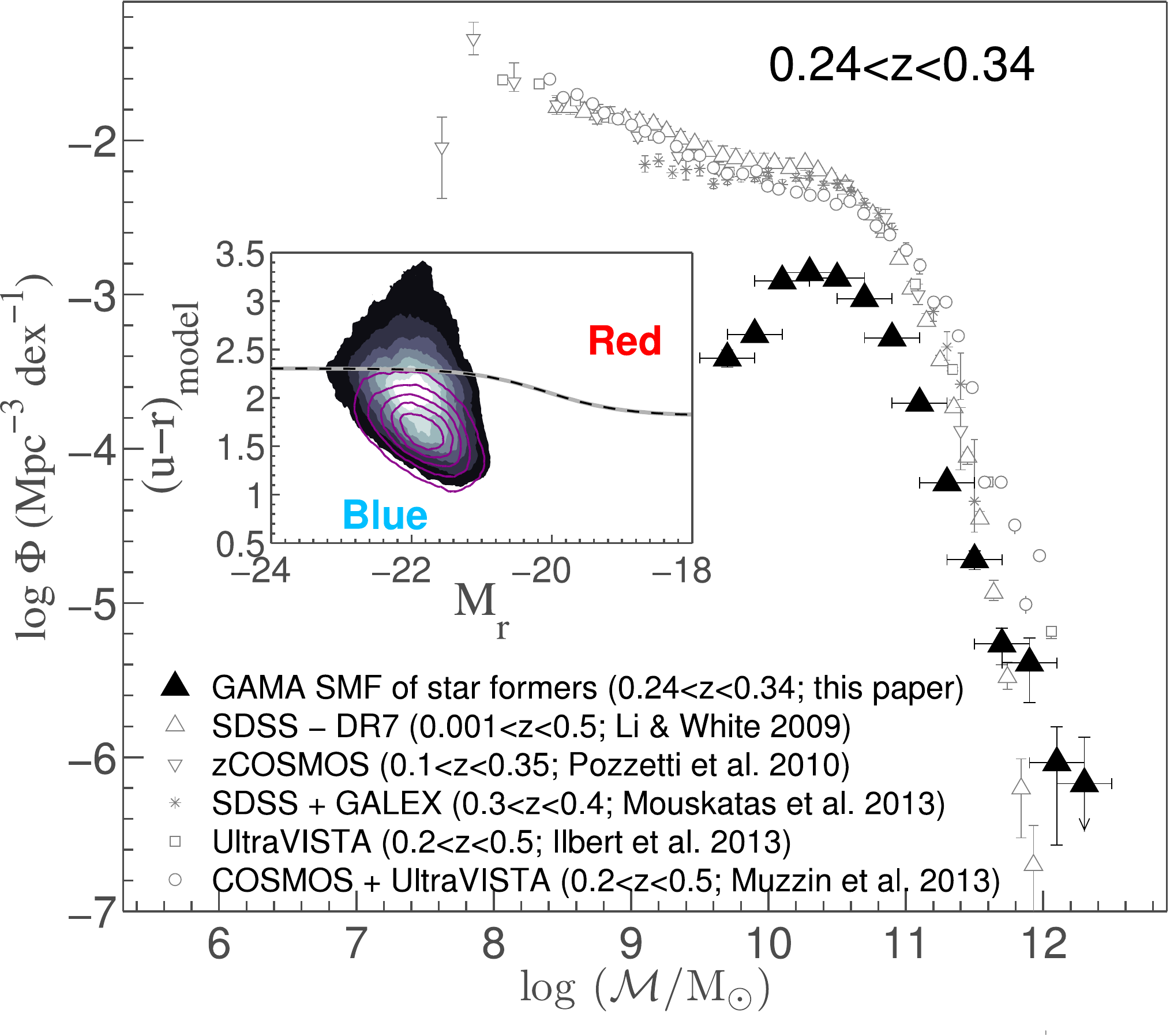}
		\caption{The SMFs derived from integrating the $z>0.1$ L$_{H\alpha}$--$\mathcal{M}$ functions shown in Figure\,\ref{fig:SMFsinz_biLF} over L$_{H\alpha}$ for all SF galaxies. The redshift increases from top--to--bottom. All published SMF measurements shown are adjusted to our assumed cosmology and to a \citet{Chabrier2003} IMF.}
		\label{fig:SMFsinz_SMFzl}
\end{figure*}

{Figure\,\ref{fig:Baldry_vs_ours_combined}a left and right panels show the bivariate  L$_{H\alpha}$--$\mathcal{M}$ functions of all SF galaxies (Figure\,\ref{fig:Baldry_vs_ours_combined}a left), and photometrically classified blue and red SF sub--populations (Figure\,\ref{fig:Baldry_vs_ours_combined}a right top and bottom panels, respectively)}. The bivariate L$_{H\alpha}$--$\mathcal{M}$ function of all $z<0.06$ SF galaxies comprise both blue and red galaxies, though dominated by blue ones.
 {The SMFs obtained from integrating the bivariate functions} {over} L$_{H\alpha}$ are shown in Figure\,\ref{fig:Baldry_vs_ours_combined}b, and the GAMA SMFs of \cite{Baldry12} are also shown for comparison. \cite{Baldry12} find that the SMF of all $z<0.06$ GAMA galaxies is well described by a double Schechter function. This functional form has also been used by other authors \citep[e.g.][]{Popesso06, Baldry08, Drory09, Pozzetti10} to describe LFs and SMFs, and its origin is related to the bimodal colour--magnitude distribution of (blue and red) galaxies. \cite{Pozzetti10} and \cite{Baldry08} find that the massive end of the SMF ($\log\,\mathcal{M}>10.5$ M$_{\odot}$) is largely dominated by red galaxies while blue galaxies mainly contribute to the faint--end of the SMF. 
{This is also evident in the left panel of Figure\,\ref{fig:Baldry_vs_ours_combined}b} where the faint--end of the SMF of all $z<0.06$ SF galaxies is well matched to that of the GAMA SMF \citep{Baldry12}, while the bright end of the SMF shows a significant discrepancy. This disagreement arises naturally from our sample consisting only of H$\alpha$ SF galaxies. 
As only a small fraction of photometrically classified $z<0.06$ red galaxies have reliably detected H$\alpha$ emission, the GAMA SMF of $z<0.06$ SF galaxies ({left panel} of Figure\,\ref{fig:Baldry_vs_ours_combined}b) disagree with the SMF of \cite{Baldry12} at the high--mass end. 
In the right top panel of Figure\,\ref{fig:Baldry_vs_ours_combined}, the SMF of blue galaxies \citep[][open blue squares]{Baldry12} and blue SF galaxies (black triangles) are in good agreement for most stellar masses. This indicates that the blue galaxies identified by the colour split advocated by \cite{Baldry12} are mostly star-forming, with F$_{H\alpha}>1\times10^{-18}$ Wm$^{-2}$.
The SMF of all SF galaxies (grey filled symbols) shows higher $\Phi$ values at higher masses than the GAMA blue SMF as a result of the contribution from the photometrically classified red SF galaxies. The SMF of red SF galaxies is shown in {the right bottom panel} of Figure\,\ref{fig:Baldry_vs_ours_combined}b. Within uncertainties the shape of the red SF SMF is similar to that of the red SMF indicating that red SF galaxies are a small and stellar mass independent fraction of all red galaxies.
Figure\,\ref{fig:Baldry_vs_ours_combined}b (left panel) inset shows the distribution of the H$\alpha$ SF sample (purple contours) compared to all $z<0.06$ galaxies regardless of star formation, and the dashed line is the colour--magnitude cut used here. Even though photometrically classified blue and red galaxy sub--populations are conventionally labelled as SF and passive galaxies, respectively, a sample selected on the basis of detected H$\alpha$ emission, which is a direct tracer of on--going star formation in a galaxy, includes both photometrically classified blue and red galaxies. Furthermore, not all galaxies with a photometric blue classification have detected H$\alpha$ emission. {The fact that some red galaxies  are SF while some blue ones are not currently forming stars raises the question: does the shape of the SMF of blue and red SF galaxies any dependence on the fraction of red star formers and blue non star formers? The evidence of the existence red H$\alpha$ SF galaxies is more pronounced in the colour--magnitude distributions shown in Figure\,\ref{fig:SMFsinz_SMFzl}, where the number statistics of red SF systems are higher than the $z<0.06$ sample.} 

Finally, it is interesting to note that approximately 20--30\% of photometrically classified red galaxies contributing to the GAMA red SMF of \cite{Baldry12} at all stellar masses are star forming. {Approximately $40\%$ of the $z<0.06$ red SF galaxies in our sample are also detected at $250\mu m$ in Herschel--ATLAS survey, and they cover over two orders of magnitude in $250\mu m$ luminosity. Given dust is a requirement to be detected at this wavelength, it is reasonable to conclude that at least $40\%$ of the red star formers in our low--redshift sample are dusty SF galaxies and that dust in these galaxies contributes to their redder colour. Moreover, these dusty systems indicate a large spread in $NUV-r$ colour, which is an indicator of recent star formation. This implies that in addition to the dusty on--going star formation these galaxies are also dominated by old stars, and these old stars likely also contribute to the redder galaxy colour.}
\begin{figure}
		\includegraphics[width=0.4\textwidth]{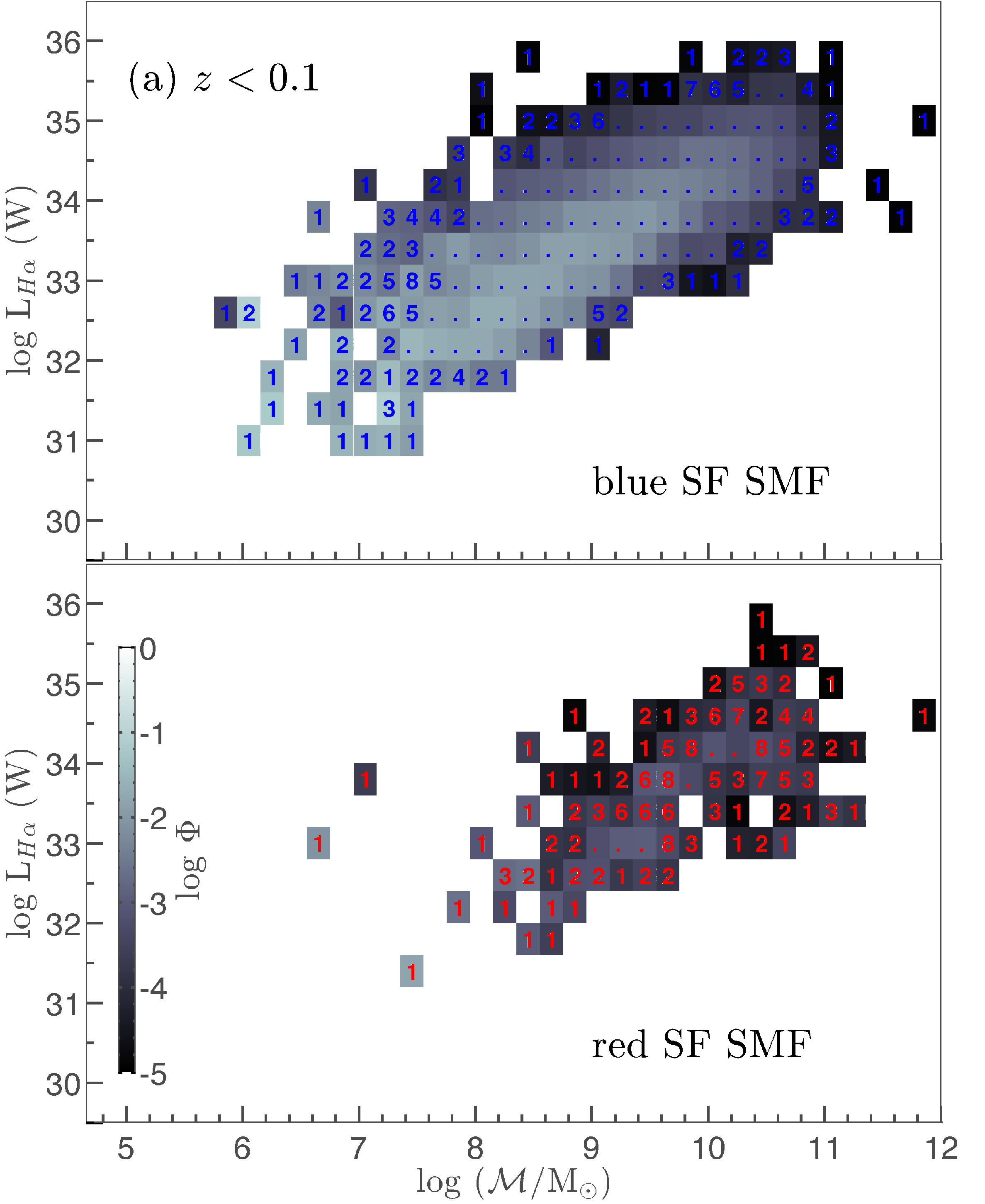}
		\caption{The $z<0.1$ bivariate L$_{H\alpha}$--$\mathcal{M}$ functions of blue and red sub--populations. The grey scale indicates the log number densities ($\Phi$) in the unit of Mpc$^{-3}$ dex$^{-2}$, and is valid for both blue and red bivariate functions corresponding to the given redshift range. }
		\label{fig:SMFsinz_biLFbr}
\end{figure}
\begin{figure*}
		\includegraphics[width=0.39\textwidth]{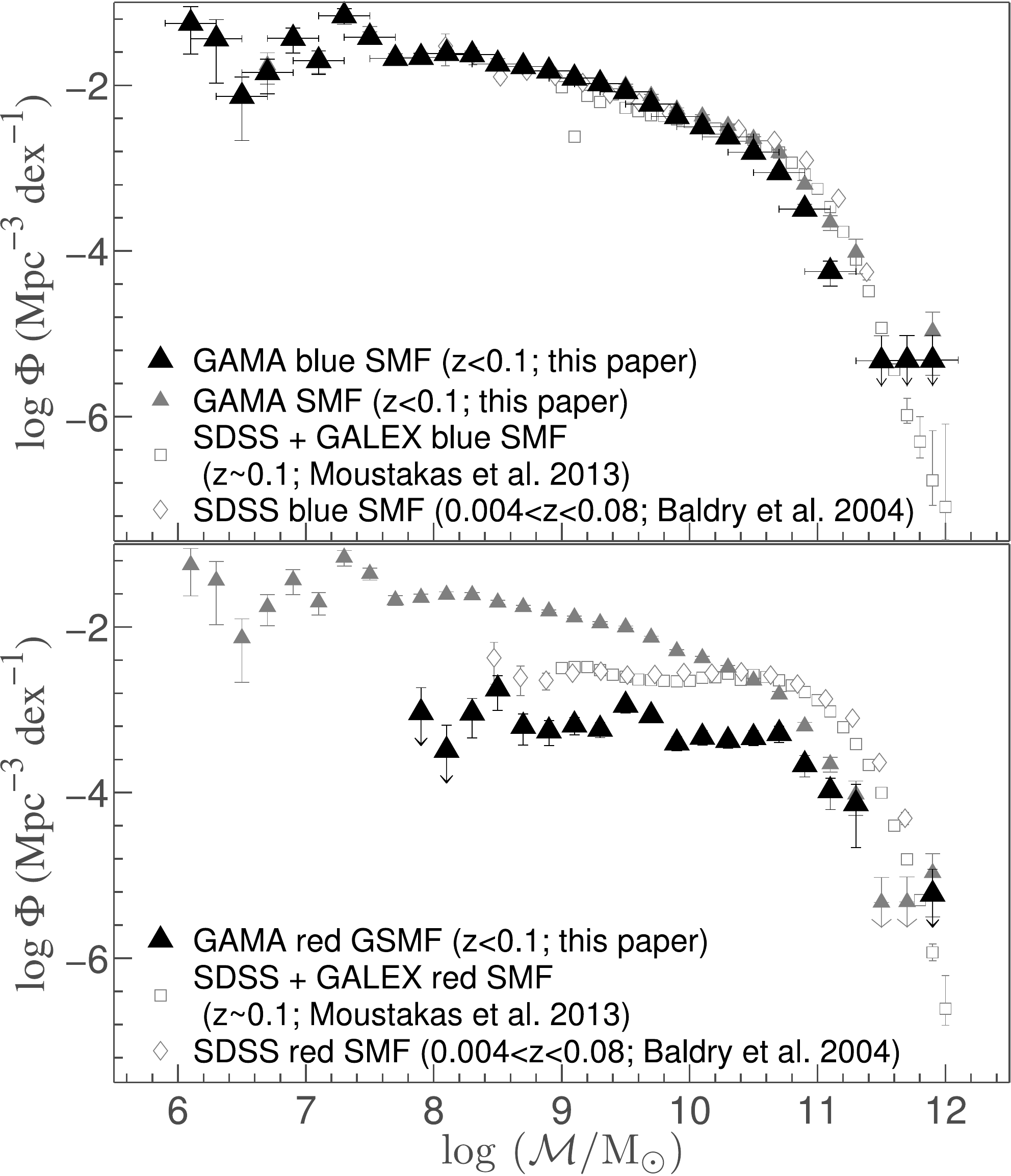}
		\includegraphics[width=0.39\textwidth]{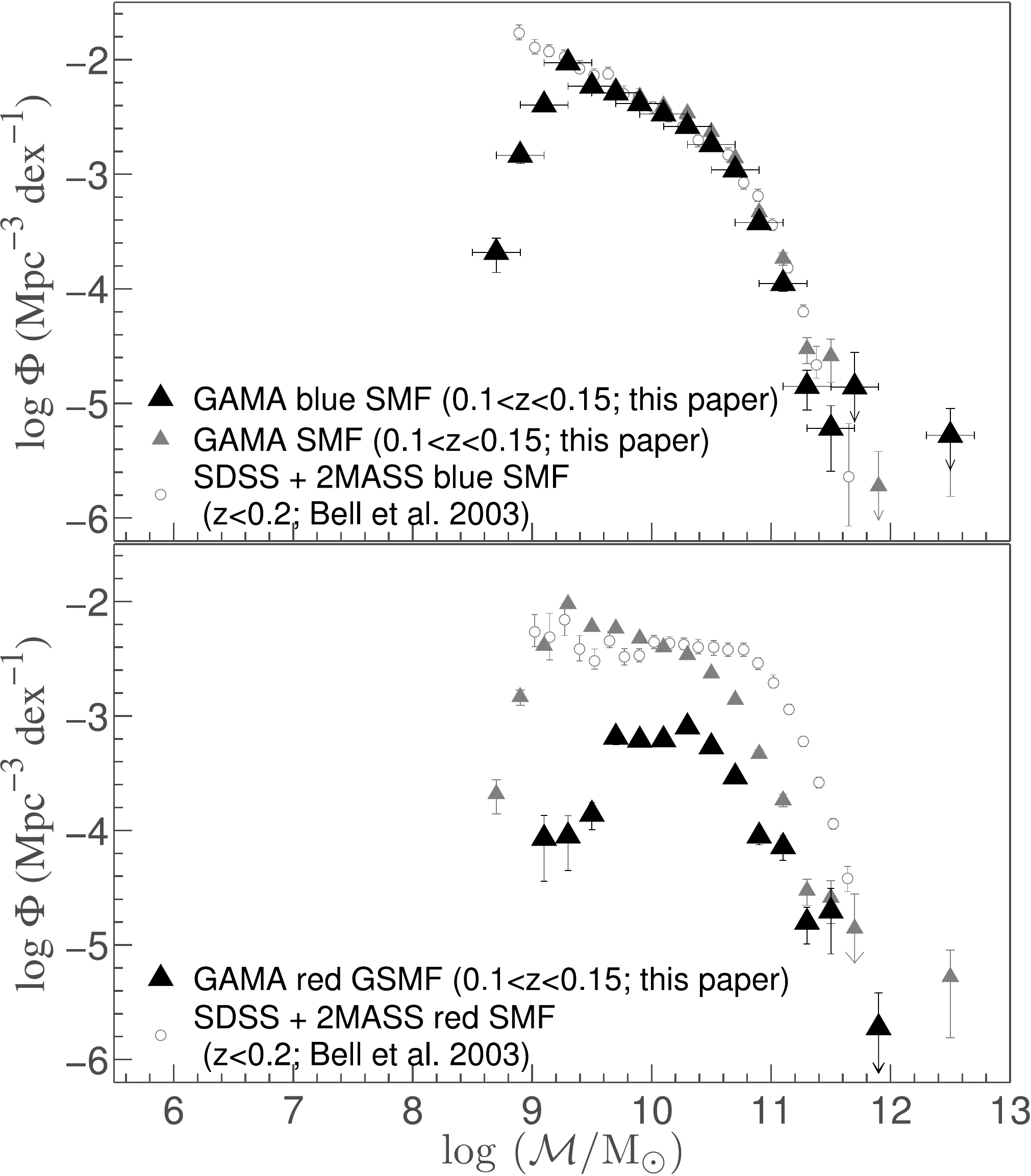}
		\includegraphics[width=0.39\textwidth]{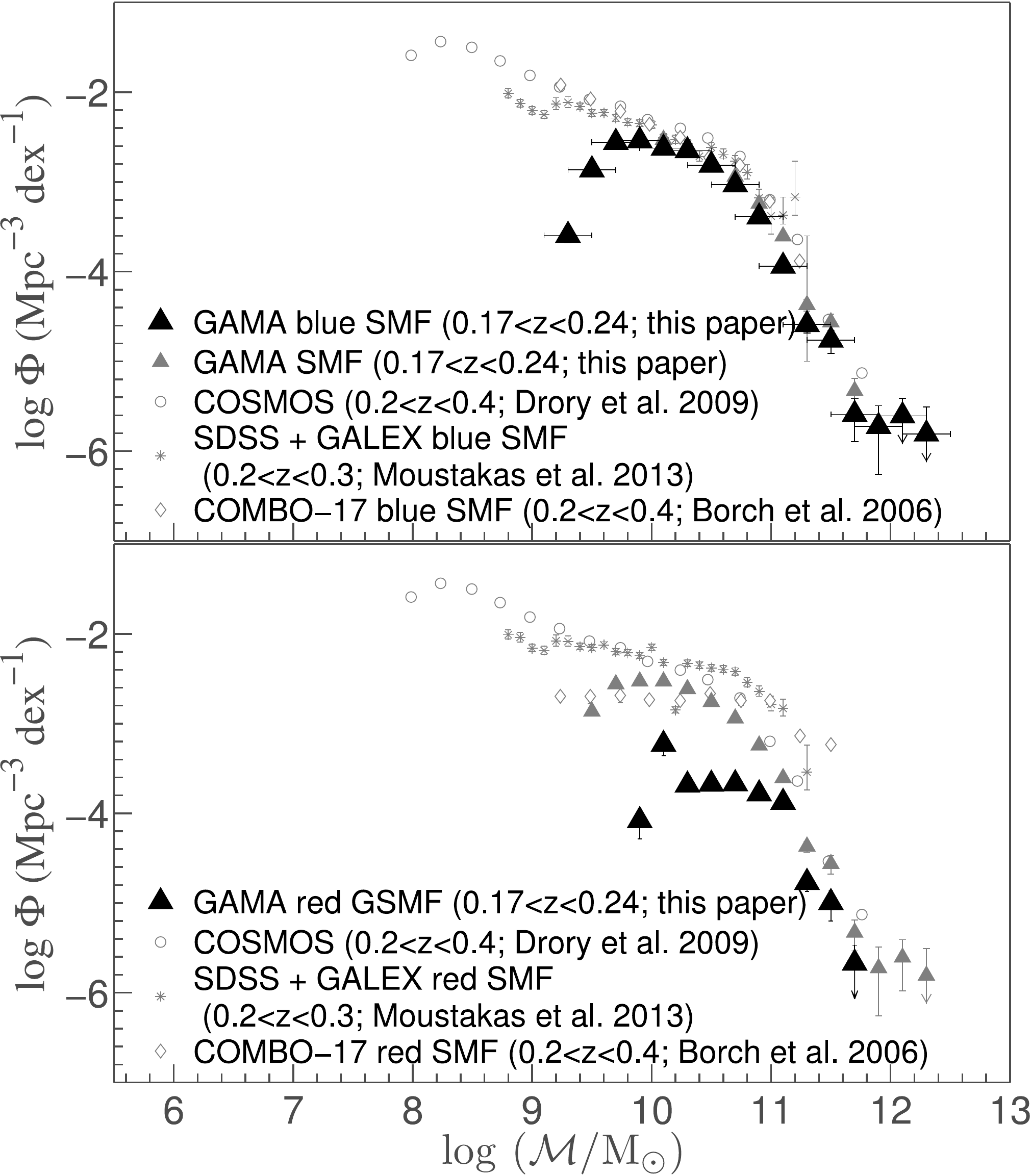}
		\includegraphics[width=0.39\textwidth]{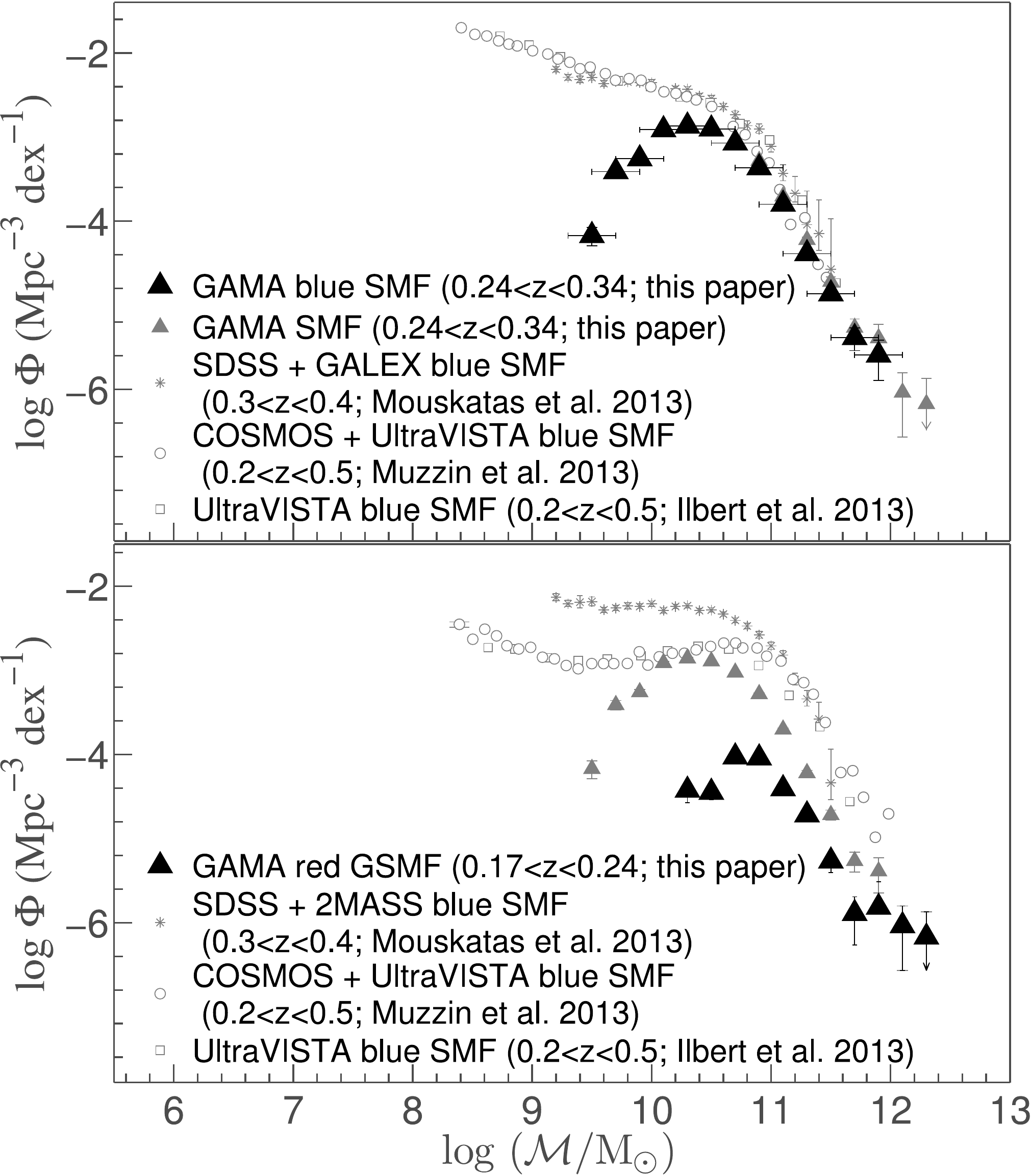}
		\caption{The blue (first panel in each two panel set) and red (second panel in each two panel set) SMFs derived from integrating the blue and red bivariate functions shown in Figures\,\ref{fig:SMFsinz_biLFbr} and \ref{fig:SMFsinz_biLFbr_app}. All published SMF measurements shown are adjusted to our assumed cosmology and to a \citet{Chabrier2003} IMF.}
		\label{fig:SMFsinz_SMFzl_app2}
\end{figure*}

\subsection{The Stellar Mass Function of Star-forming Galaxies Across Time} \label{subset:higher_smfs}

{We divide our sample into four redshift bins to investigate the evolution of the bivariate L$_{H\alpha}$--$\mathcal{M}$ functions. For each redshift range, we construct the all H$\alpha$ SF (Figure\,\ref{fig:SMFsinz_biLF}), and the respective photometrically classified blue and red SF bivariate functions (Figure\,\ref{fig:SMFsinz_biLFbr}). The lowest redshift bivariate L$_{H\alpha}$--$\mathcal{M}$ functions (i.e.\,all SF, and blue and red SF) probe the largest range in both L$_{H\alpha}$ ($30.5\lesssim\log$ L$_{H\alpha}$ W $\lesssim36$) and $\mathcal{M}$ ($6\lesssim\log \frac{\mathcal{M}}{M_{\odot}}\lesssim12$), while the survey and sample selection effects dominate the higher--$z$ ($0.1\lesssim z \lesssim0.34$) bivariate L$_{H\alpha}$--$\mathcal{M}$ functions as evident from the decrease in L$_{H\alpha}$ and $\mathcal{M}$ ranges.

The SMFs computed from integrating each bivariate function over L$_{H\alpha}$ are compared with the published measurements that cover similar redshift ranges. The results for the $z<0.1$ range is shown in Figure\,\ref{fig:SMFsinz_SMFzl}}, {\color{black} and \ref{fig:SMFsinz_SMFzl_app2} show the higher--$z$ ($0.1\lesssim z\lesssim0.34$) results}. The $z<0.1$ SMF of all SF galaxies (black filled symbols in all the panels of Figure\,\ref{fig:SMFsinz_SMFzl}) {is mostly in agreement, particularly at the low stellar mass end, with the photometrically classified blue SMFs of \cite{Baldry04} and \cite{Moustakas13}. The differences seen at higher masses, where the $z<0.1$ blue SF SMF shows relatively lower amplitudes than other published measurements, highlight that not all photometrically classified blue sources are in fact SF. The same trend can be seen in the higher--$z$ SMFs shown in Figure\,\ref{fig:SMFsinz_SMFzl_app2}.} 

Overall, the trends discussed in \S\,\ref{subsec:Baldryvsus} with regard to the $z<0.06$ SMF {(i.e.\,the difference between the SMFs of all H$\alpha$ SF galaxies and all galaxies at the high mass end that arises from the lack of many red SF galaxies, and} that $\sim20-30\%$ of photometrically classified red galaxies contributing to the red SMFs are likely SF) are also evident in the higher--$z$ SMFs shown in {Figures\,\ref{fig:SMFsinz_SMFzl} and \ref{fig:SMFsinz_SMFzl_app2}}.

\subsection{Mass--dependent evolution of the cosmic star formation history} \label{sec:SFH_mass}

\begin{figure*}
		\includegraphics[trim=1.cm 0.05cm 6.5cm .1cm, clip=true, width=0.45\textwidth]{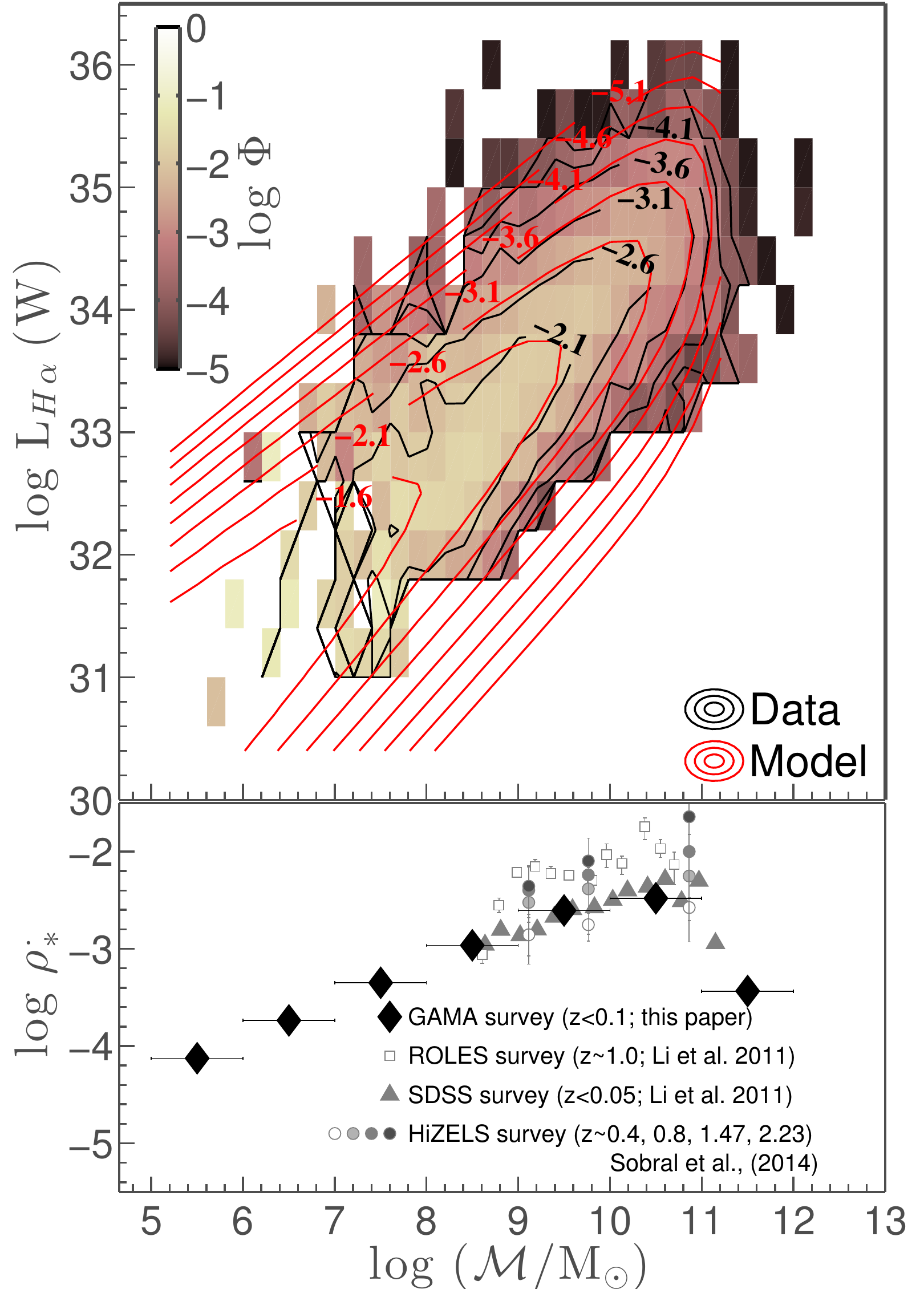}
		\includegraphics[trim=1.cm 0.05cm 6.5cm .1cm, clip=true, width=0.454\textwidth]{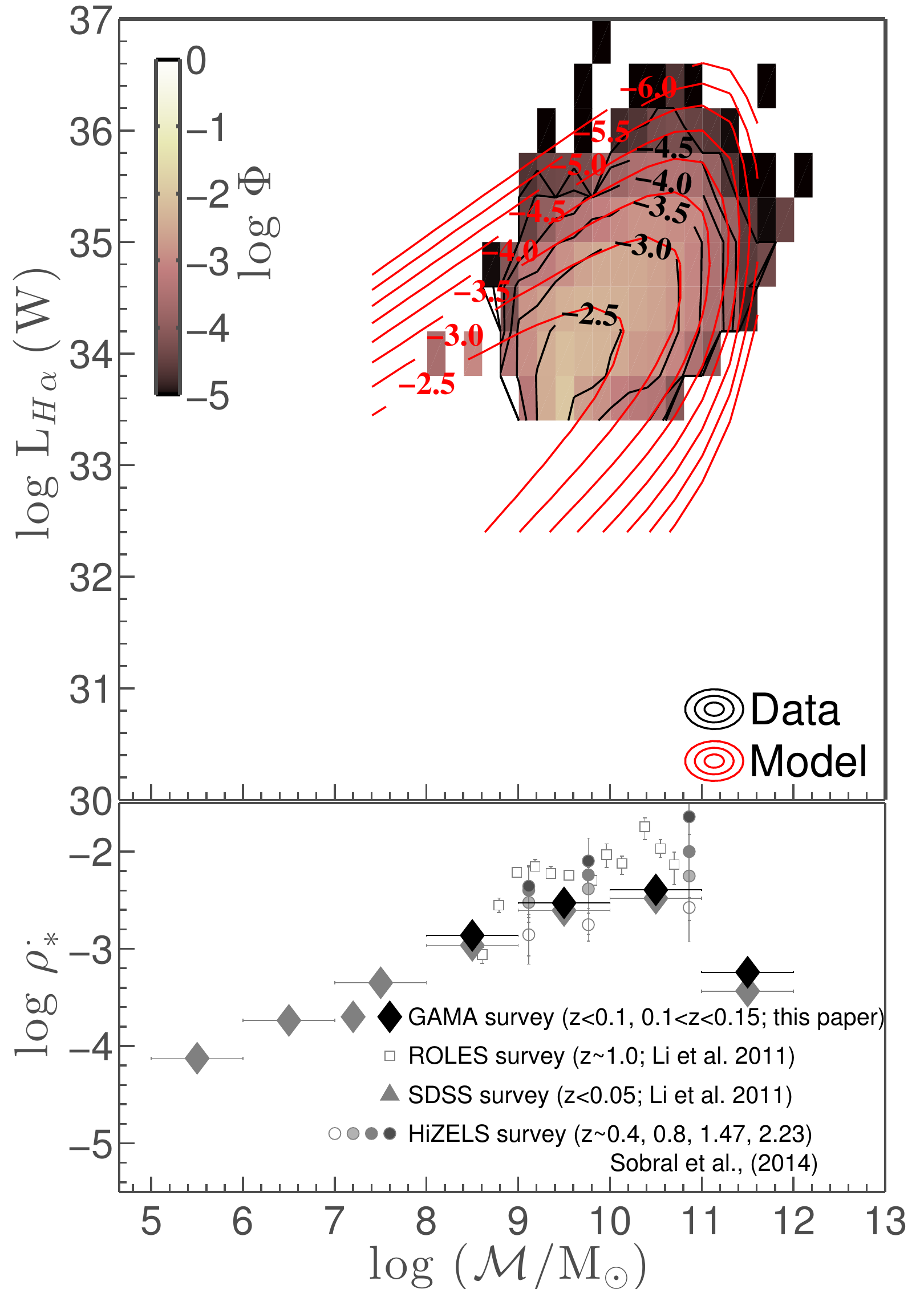}
		\caption{Top panels: Best--fitting analytic models (red lines) to the $z<0.1$ (left) and $0.1<z<0.15$ (right) bivariate L$_{H\alpha}$--$\mathcal{M}$ functions. The colour scale and black contours indicate the log number densities ($\Phi$) in the unit of Mpc$^{-3}$ dex$^{-2}$. Bottom panels:  SFRD (M$_{\odot}$yr$^{-1}$Mpc$^{-3}$dex$^{-1}$) versus stellar mass for the $z<0.1$ (left) and $0.1<z<0.15$ (right). For reference, we show the \citet{Li11} results based on SDSS (filled triangles) and ROLES (open squares) surveys and the \citet{Sobral2014} results based on HiZELS (circles) survey. All measurements are adjusted to the \citet{Baldry2003} IMF.}
		\label{fig:funcFit_SMD}
\end{figure*}

\begin{figure*}
		\includegraphics[scale=0.355]{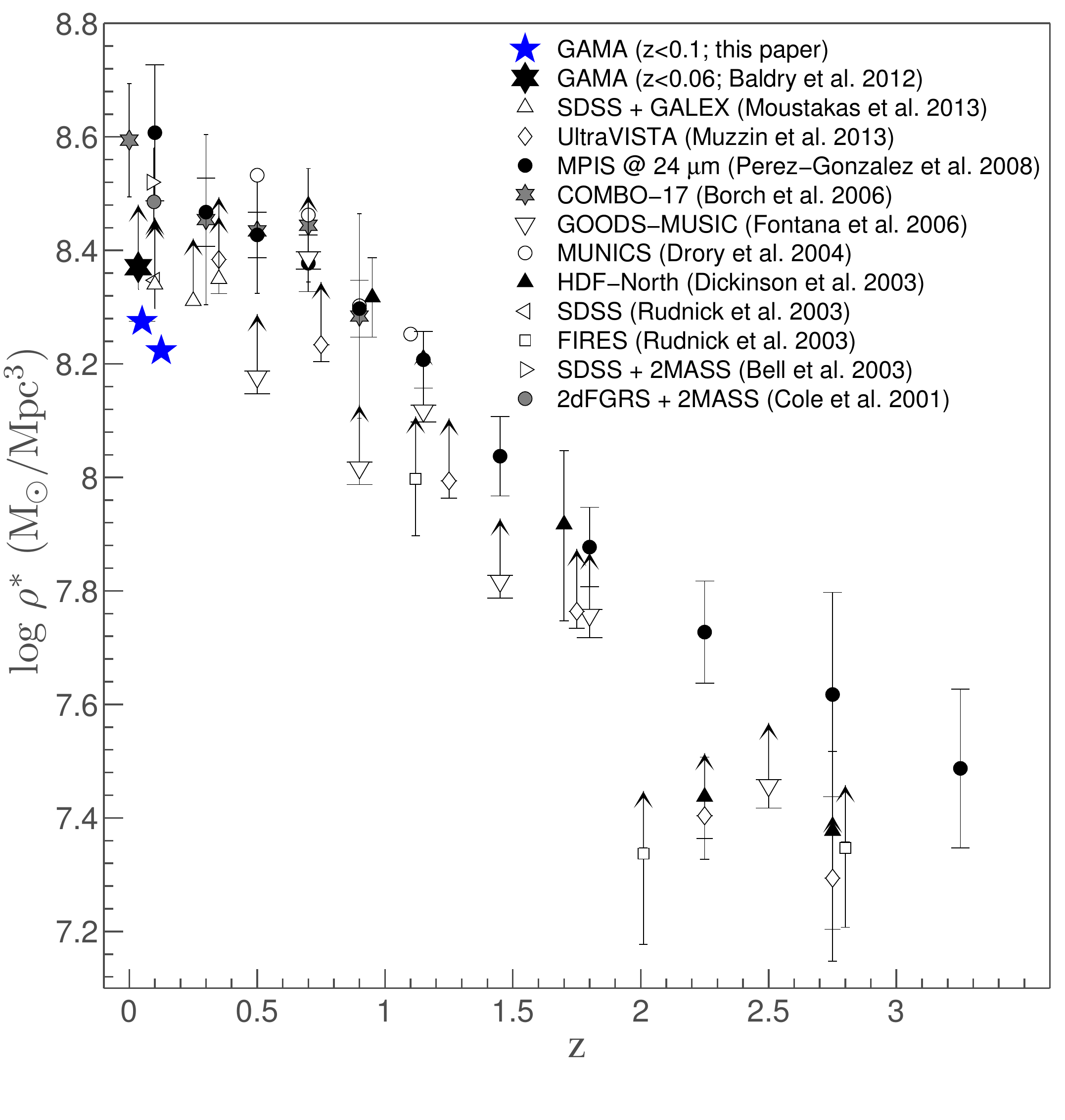}
		\includegraphics[scale=0.375]{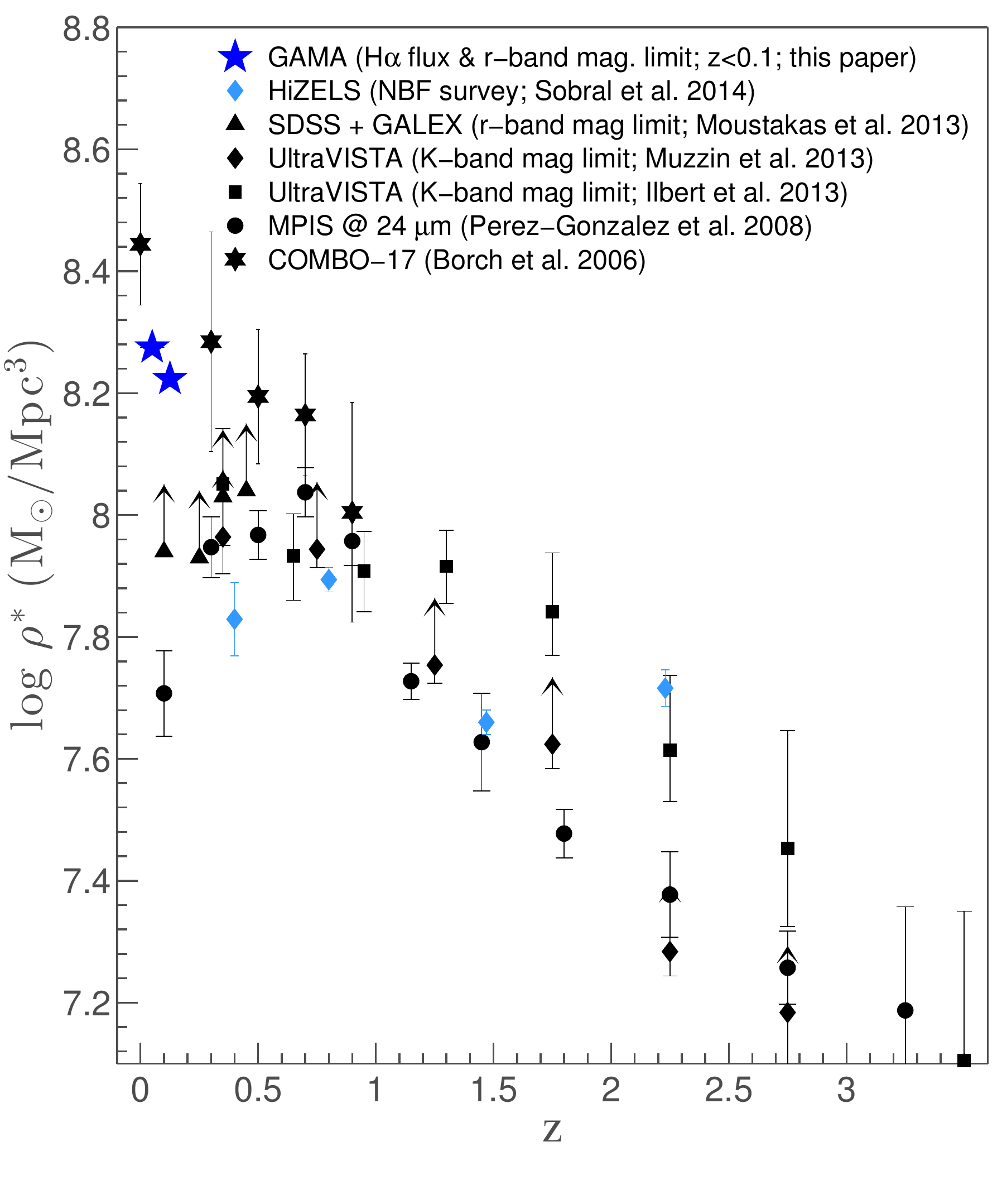}
		\caption{The evolution of the $\rho^*$, as traced by all galaxies (left) and blue/emission line galaxies (right). GAMA measurements are derived from integrating the analytic forms shown in the top panels of Figure\,\ref{fig:funcFit_SMD}. In the right panels photometrically classified blue galaxies are shown with black symbols, while coloured symbols refer to emission line samples. The cases where the authors report a density value evaluated assuming a lower mass limit are shown as lower limits.  All measurements are adjusted for our assumed cosmology and for a \citet{Chabrier2003} IMF.}
		\label{fig:SMDcomparison}
\end{figure*}

To estimate an integrated stellar mass density ($\rho^*$), {we fit the analytic form introduced in \S\,\ref{sec:funcfits}} to the bivariate  L$_{H\alpha}$--$\mathcal{M}$ functions. 
{We assume here that the relationship between $\log$ L$_{H\alpha}$ and M$_r$ derived in \S\,\ref{sec:funcfits} is also valid for $\log$ L$_{H\alpha}$ and $\log$ $\mathcal{M}$, which is not an unreasonable assumption given the observed tight relationship between M$_r$ and  $\log$ $\mathcal{M}$. In this interpretation of Eq.\,\ref{eq:bivariate}, $\log$ $\mathcal{M}^0$, the equivalent of $M^0_r$, is fixed at $\log$ $\mathcal{M}/$M$_{\odot}=9$.} The best fitting functions for the $z<0.1$ and $0.1<z<0.15$ bivariate  L$_{H\alpha}$--$\mathcal{M}$ functions are shown in the top panels of Figure\,\ref{fig:funcFit_SMD}. Our model provides a good description of the observed low redshift bivariate L$_{H\alpha}$--$\mathcal{M}$ function, especially the faint--end distribution of the function. As the ranges in L$_{H\alpha}$ and $\mathcal{M}$ sampled by the observed $0.1<z<0.15$ bivariate L$_{H\alpha}$--$\mathcal{M}$ function is significantly less than that of the $z<0.1$ function, the two model parameters, $\alpha_L$ and $\alpha_M$, that describe the faint--end shape a bivariate function are assumed to be equal to the best--fitting $z<0.1$ model values for the $0.1<z<0.15$ bivariate function. The best--fitting parameters for the $z<0.1$ and $0.1<z<0.15$ redshift bins are given in Table\,\ref{table:model_param_stellar_mass}.

\begin{table}
	\begin{center}
%		\begin{minipage}{14cm}
			\caption{ The best fitting Schechter--Saunders parameters for the $z<0.1$ and $0.1<z<0.15$ bivariate L$_{H\alpha}$--$\mathcal{M}$ functions.}
			\begin{tabular}{lll}
\hline
	Parameter       		 		&     $z<0.1$                      			&    $0.1<z<0.15$          \\
\hline
\hline
         $\log\,\mathcal{M}^*$/M$_{\odot}$	& 	$10.77\pm0.07$			        & 	$10.84\pm0.04$  		\\
         $\alpha_{\mathcal{M}}$        	& 	$-1.31\pm0.04$   				& 	-   		\\
         $\log\,\Psi^*$ (Mpc$^{-3}$)    	& 	$-4.50\pm1.23$   				& 	$-4.77\pm0.20$  		\\
         $\log$ L$^*_{H\alpha}$ (W)      & 	$32.00\pm0.58$   				& 	$32.06\pm0.20$  		\\
         $\beta$        	 	 		& 	$0.70\pm0.02$   				& 	$0.67\pm0.05$  		\\
         $\sigma$        	 	 		& 	$0.50\pm0.02$   				& 	$0.52\pm0.03$  		\\
         $\alpha_L$        	 		& 	$1.20\pm0.93$   				&      - 	\\                               
\hline
			\end{tabular}
			\label{table:model_param_stellar_mass}
%		\end{minipage}
	\end{center}
\end{table}

The relationship between $\mathcal{M}$ and $\rho^*$ for a fixed redshift range is shown in the bottom panel of Figure\,\ref{fig:funcFit_SMD}. {Overplotted in Figure\,\ref{fig:funcFit_SMD} are the $\mathcal{M}$ and $\rho^*$ relationships based on SDSS data at $z\sim0.05$ and ROLES data at $z\sim1$ \citep{Li11} {and \cite{Sobral2014} HiZELS data at $z\sim0.4, 0.8, 1.47$ and $2.23$}}. The GAMA $\mathcal{M}$ and $\rho^*$ relationship agrees well with that of \cite{Li11} {and \cite{Sobral2014}}. Also, a comparison between GAMA, SDSS, ROLES {and HiZELS} results indicates that the shape of the relationship does not vary as a function of redshift, rather it is the normalisation of the relationship that change. In the context of galaxy downsizing \citep{Cowie99}, which states that high--mass galaxies formed their stars early and rapidly while low--mass counterparts formed stars at a slower rate and later times, the peak of the $\log\,\mathcal{M}$ and $\rho^*$ relationship is expected to shifts towards lower masses with decreasing redshift. The lack of such a change contradicts the downsizing scenario, however, as \cite{Gilbank2010} point out high SFR galaxies are also likely to be high--mass systems that both dominate the galaxy numbers and $\rho^*$ at high $z$, which could be considered `downsizing'. 

The $\rho^*$ derived from integrating the best--fitting analytic functions to the bivariate L$_{H\alpha}$--$\mathcal{M}$ functions (Figure\,\ref{fig:funcFit_SMD}) are shown in Figure\,\ref{fig:SMDcomparison}. Also shown for comparison are the $\rho^*$ measurements from recent studies at various redshifts up to $z\sim3$. Note that all published measurements shown in Figure\,\ref{fig:SMDcomparison} are calculated from univariate SMFs based on galaxy samples drawn from broadband surveys. The data indicated as lower limits show the cases where the authors have provided a $\rho^*$ measurement by integrating the univariate SMF down to a limiting $\mathcal{M}$, rather than to zero. 

The left panel of Figure\,\ref{fig:SMDcomparison} compares the GAMA $\rho^*$ measurements based on bivariate L$_{H\alpha}$--$\mathcal{M}$ functions with those derived from total galaxy samples, and the right panel of the same figure compares our results with the $\rho^*$ based on photometrically classified blue galaxy populations {and emission line samples}. Our results agree well with the published measurements based on photometrically classified blue galaxy populations {and emission line samples}, as might be expected given the close agreement between our SF SMFs and the blue galaxy SMF of \cite{Baldry12}. 

{
\section{Discussion}\label{sec:discuss}

In this followup paper to paper~I, we have explored the bivariate L$_{H\alpha}$--M$_r$ and  L$_{H\alpha}$--$\mathcal{M}$ functions of GAMA galaxies. One of the main aims of this analysis is to investigate whether we can reliably recover a correction for the incompleteness introduced in selecting H$\alpha$ detected objects from a $r$--band limited survey by modelling the low redshift bivariate function, which can then be used as a reference to account for the missing optically faint SF galaxies at higher--$z$ ($0.1\lesssim z\lesssim0.34$). Other goals of this investigation include exploring the evolution of bivariate and univariate functions of all H$\alpha$ SF and photometrically classified blue and red H$\alpha$ SF sub--populations relative to the evolution of all, blue and red populations regardless of star star formation, and the mass dependence of the SFR history.  

In order to compare our results more directly with the previous GAMA LF \citep{Loveday12} and SMF \citep{Baldry12} studies, we adopt the LF estimators used in their studies to construct the bivariate functions. A discussion on the formulation of three LF estimators (e.g.\,the classical method, density corrected 1/V$_{\rm max}$ method and SWML estimator) for constructing bivariate functions is presented in \S\,\ref{sec:construction} and in Appendix\,\ref{sec:swml_lf}.  
A comparison between the bivariate functions based on these methods (\S\,\ref{subsec:comparison}) shows that the differences in number densities are largely limited to the faint L$_{H\alpha}$--M$_r$ end of the bivariate functions as expected given the relatively small volumes sampled. The bivariate L$_{H\alpha}$--M$_r$ functions and LFs presented in \S\,\ref{sec:biLF} and  L$_{H\alpha}$--$\mathcal{M}$ functions and SMFs presented in \S\,\ref{subset:higher_smfs} are based on the classical 1/V$_{\rm max}$ method, and are compared with the GAMA LFs of \cite{Loveday12} and other published measurements that are mostly based on the classical method. The bivariate L$_{H\alpha}$--$\mathcal{M}$ functions and SMFs presented in \S\,\ref{subsec:Baldryvsus} are based on the density corrected 1/V$_{\rm max}$, and are compared with the GAMA SMFs \citep{Baldry12} based on the same method.

As a consequence of the magnitude limited nature of the GAMA survey, the bivariate L$_{H\alpha}$--M$_r$ functions (Figure\,\ref{fig:vmax_bilf} and \ref{fig:vmax_bilf_photom}) show a progressive decline in number density ($\Phi$) towards fainter L$_{H\alpha}$ and M$_r$, and the range in  L$_{H\alpha}$ and M$_r$ probed decrease with increasing redshift. In each of the four redshift bins considered, the L$_{H\alpha}$--M$_r$ range, and the $\Phi$ values of the blue H$\alpha$ SF bivariate functions (Figure\,\ref{fig:vmax_bilf_photom}) are similar to that of the total H$\alpha$ SF functions (Figure\,\ref{fig:vmax_bilf}). This result indicates that the galaxies contributing to the total H$\alpha$ SF bivariate functions are mostly drawn from the photometrically classified blue sub--populations at each redshift. This is further collaborated by both the relatively smaller range in L$_{H\alpha}$ and M$_r$ probed by the red bivariate L$_{H\alpha}$--M$_r$ functions (Figure\,\ref{fig:comprison_app}), and the close agreement between the \cite{Loveday12} blue LFs and those computed by integrating the blue bivariate L$_{H\alpha}$--M$_r$ functions at all redshifts (Figures\,\ref{fig:comprison1} and \ref{fig:comprison_app}). While the red H$\alpha$ SF bivariate number density contribution to the total H$\alpha$ SF bivariate function is relatively low, Figures\,\ref{fig:comprison1} and \ref{fig:comprison_app} show that a fraction of galaxies classified as red at all M$_r$ values are in fact SF galaxies. The same conclusion is drawn the bivariate L$_{H\alpha}$--$\mathcal{M}$ functions (Figures\,\ref{fig:Baldry_vs_ours_combined}, \ref{fig:SMFsinz_biLF} and \ref{fig:SMFsinz_biLFbr}), and the SMFs (Figures\,\ref{fig:SMFsinz_SMFzl}, \ref{fig:SMFsinz_SMFzl_app2}) computed from integrating the bivariate functions. While the H$\alpha$ SF galaxy population at each redshift is primarily drawn from the blue sub--populations at that redshift, approximately $20-30\%$ of those photometrically classified as red and conventionally called passive are forming massive stars at all stellar masses at each redshift range probed.  We find that $\sim40\%$ of the red H$\alpha$ SF galaxies at $z<0.06$ (i.e.\,those contributing to the red bivariate L$_{H\alpha}$--$\mathcal{M}$ function and the SMF shown in Figure\,\ref{fig:Baldry_vs_ours_combined}) is also reliably detected at $250\mu m$. As dust is a requirement to be able to be detected at this wavelength, it is likely that some of the red H$\alpha$ SF galaxies are dusty (and therefore red) SF systems. Moreover, those detected at $250\mu m$ cover a large range in $NUV-r$ colour, which is an indicator of recent star formation in galaxies, and most lie below the $NUV-r=5.4$ cutoff for recent star formation from \cite{Schawinski2007}. This points to red H$\alpha$ SF galaxies having a non-negligible fraction of underlying old stellar population that likely also contribute to their redder colour. 

Motivated by the analytic formalism widely used to model bivariate brightness profiles \citep[e.g.][]{deJong00, Cross01, Driver05, Ball06}, in \S\,\ref{subsubsec:simple_model} we introduce a simple analytic model to describe the observed bivariate functions presented in \S\,\ref{sec:biLF} and \ref{sec:biLFs_SMF}. This model assumes that the bivariate function can be written as a product of two functions \citep{Choloniewski85, Corbelli91}, a \cite{Schechter76} function representative of  $r$--band LFs (or SMFs) and a \cite{Saunders90} function representative of H$\alpha$ LFs (paper~I). The two functional forms are linked in the bivariate analytic relation through the observed  relationship between L$_{H\alpha}^*$ and M$_r$ (Figure\,\ref{fig:normalised_LFs}). The resultant best--fitting bivariate models to the $z<0.1$ L$_{H\alpha}$--M$_r$, and the $z<0.1$ and $0.1<z<0.15$ L$_{H\alpha}$--$\mathcal{M}$ functions (Figures\,\ref{fig:simple_model} and \ref{fig:funcFit_SMD}, respectively) provide a good description of the data, and from integrating the best--fitting $z<0.1$ bivariate L$_{H\alpha}$--M$_r$ model we were able to recover the $z<0.1$ SFRD reported in paper~I. The same model is used to fit the rest of the higher--$z$ bivariate functions by fixing the faint L$_{H\alpha}$--M$_r$ (or $\mathcal{M}$) end slope of the bivariate model to be equal to that of the lowest redshift best--fitting model. As the $0.1<z<0.15$ bivariate functions sample a relatively large range in both L$_{H\alpha}$ and M$_r$ (or $\mathcal{M}$), the $0.1<z<0.15$ models can be reasonably well constrained, however, the two higher--$z$ models cannot be constrained accurately even with assuming a constant faint--end slope (i.e.\,assuming no evolution in the faint--end slope of the bivariate function) for the model. This is mainly due to the observed higher--$z$ bivariate L$_{H\alpha}$--M$_r$ and L$_{H\alpha}$--$\mathcal{M}$ functions being incomplete at around the characteristic L$_{H\alpha}$, M$_r$ and $\mathcal{M}$ values. By integrating the best--fitting higher--$z$ bivariate L$_{H\alpha}$--M$_r$ and L$_{H\alpha}$--$\mathcal{M}$ models, particularly the $0.1<z<0.15$ models which we were able to constrain more accurately than the rest, we were able to recover an approximate correction for the missing optically faint SF galaxies. The SFRDs estimated this way (Figure\,\ref{fig:SFRD_density}) show a strong evolution with redshift up to $z\sim0.2$ and a flattening thereafter, though, we caution against using the last two points (i.e.\,the two higher--$z$ $\rho^*$ measurements) in understanding the evolution of $\rho^*$ with redshift as their best--fitting bivariate models are not well constrained. 

The mass dependence of the SFR history and the evolution of the stellar mass density of H$\alpha$ SF galaxies are explored in \S\,\ref{sec:SFH_mass}. The $\mathcal{M}$ and $\rho^*$ relationship for the $z<0.1$ and $0.1<z<0.15$ redshift ranges derived from integrating the respective GAMA bivariate L$_{H\alpha}$--$\mathcal{M}$ functions is shown in Figure\,\ref{fig:funcFit_SMD}. Also shown for comparison are the $\mathcal{M}$--$\rho^*$ relations derived using SDSS data at $z\sim0.05$, ROLES data at $z\sim1$ \citep{Li11} and HiZELS data at $z\sim0.4, 0.8, 1.47, 2.23$ \citep{Sobral2014}. A comparison between GAMA, SDSS, ROLES and HiZELS $\mathcal{M}$--$\rho^*$ results show that while the normalisation of the $\mathcal{M}$--$\rho^*$ relation increases modestly with redshift, the shape remains the same. This result points towards a scenario where SFRs in all galaxies, regardless of stellar mass, decline with decreasing redshift \citep{Karim2011, Sobral2014}, contradicting the `galaxy downsizing' scenario of \cite{Cowie99}, which states that massive galaxies formed their stars early at a rapid rate while the low--mass systems formed stars late at a slower rate. However, as \cite{Gilbank2010} point out the high SFR systems are also likely to be high--mass galaxies that dominate SFRD at high--$z$, which could be considered as downsizing. 

Finally, the evolution of the $\rho^*$ of H$\alpha$ SF population in comparison to the evolution of $\rho^*$ of all galaxies regardless of star formation, and of photometrically classified blue galaxies (black markers) and other H$\alpha$ galaxy samples (coloured markers) is shown in the left and right panels of Figure\,\ref{fig:SMDcomparison}, respectively. The SMDs based on the H$\alpha$ SF sub sample of GAMA galaxies lie low in the left panel of Figure\,\ref{fig:SMDcomparison}. This is expected as other stellar mass densities shown in that panel are based on all galaxies regardless of star formation. Our results are in good agreement with the SMDs based only on either the blue sub--population of galaxies or other emission-line samples (right panel of Figure\,\ref{fig:SMDcomparison}). The scatter in data points is most likely due to cosmic (sample) variance. 
}

\section{Summary} \label{sec:conclude}

{We have explored the GAMA bivariate L$_{H\alpha}$--M$_r$ and  L$_{H\alpha}$--$\mathcal{M}$ functions in the paper, and the key results of the analysis are as follows:}

\begin{enumerate}[label=\roman*., leftmargin=*]
	\itemsep4pt
	
	\item By modelling the low redshift distribution of the  bivariate L$_{H\alpha}$--M$_r$ function, we estimate a correction for the missing optically faint star forming galaxies at higher--$z$. The corrected stellar mass densities presented in Figure\,\ref{fig:SFRD_density} show high level of consistency with earlier published results \citep[e.g.][]{HB06}, suggesting that this approach is reasonable. The implication is that the shape of the faint--end of the bivariate functions does not evolve strongly over this redshift range.  
	
	\item The H$\alpha$ SF sample used for this study consists of both photometrically classified blue and red galaxies, though dominated by blue galaxies. This allows us to construct not only the bivariate and univariate LFs and SMFs of H$\alpha$ SF galaxies, but also the LFs and SMFs of photometrically classified blue and red SF galaxies. 
	
	\item The M$_r$ LFs obtained from integrating the bivariate L$_{H\alpha}$--M$_r$ functions are in agreement with the GAMA M$_r$ LFs of photometrically classified blue galaxies from \cite{Loveday12}. Also, the $z<0.06$ SMF of SF galaxies obtained from integrating the bivariate L$_{H\alpha}$--$\mathcal{M}$ function is consistent with the blue SMF of \cite{Baldry12}.
	
	\item The low redshift ($z<0.15$) $\rho^*$ and $\mathcal{M}$ relationship derived using GAMA data agrees well the $z<0.05$ results from \cite{Li11}. The shape of the $\rho^*$ and $\mathcal{M}$ relationship at low redshift is similar to that obtained at $z\sim1$ from \cite{Gilbank2010}. The comparison of results between different redshifts indicate that the shape of the $\rho^*$ and $\mathcal{M}$ relationship does not change with redshift over the redshift range probed by the bivariate functions, rather it is the normalisation of the relationship at all masses that varies with redshift.   
	
	\item The GAMA stellar mass densities based on bivariate L$_{H\alpha}$--$\mathcal{M}$ functions, i.e.\,based on a sample of H$\alpha$ SF galaxies, is in good agreement with the published stellar mass densities based on photometrically classified blue galaxy populations. 
		
\end{enumerate}

\nocite{Muzzin2013}
\nocite{Drory2004}
\nocite{Fontana2006}
\nocite{Dickinson2003}
\nocite{Rudnick2003}
\nocite{Cole2001}
\nocite{Ilbert2013}
\nocite{Panter2007}
\nocite{Borch2006}

\section*{Acknowledgments}

We thank the anonymous referee for extensive comments that have helped us to refine the analysis presented in this paper. We also thank Qi Guo and Loretta Dunne for their assistance with Herschel ATLAS data. 
M.L.P.G.\ acknowledges support provided through the Australian Postgraduate Award and Australian Astronomical Observatory Ph.D scholarship. M.L.P.G.\,and P.N.\,acknowledge support from a European Research Council Starting Grant (DEGAS-259586). 

GAMA is a joint European-Australasian project based around a spectroscopic campaign using the Anglo-Australian Telescope. The GAMA input catalogue is based on data taken from the Sloan Digital Sky Survey and the UKIRT Infrared Deep Sky Survey. Complementary imaging of the GAMA regions is being obtained by a number of independent survey programs including GALEX MIS, VST KIDS, VISTA VIKING, WISE, Herschel-ATLAS, GMRT and ASKAP providing UV to radio coverage. GAMA is funded by the STFC (UK), the ARC (Australia), the AAO, and the participating institutions. The GAMA website is http://www.gama-survey.org/. 

Data used in this paper will be available through the GAMA website (\url{http://www.gama-survey.org/}) once the associated redshifts are publicly released. 

\footnotesize
{
\bibliographystyle{apsrmp}
\bibliography{references}
}
\bsp

\appendix
{

\section{Bivariate stepwise maximum likelihood method}\label{sec:swml_lf}

In order to overcome the inconvenience of not being able to adequately visualise whether the chosen parameterisation represents a good fit to the data \citep{Willmer97}, a ``nonparametric" LF stepwise maximum likelihood (SWML) LF estimator is introduced by \cite{Efstathiou88}. The SWML method does not suffer from the same biases that affect 1/V$_{\rm max}$ technique. Unlike 1/V$_{\rm max}$ method, the luminosity bins in SWML are highly correlated, so any issue that occurs affect the whole luminosity function.

 \cite{Efstathiou88} describes the SWML formulation for univariate LFs. An extension of the SWML LF estimator for bivariate LFs is discussed in \cite{Sodre93} for the bivariate diameter--luminosity function and in \cite{Driver05} for the bivariate brightness distribution. Following \cite{Efstathiou88, Sodre93} and \cite{Driver05}, we construct a bivariate SWML estimator for H$\alpha$/M$_r$ and H$\alpha$/stellar mass LFs. 

The probability of observing a galaxy $i$ with a H$\alpha$ luminosity $L_i$ and an absolute $r$--band magnitude $M_{r_i}$ at a redshift $z_i$, inclusive of the completeness/selection function ($f$), is defined to be \citep{Sodre93, Loveday00, Ball06}
\begin{equation} 
    	p_i \propto \frac{\Phi(M_i, L_i)\,f(M_i, L_i, z_i)}{\int_{L_{faint}(z_i)}^{L_{bright}(z_i)=\infty}\int_{M_{faint}(z_i)}^{M_{bright}(z_i)} \Phi(M,L) f(M,L,z) dM dL}.
	\label{math:swml_prob}
\end{equation}
The probability, $p_i$, is directly proportional to the differential LF at $M_i$ and $L_i$, and inversely proportional to the faintest and brightest absolute magnitudes ($M_{faint}$, $M_{bright}$) and H$\alpha$ luminosities ($L_{faint}$, $L_{bright}$) visible at $z_i$ \citep{Willmer97, Heyl97}. 
Note that we do not account for survey incompleteness via $f$. Instead the incompleteness corrections defined in paper~I are incorporated into the bivariate LF through a weighting function \citep{Driver05} as explained below. 

The SWML LF is derived by maximising the likelihood function, $\mathcal{L}  = \prod_{i=1}^{Ng} p_i$, generally log $\mathcal{L}$, with respect to the discretised luminosity distribution. 
The bivariate LF, $\Phi(M,L)$, is parameterised as N$_M$ and N$_L$ steps \citep{Efstathiou88}: 
\begin{equation}
    \small{\Phi (M,L)\,=\,\Phi_{j,k} \ \ \small{\text{j = 1......$N_M$, and k = 1......$N_L$}}}
\end{equation}
where $N_M$ and $N_L$ are evenly spaced bins in M$_r$ and H$\alpha$ luminosity with $M_j - \frac{\Delta M}{2} \leqslant M_i \leqslant M_j + \frac{\Delta M}{2}$, and $L_k - \frac{\Delta L}{2} \leqslant L_i \leqslant L_k + \frac{\Delta L}{2}$. 

Rewriting the denominator of Eq.\,\ref{math:swml_prob} in summation notation gives the following log-likelihood function
\begin{equation}
	\begin{aligned}
		\ln\, \mathcal{L} & =  \sum_{i=1}^{N_g}  \sum_{j=1}^{N_M}  \sum_{k=1}^{N_L} W[M_i - M_j, L_i - L_k]\, \ln\, \Phi_{jk} - \sum_{i=1}^{N_g} \ln\\
					  & \left\{ \sum_{a=1}^{N_M}  \sum_{b=1}^{N_L}\, \Phi_{ab} \Delta M \Delta L H[M_a - M_{faint}(z_i), L_b - L_{faint}(z_i)]  \right\} \\
					  & + \,C, 
	\end{aligned}
	\label{math:log_like}
\end{equation}
where $C$ is a constant, $W[M_i - M_j, L_i-L_k]$ is the weighting function defined as \citep{Driver05}
\begin{equation}
	W[M_i - M_j, L_i-L_k] = \left\{
  	\begin{array}{l l}
    		1 & \quad \small{\text{if} \ \   M_j -  \frac{\Delta M}{2}\leqslant M_i \leqslant M_j +  \frac{\Delta M}{2}}\\
         	      & \quad \small{\text{and} \ \ L_k -  \frac{\Delta L}{2}\leqslant L_i \leqslant L_k +  \frac{\Delta L}{2}} \\
          	      & \\
   		 0   & \quad \small{\text{otherwise,}}\\
  	\end{array} \right.
\end{equation}
and $H[M_j- M_{faint}(z_i), L_k-L_{faint}(z_i)]$ is the ramp function, inclusive of incompleteness corrections ($c$; paper~I), defined as, 
\begin{equation}
	\begin{aligned}
  		& H[M_j- M_{faint}(z_i), L_k-L_{faint}(z_i)]  =  \\ 
		& \frac{1}{\Delta M \Delta L} \int \limits_{M_j- \frac{\Delta M}{2}}^{M'} dM \int \limits_{L'}^{L_k + \frac{\Delta  L}{2}} dL\, O_i(M,L)\,c. 
	\end{aligned}
	\label{eq:ramp}
\end{equation}
The definitions of $M'$ and $L'$ are: 
\[
\begin{array}{lll}
  M'  & =  &  \max\{M_j - \frac{\Delta M}{2}, \min[M_j + \frac{\Delta M}{2}, M_{faint}(z_i)]\} \\
  L'   & =  &  \max\{L_k - \frac{\Delta L}{2}, \min[L_k + \frac{\Delta L}{2}, L_{faint}(z_i)]\}
\end{array}
\]
Simply setting $M'$ or $L'$ to be equal to the faint magnitude or luminosity bin boundary will result in the LF being underestimated in incompletely sampled bins. If the faint bin boundaries are used in the ramp function, then the incomplete bins should be excluded \citep{Loveday12}. Finally, the function, $O_i(M,L)$, in Eq.\,\ref{eq:ramp} is the observable window function for each galaxy at $z_i$ \citep{Driver05}, and has the following form. 
\begin{equation}
 	O_i(M, L) = \left\{
  		\begin{array}{l l}
    		    1  & \quad \small{\text{if} \ \   M_{bright, i} \leqslant M_i \leqslant M_{faint, i}}\\
          		& \quad \small{\text{and} \ \  L_{faint, i} \leqslant L_i }\\
          		& \\
    		     0 & \quad \small{\text{otherwise,}}\\
  		\end{array} \right.
\end{equation}
where, 
\[
\begin{array}{lll}
  M_{bright, i} & =  &  m_{bright} - 5\log d_{L(z_i)} - 25 - K(z_i) \\
  M_{faint, i}   & =  &  m_{faint} - 5\log d_{L(z_i)} - 25 - K(z_i)\\
  L_{faint, i}    & =  &   4\pi d_{L(z_i)}^2 f_{H\alpha, faint}
\end{array}
\]
and
\[
\begin{array}{lll}
  m_{bright}  & =  &  14.65  \quad \small{\text{(SDSS bright magnitude limit)}} \\
  m_{faint}    & =  &  19.4 \quad \small{\text{or  }} 19.8  \quad \small{\text{(GAMA faint magnitude limits)}}\\
  f_{faint}      & =  &   1\times 10^{-18} W/m^2 \quad \small{\text{(H$\alpha$ flux limit; paper~I)}}
\end{array}
\]

As discussed in \cite{Efstathiou88} a constraint must be imposed on the likelihood to fix the normalization constant in Eq.\,\ref{math:log_like}, by using a Lagrangian multiplier ($\lambda$). We adopt the constraint used by \cite{Ball07} and \cite{Sodre93}. 
\begin{equation}
    	g = \sum_{j=1}^{N_M}  \sum_{k=1}^{N_L} \Phi_{jk} \Delta M \Delta L - 1 = 0.
	\label{math:swml_constraint}
\end{equation}

The likelihood with the constraint applied, $\ln\, \mathcal{L}'  = \ln\,\mathcal{L} + \lambda g(\Phi_{jk})$, is maximised with respect to $\Phi_{jk}$ and $\lambda$, requiring $\lambda=0$. The constraint, although it does not affect the shape of the LF determined by $\Phi_{jk}$, plays a role in the error determination \citep{Efstathiou88}. The maximum likelihood (i.e.\,$\partial\, \ln \,\mathcal{L}' / \partial \Phi_{jk} = 0$) is then given by, 
\begin{equation} 
	\begin{aligned}
    		& \Phi_{jk} \Delta M \Delta L  = \\
		& \frac{\sum_{i=1}^{Ng}{W[M_i - M_j, L_i-L_k]}} {\sum_{i=1}^{Ng}\left\{\frac{H[M_j-M_{faint}(z_i), L_k - L_{faint}(z_i)]}{\sum_{a=1}^{Nm} \sum_{b=1}^{Nl}\Phi_{ab} \Delta M \Delta L H[M_a - M_{faint}(z_i), L_b - L_{faint}(z_i)]} \right\}},
	\end{aligned}
	\label{eq:phi_det}
\end{equation} 
where, $\Phi_{ab}$ is from the previous iteration. 

The bivariate SWML LF ($\Phi_{jk}$), by construction, loses the information regarding the absolute normalisation \citep{Efstathiou88}. We achieve a normalisation for the bivariate SWML LF results presented in this paper by matching the bright--end of the bivariate SWML LFs to their 1/V$_{\rm max}$ LFs. While the shape of the faint--end 1/V$_{\rm max}$ LF can be affected by the over/under densities, this estimator provides reliable abundances at higher luminosites, where it probes a relatively larger volume \citep{Driver05, Eke05}. Therefore, matching to the bright end of the LFs is a robust approach to fix the normalisation of the bivariate SWML LF. 

The LF errors can be determined using the fact that maximum likelihood estimates ($\Phi_{jk}$) are asymptotically normally distributed with the covariance matrix. 
\begin{equation}
	cov(\Phi_{jk}) = {\bf I}^{-1}(\Phi_{jk}), 
\end{equation}
where {\bf I} is the information matrix,  
\[ 
{\bf I}(\Phi_{jk}) = - \left( \begin{array}{cc}
\frac{\partial^2 \ln \mathcal{L}}{\partial \Phi_{ik} \Phi_{jk}} + \frac{\partial g}{\partial \Phi_{ik}}\frac{\partial g}{\partial \Phi_{jk}} & \frac{\partial g}{\Phi_{jk}} \\
\frac{\partial g}{\partial \Phi_{ik}} & 0  \end{array} \right).
\] 

The H$\alpha$ and M$_r$ LFs can be recovered from Eq.\,\ref{eq:phi_det} by summing over M$_r$ and L$_{H\alpha}$ respectively. For example, to recover the H$\alpha$ univariate LF, 
\begin{equation}
	\phi_{H\alpha} = \sum_{j=1}^{N_M} \phi_{jk}\times\Delta M. 
	\label{eq:recover_uniLF}
\end{equation}

\begin{figure}
	\includegraphics[width=0.5\textwidth]{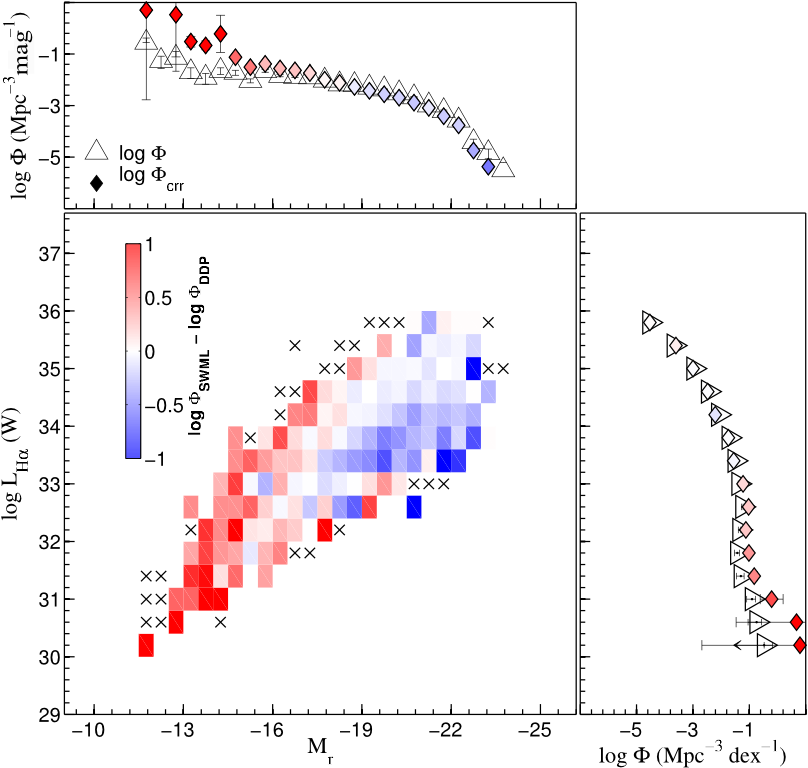}
	\caption{he residual map showing the difference in number densities calculated using the density corrected 1/V$_{\rm vmax}$ ($\log\,\Phi_{DDP}$) and SWML ($\log\,\Phi_{SWML}$)  methods for the $z<0.1$ bivariate functions. The crosses indicate where the galaxy number statistics are low. Top and right panels show the univariate LFs inferred from the bivariate functions.}
	\label{fig:res_map4}
\end{figure}
{Figure\,\ref{fig:res_map4} shows the residual between the bivariate LFs based on the density corrected V$_{\rm max}$ and bivariate SWML methods}. Again the differences are primarily in the faintest bins of the lowest redshift sample. {The differences here are greater than seen with the density corrected 1/V$_{\rm max}$ method. Given the small number of galaxies contributing to these faintest bins, the discrepancy here should be interpreted primarily as a limit on the systematic uncertainty of the measurement of the faint--end of the bivariate function, from all of these methods. }

\section{Bivariate and univariate functions}

The higher--$z$ bivariate L$_{H\alpha}$--$\mathcal{M}$ functions and the univariate LF are shown in Figures\,\ref{fig:SMFsinz_biLFbr_app} and \ref{fig:comprison_app}.

\begin{figure}
		\includegraphics[width=0.4\textwidth]{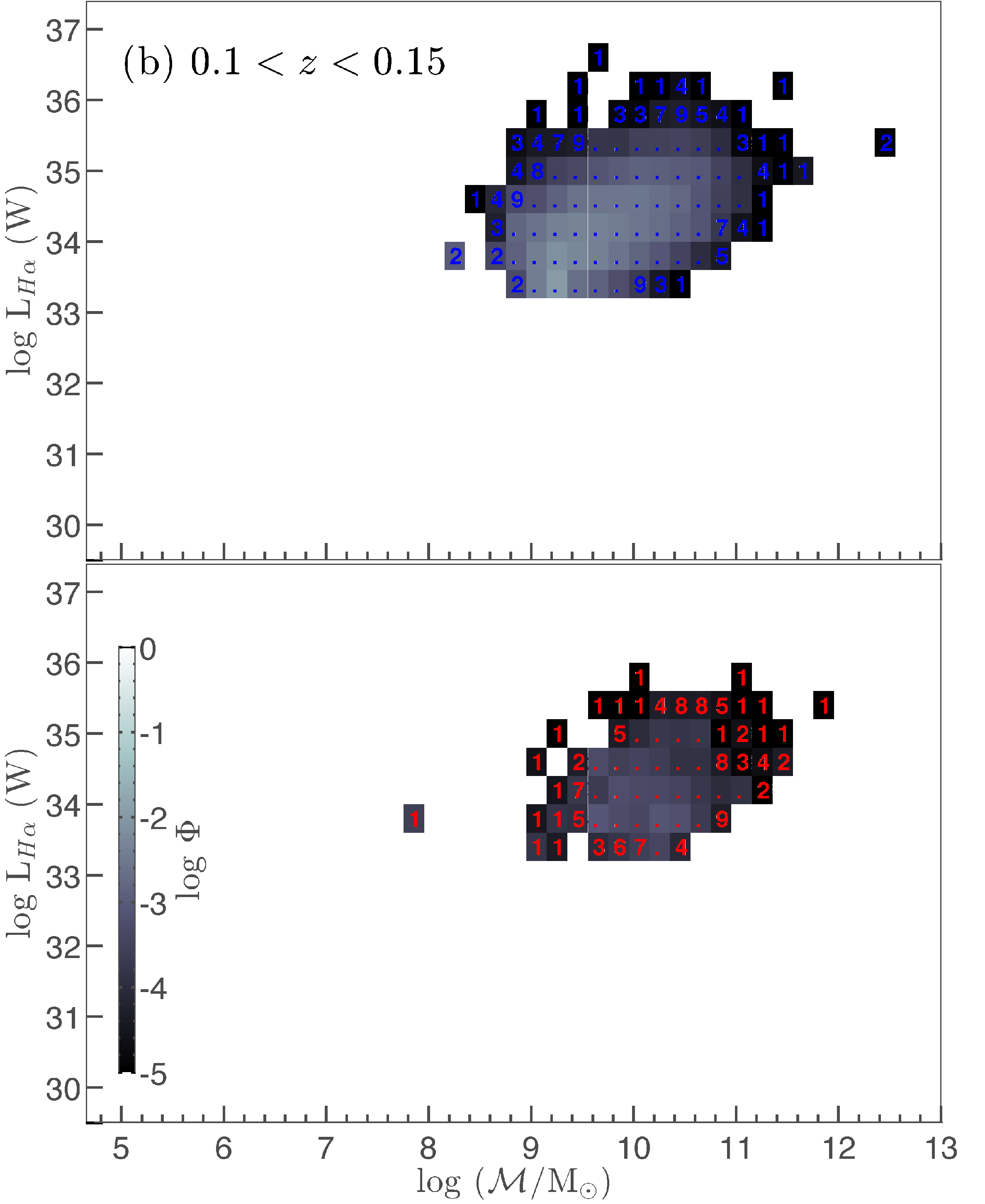}\\
		\includegraphics[width=0.4\textwidth]{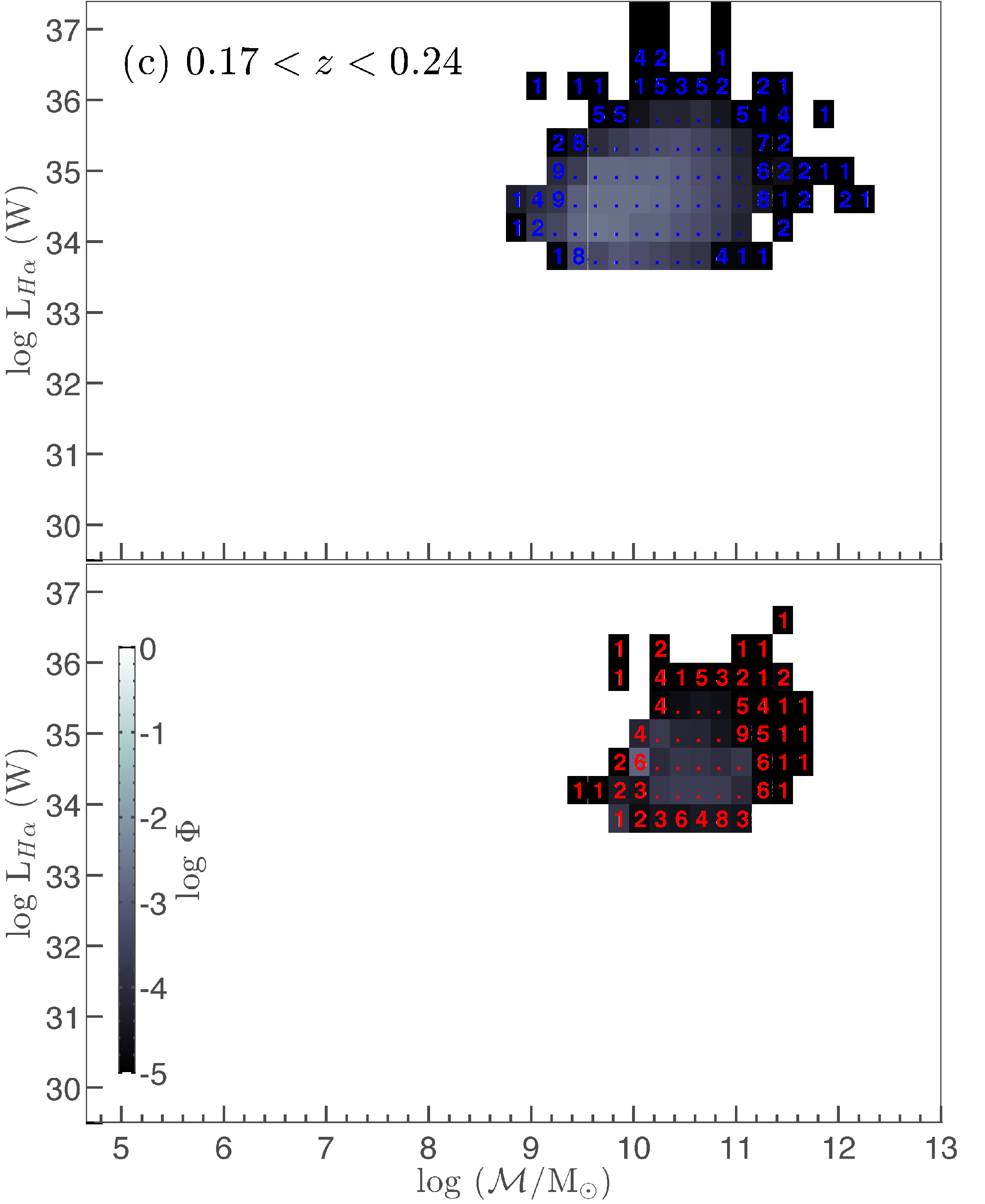}\\
		\includegraphics[width=0.4\textwidth]{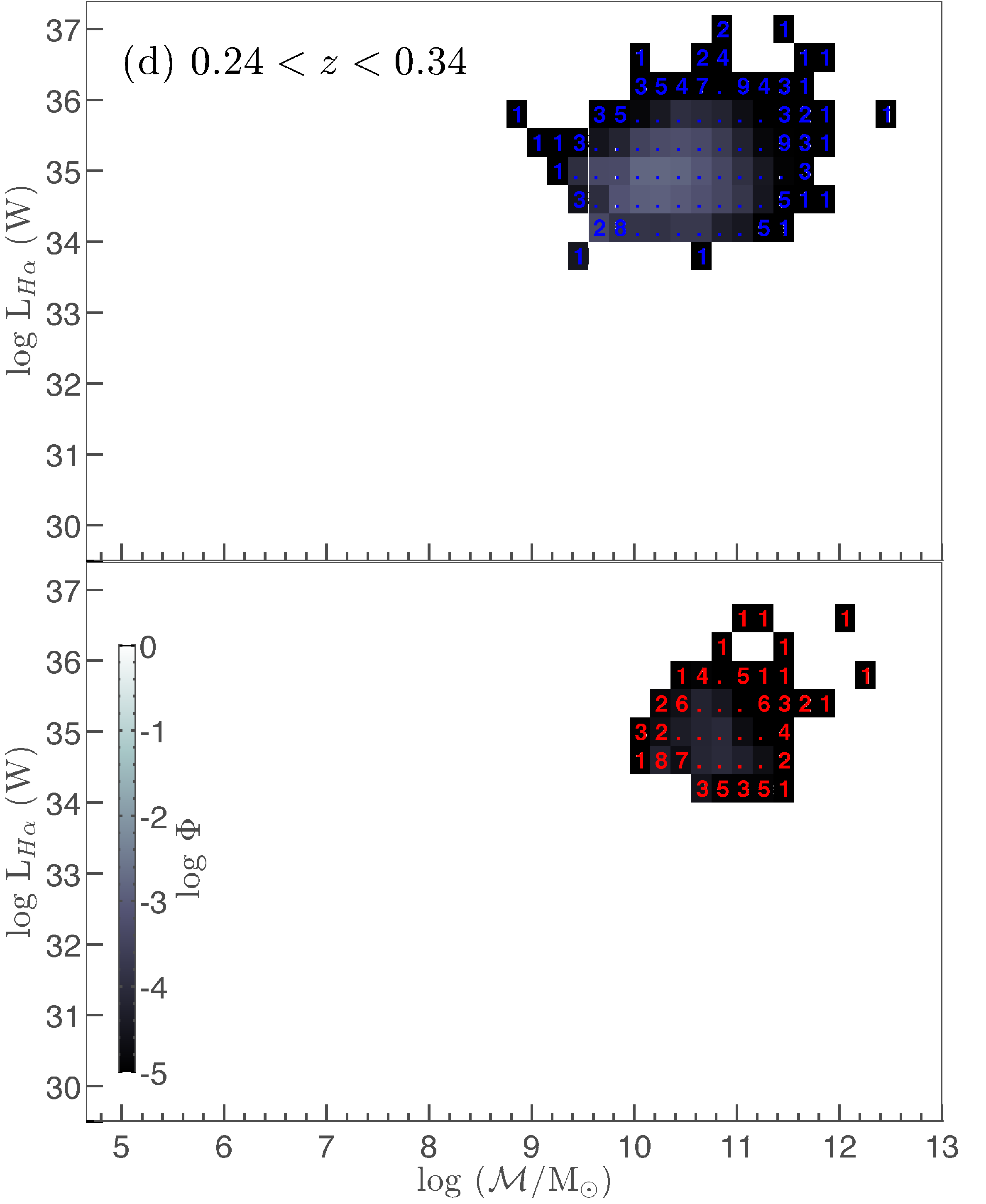}
		\caption{\color{black}The bivariate L$_{H\alpha}$--$\mathcal{M}$ functions of blue and red sub--populations split into four redshift ranges. The grey scale indicates the log number densities ($\Phi$) in the unit of Mpc$^{-3}$ dex$^{-2}$, and is valid for both blue and red bivariate functions corresponding to the given redshift range. }
		\label{fig:SMFsinz_biLFbr_app}
\end{figure}

\begin{figure}
	\begin{center}
		\includegraphics[trim=.7cm .05cm 3.2cm .1cm, clip=true, width=0.4\textwidth]{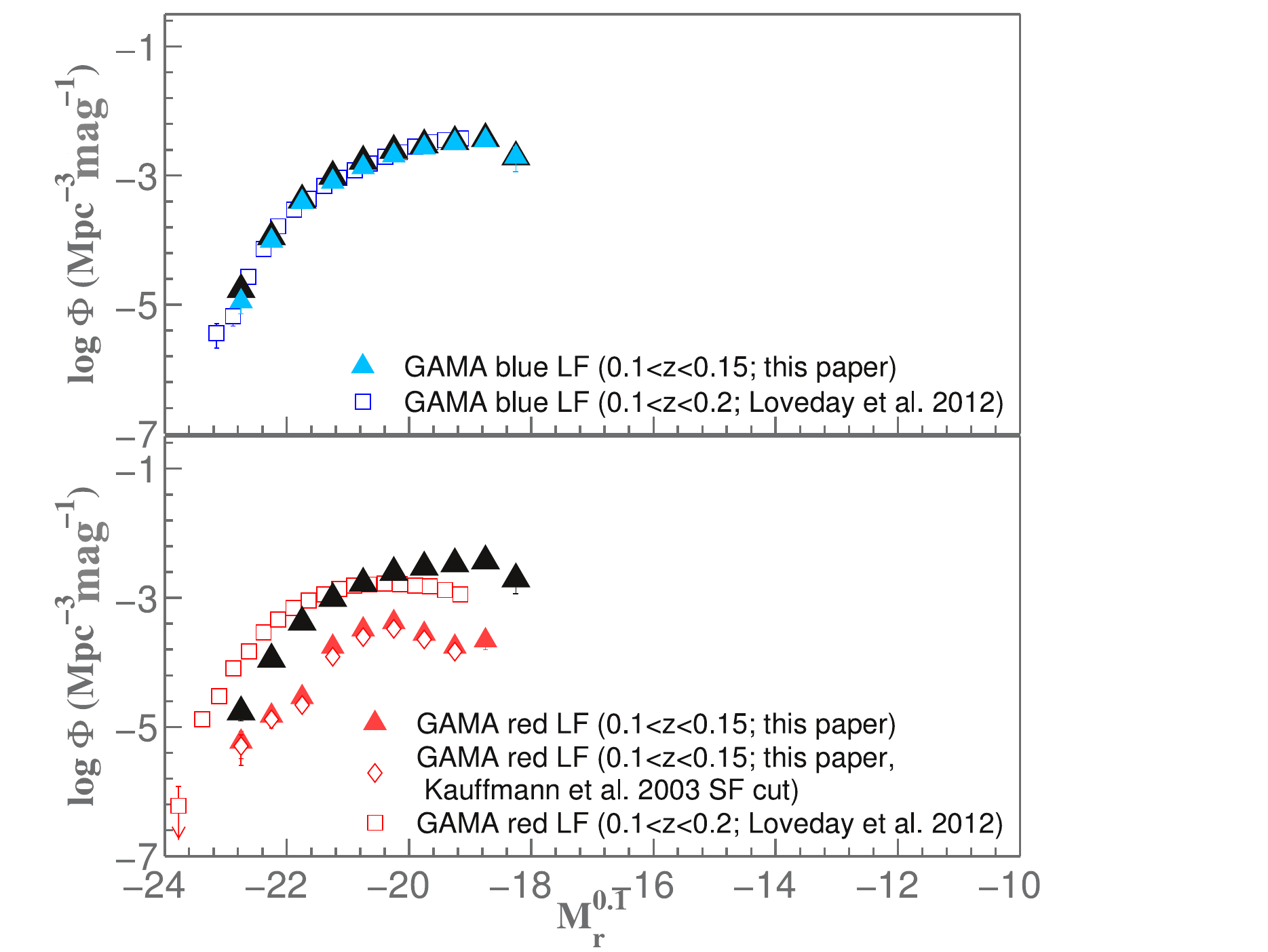}		
		\includegraphics[trim=.7cm .05cm 3.2cm .1cm, clip=true, width=0.4\textwidth]{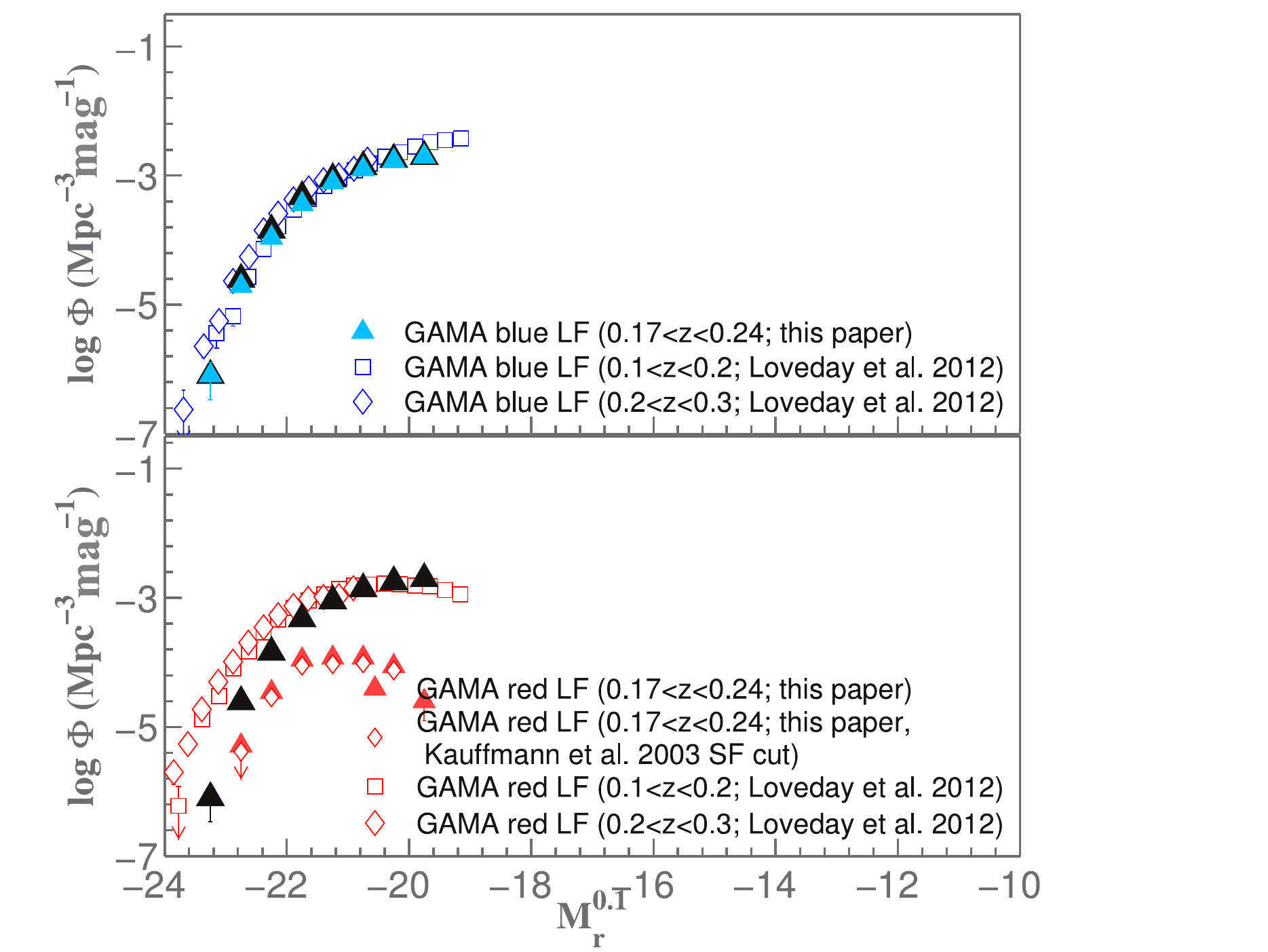}		
		\includegraphics[trim=.7cm .05cm 3.2cm .1cm, clip=true, width=0.4\textwidth]{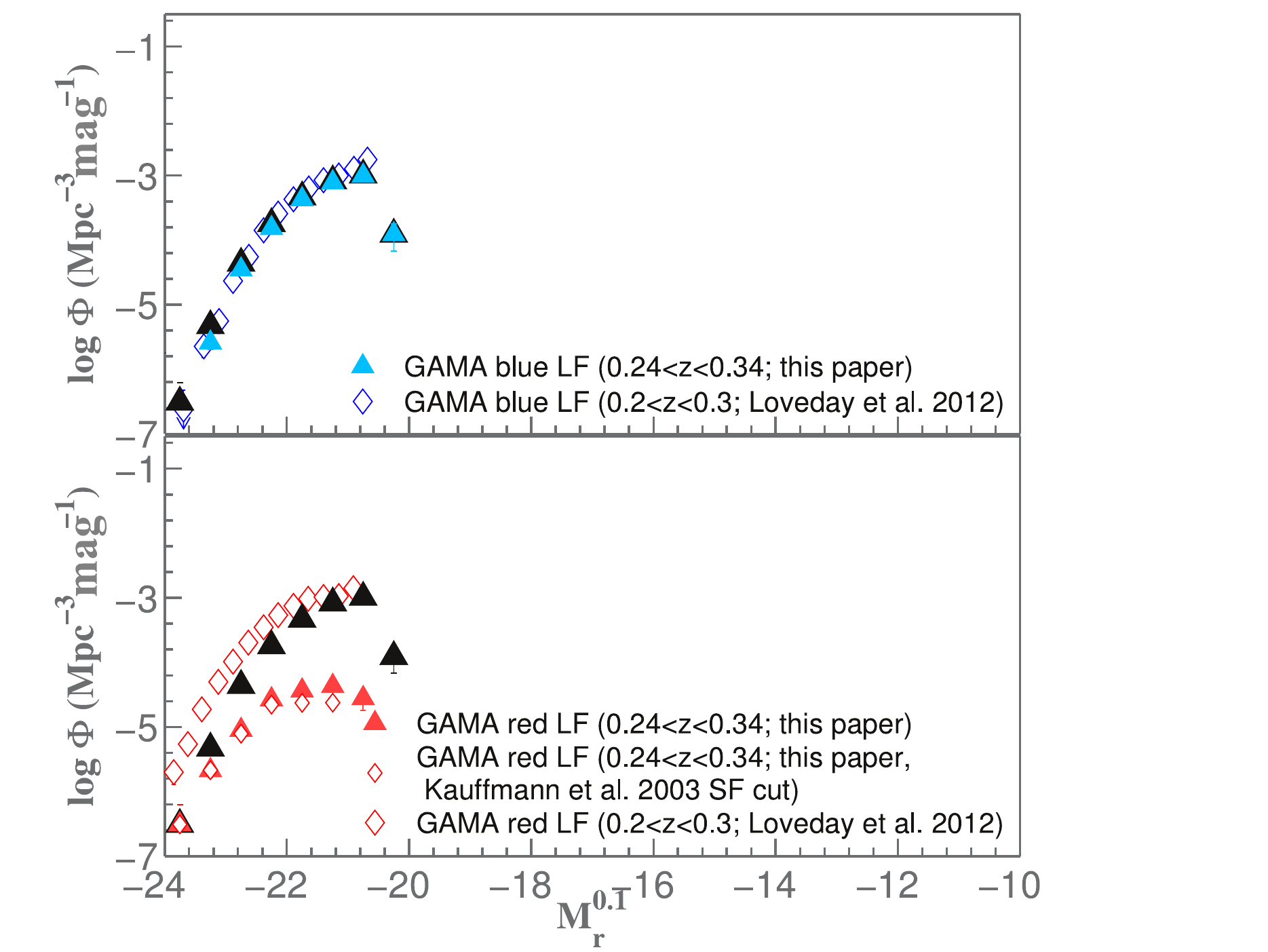}			
		\caption{Each two sets of panels from top to bottom show the blue and red SF M$_r$ functions derived from integrating the bivariate L$_{H\alpha}$--M$_r$ functions shown in Figure\,\ref{fig:vmax_bilf_photom} over L$_{H\alpha}$. The key indicate the redshift ranges probed. The all SF M$_r$ functions are shown in Figures\,\ref{fig:comprison1} and \ref{fig:comprison2}. The colour cut given in Eq.\,\ref{eq:color} used to classify galaxies as blue and red. The open red diamonds in each red SF M$_r$ functions panels indicate the red SF M$_r$ function constructed using the \citet{Kauffmann03} SF/AGN prescription instead of that of \citet{Kewley02}.}
		\label{fig:comprison_app}
	\end{center}
\end{figure}

}
\label{lastpage}

\end{document}